%% file: main.tex
\documentclass[journal]{IEEEtran}
\IEEEoverridecommandlockouts
\usepackage{newtxtext}
\usepackage[utf8]{inputenc}
\usepackage{amsmath,amssymb,amsfonts}
\usepackage{bm}
\usepackage{mathtools}
\usepackage[nocompress]{cite}
\usepackage{graphicx}
\usepackage{xcolor}
\usepackage{booktabs}
\usepackage{array}
\usepackage{hyperref}
\hypersetup{hidelinks}

\renewcommand{\IEEEQED}{}
\makeatletter
\g@addto@macro\normalsize{%
	\setlength{\abovedisplayskip}{3pt plus 1pt minus 2pt}%
	\setlength{\belowdisplayskip}{3pt plus 1pt minus 2pt}%
	\setlength{\abovedisplayshortskip}{0pt plus 1pt}%
	\setlength{\belowdisplayshortskip}{3pt plus 1pt minus 2pt}%
}
\input{tex/macros}

\usepackage{caption}
\newtheorem{remark}{Remark}
\usepackage{bm}
\usepackage{algorithm}
\usepackage{algorithmic}

\usepackage{tikz}
\usetikzlibrary{arrows.meta,positioning,fit,calc}
\usepackage{placeins}
\usepackage{caption}
\newtheorem{prop}{Proposition}
\definecolor{red}{RGB}{0,0,0}
\begin{document}
\bstctlcite{IEEEReferenceFormat2025}
\nocite{Liu2018TowardDFRC,Liu2022ISAC6G,Mao2022WaveformMillimeterTHz,Zhang2021OverviewJCR,Liu2022FundamentalLimitsISAC,Zhou2022ISACWaveformSurvey,Blunt2010IntrapulseREC,Ciuonzo2015IntrapulseMOO,Hassanien2016SidelobeWaveformDiversity,Yang2020DualUseAF,Sturm2011WaveformFusion,Braun2014OFDMRadarAlgorithms,Cui2022NRWaveformISAC,Xiao2022FullDuplexISAC,Li2007MIMORadarColocated,Yu2022SpatioSpectral,Yao2023IntrapulseIM,Yao2024NarrowbandOrthogonal,Chen2008FHMIMOAmbiguity,Baxter2018FHCS,Wu2021ChannelEstimation,Baxter2022JointRadarComFH,Wu2022Overview,Baxter2020DictionarySelection,Aboutanios2023ImprovedFHCS,Wu2021Reliable,Wu2022Secure,Eedara2019DPSK,Eedara2022CSK,Xu2023HIM,Liu2025CRM,Wang2020PhaseModulatedFHMIMO,Eedara2021TargetDetection,Chen2025Processing,Eedara2020GA,Eedara2020Clutter,Alabaster2012,Barton2004,Wang2025HybridFHPC,Li2024InterPulseAgile,Lellouch2015SteppedOFDM,Anderson1986DigitalPhaseModulation,Sundberg1986CPM,McCormick2017WidebandMIMO,Scharf1994MatchedSubspace,Graham1994ConcreteMathematics,Shoup2008NumberTheory}

\title{Continuous-Phase Frequency-Hopping MIMO Signaling and Processing for ISAC}
\author{%
	Yi Wang, Xianxiang Yu${}^{*}$, Jing Yang, Tao Fan, Linlong Wu, Xiangrong Wang, and Guolong Cui%
	\thanks{This work was supported in part by the National Natural Science Foundation of China under Grant 62571099, Grant U24B20188, and Grant 62271126; in part by the 111 Project under Grant B17008; in part by the 9th Youth Talent Support Project of Chinese Institute of Electronics under Grant M11NOYESS20230488; and in part by the Natural Science Foundation of Sichuan Province under Grant 2026NSFSC1418. (Corresponding author: Xianxiang Yu.)}
	\thanks{Yi Wang, Xianxiang Yu, Tao Fan, Linlong Wu, and Guolong Cui are with the School of Information and Communication Engineering, University of Electronic Science and Technology of China, Chengdu 611731, China (e-mail: yiwang@std.uestc.edu.cn; xianxiangyu@uestc.edu.cn; thaumielfan@gmail.com; linlong.wu@uestc.edu.cn; cuiguolong@uestc.edu.cn).}
	\thanks{Jing Yang is with the School of Electrical and Information Engineering, Zhengzhou University, Zhengzhou 450001, China (e-mail: yangjinguestc@163.com).}
	\thanks{Xiangrong Wang is with the School of Electronic and Information Engineering, Beihang University, Beijing 100191, China (e-mail: xrwang@buaa.edu.cn).}
}

\maketitle

\begin{abstract}
\textcolor{red}{Frequency-hopping multiple-input multiple-output integrated sensing and communications (FH-MIMO-ISAC) supports waveform agility and multidimensional information embedding, but still faces challenges in phase-continuous transmission, demodulation without channel-phase prior information, and coherent processing of multi-frequency echoes. This paper proposes a continuous-phase frequency-hopping (CP-FH) MIMO ISAC signal and corresponding communication and radar receive processing methods. The signal jointly conveys information through frequency combinations and continuous-phase-modulation (CPM) symbols. CPM phase accumulation and frequency-hopping linear-frequency-modulation (FH-LFM) phase compensation ensure phase continuity between subpulses, reducing spectral leakage caused by boundary phase jumps. Theoretical derivations establish exact inter-element waveform orthogonality conditions, residual cross-correlation bounds, and an upper bound on the averaged transmit-beampattern ripple. For communication reception, projection-energy demodulation (PED) requires no accurate channel-phase prior information and approximates the least-squares detection metric by the sum of matched-projection energies under weak inter-branch correlation. Decoupling the frequency-combination search from CPM-symbol decisions reduces complexity while yielding the same PED decisions as direct joint search. For radar reception, structure-matched coherent (SMC) processing forms a three-dimensional coarse-range--velocity--fine-range response through space--time--frequency joint matched filtering, velocity-consistent Doppler focusing, and frequency-aperture focusing, reducing Doppler-mismatch-induced coherent signal loss and extra detections caused by periodic range grating lobes in conventional pulse-Doppler (PD) processing. The joint range response, output signal and noise powers, and effective receive beampattern are further derived. The theoretical analysis characterizes target-discrimination limitations under periodic ambiguity and shows that SMC preserves the range-resolving capability of the frequency aperture, maintains the same output noise power as PD under the applicable branch-noise conditions, and preserves the angular-resolving capability of the full aperture under colocated-array and matched-beamforming assumptions.}

\end{abstract}

\begin{IEEEkeywords}
	Integrated sensing and communications (ISAC), frequency-hopping MIMO,
	continuous-phase frequency hopping (CP-FH), noncoherent demodulation,
	 structure-matched coherent processing.
\end{IEEEkeywords}

\input{sections/00_introduction}

\input{sections/01_signal_model}

\input{sections/02_communication}

\input{sections/03_radar}

\input{sections/04_simulation}

\input{sections/99_conclusion}
\flushbottom 


\bibliographystyle{IEEEtran}
\begingroup
\makeatletter
\expandafter\def\csname revisedref@Wang2025HybridFHPC\endcsname{}
\expandafter\def\csname revisedref@Li2024InterPulseAgile\endcsname{}
\expandafter\def\csname revisedref@Lellouch2015SteppedOFDM\endcsname{}
\expandafter\def\csname revisedref@McCormick2017WidebandMIMO\endcsname{}
\expandafter\def\csname revisedref@Scharf1994MatchedSubspace\endcsname{}
\expandafter\def\csname revisedref@Shoup2008NumberTheory\endcsname{}
\let\revision@bibitem\@bibitem
\def\@bibitem#1{%
  \par\ifcsname revisedref@#1\endcsname\color{red}\else\color{black}\fi
  \revision@bibitem{#1}}
\makeatother
\bibliography{bib/refs}
\endgroup
\clearpage
\begingroup
\renewcommand{\baselinestretch}{0.9}
\let\appendixnormalsize\normalsize
\renewcommand{\normalsize}{%
  \appendixnormalsize
  \setlength{\jot}{0pt}%
  \setlength{\abovedisplayskip}{1.5pt plus 0.5pt minus 0.5pt}%
  \setlength{\belowdisplayskip}{1.5pt plus 0.5pt minus 0.5pt}%
  \setlength{\abovedisplayshortskip}{0pt}%
  \setlength{\belowdisplayshortskip}{1.5pt plus 0.5pt minus 0.5pt}%
}
\normalsize
\input{sections/appendices}
\par
\endgroup
\end{document}

%% file: tex/macros.tex
\providecommand{\CM}{\mathcal{M}}

\providecommand{\CU}{\mathcal{U}}


%% file: sections/00_introduction.tex
\section{Introduction}
\label{sec:intro}

\begingroup\color{red}
Integrated sensing and communications (ISAC) has attracted extensive attention
by integrating sensing and communication functions on a common platform
through shared spectrum and hardware resources
\cite{Liu2018TowardDFRC,Liu2022ISAC6G,Mao2022WaveformMillimeterTHz,
Zhang2021OverviewJCR,Liu2022FundamentalLimitsISAC,Zhou2022ISACWaveformSurvey}.
Existing waveform designs mainly follow radar-centric, communication-centric,
and joint design approaches. The first two integrate the functions by embedding
communication information into radar signals
\cite{Blunt2010IntrapulseREC,Ciuonzo2015IntrapulseMOO,
Hassanien2016SidelobeWaveformDiversity,Yang2020DualUseAF}
or enabling sensing with communication waveforms
\cite{Sturm2011WaveformFusion,Braun2014OFDMRadarAlgorithms,Cui2022NRWaveformISAC},
respectively, whereas joint designs balance sensing and communication requirements
\cite{Liu2018TowardDFRC,Mao2022WaveformMillimeterTHz,Xiao2022FullDuplexISAC}.
Building on these approaches, multiple-input multiple-output (MIMO) technology
introduces spatial degrees of freedom (DoFs), providing greater flexibility
for waveform design, beampattern control, and information embedding
\cite{Li2007MIMORadarColocated,Yu2022SpatioSpectral,
Yao2023IntrapulseIM,Yao2024NarrowbandOrthogonal}.
\endgroup

In this context, frequency-hopping MIMO (FH-MIMO) has been widely investigated as a waveform framework for ISAC systems because it combines frequency
agility, MIMO spatial DoFs, and flexible fast-time resources for information
embedding
\cite{Chen2008FHMIMOAmbiguity,Baxter2018FHCS,
	Wu2021ChannelEstimation,Baxter2022JointRadarComFH,Wu2022Overview}.
Early work mainly investigated the ambiguity properties of FH-MIMO waveforms
and the associated hopping-code optimization problem from a radar perspective
\cite{Chen2008FHMIMOAmbiguity}. Frequency-hopping code selection (FHCS) was
then proposed to map hopping codes to communication symbols and realize
subpulse-level information embedding
\cite{Baxter2018FHCS,Baxter2022JointRadarComFH}. Related studies further
examined the effects of symbol-dictionary selection and code constraints on
ambiguity-function characteristics and waveform degradation
\cite{Baxter2020DictionarySelection,Aboutanios2023ImprovedFHCS}. These studies
established the basic construction of FH-MIMO dual-function signals and provided
the foundation for subsequent developments in communication embedding and
receiver design.

Research on FH-MIMO communication design has mainly advanced along two related directions. One direction focuses on reliable reception, channel estimation, multi-antenna reception, interference suppression, timing-offset estimation, and secure hopping permutation design
\cite{Wu2021ChannelEstimation,Wu2021Reliable,Wu2022Secure,Wu2022Overview}.
The other direction increases the information payload by introducing additional
embedding resources, such as phase, codeword, antenna index, and
constellation-rotation based mechanisms
\cite{Eedara2019DPSK,Eedara2022CSK,Xu2023HIM,Liu2025CRM}. These designs
improve the communication capability of FH-MIMO dual-function signals.
\textcolor{red}{Among these designs, some phase-modulation schemes employing coherent
detection require accurate channel-phase information or prior calibration
\cite{Wang2020PhaseModulatedFHMIMO,Baxter2022JointRadarComFH}.} In addition,
subpulse-wise frequency and phase switching may introduce phase discontinuities,
which can aggravate sidelobe modulation and spectral leakage
\cite{Eedara2021TargetDetection,Chen2025Processing}.

For radar sensing, existing FH-MIMO dual-function studies have mainly examined
the sensing impact of communication embedding and the applicability of
conventional radar processing chains. Optimized hopping codes and modulation
sequences have been used to suppress range sidelobes and sidelobe modulation
and to analyze clutter responses
\cite{Eedara2020GA,Eedara2020Clutter}. Target detection and subpulse-wise
log-likelihood processing have also been studied for specific FH-MIMO
dual-function waveforms
\cite{Eedara2021TargetDetection,Chen2025Processing}. 
However, a separate receiver-side issue arises from the discrete multi-frequency
aperture itself. Although conventional pulse-Doppler (PD) processing can be
applied \cite{Alabaster2012,Barton2004,Chen2025Processing}, it mainly provides
a coarse range--velocity response and does not explicitly synthesize the
carrier-dependent phase structure across frequency branches. As a result, the
periodic range response of the discrete frequency aperture may produce range
grating lobes around a physical target, leading to strong-to-weak target masking
and unreliable detection.

\textcolor{red}{To address these issues, a continuous-phase frequency-hopping (CP-FH) MIMO signal and corresponding communication demodulation and radar processing methods are proposed, extending the preliminary conference work in \cite{Wang2025HybridFHPC}.} The main contributions are summarized as follows.

\begin{enumerate}
	\item \textcolor{red}{A CP-FH MIMO ISAC signal construction mechanism is proposed in which frequency combinations and CPM symbols jointly carry communication information, while the LFM bandwidth controls the coarse-range resolution. CPM phase accumulation and FH-LFM phase compensation ensure phase continuity between subpulses, suppressing spectral leakage caused by boundary phase jumps. The exact subpulse-level orthogonality conditions for waveforms transmitted by different antennas and the frequency-spacing condition for controlling residual cross-correlation are further established, and a bound on the resulting ripple in the averaged transmit beampattern is derived.}
	
	\item \textcolor{red}{A projection-energy demodulation (PED) method is proposed, requiring no accurate channel-phase prior information. By projecting the received signal onto candidate frequency--CPM waveforms, the method approximates the least-squares detection metric with the sum of branch-wise matched-projection energies under weak inter-branch correlation. The frequency-combination search is further decoupled from CPM-symbol decisions, reducing demodulation complexity while yielding the same PED decisions as direct joint search.}
	
	\item \textcolor{red}{A structure-matched coherent (SMC) method is proposed for multi-frequency FH-MIMO echoes. Space--time--frequency joint matched filtering, velocity-consistent Doppler focusing, and frequency-aperture focusing form a three-dimensional coarse-range--velocity--fine-range response, reducing extra detections caused by PD's periodic range grating lobes. Unlike range-aided \cite{Li2024InterPulseAgile} or velocity-search-based \cite{Lellouch2015SteppedOFDM} methods, SMC needs no target-range or velocity priors and reuses velocity-cell coordinates for frequency focusing without additional search. The derived approximate joint range response gives the widths and number of 3-dB unresolved intervals, characterizing frequency-aperture resolution and target-discrimination limits under periodic ambiguity. Output signal and noise comparisons show that branch-dependent Doppler compensation reduces PD's coherent signal loss with equal output noise power under the stated branch-noise conditions. The derived receive beampattern restores full-aperture coherent spatial response and preserves angular resolution under colocated-array and matched-beamforming assumptions. Quantified residual-phase and grid-mismatch losses assess model approximation errors and guide processing-grid selection.}
\end{enumerate}

Notations: Scalars, vectors, and matrices are denoted by italic letters, bold
italic lower-case letters, and bold italic upper-case letters, respectively.
The superscripts \((\cdot)^{\mathsf T}\), \((\cdot)^{\mathsf H}\), and
\((\cdot)^*\) denote transpose, conjugate transpose, and complex conjugate.
For two continuous-time signals \(x(t)\) and \(y(t)\), their inner product over
an interval \(\mathcal I\) is defined as
\(\langle x,y\rangle_{\mathcal I}=\int_{\mathcal I}x(t)y^*(t)\,dt\). The symbols \(|\cdot|\), \(\Re\{\cdot\}\), \(\mathrm{tr}(\cdot)\), and
\(\mathbb{E}\{\cdot\}\) denote modulus, real part, trace, and expectation.
\textcolor{red}{The indicator function \(\mathbf{1}\{\mathcal A\}\) equals
one if the condition \(\mathcal A\) holds and zero otherwise.}
\textcolor{red}{For \(x\in\mathbb R\), \([x]_+=\max(x,0)\) denotes its positive part.}
The sets of real and complex numbers are denoted by \(\mathbb{R}\) and
\(\mathbb{C}\). \textcolor{red}{The spaces \(\mathbb R^{M\times N}\) and
\(\mathbb C^{M\times N}\) consist of real and complex \(M\times N\)
matrices, respectively.} For a set \(\mathcal S\), \(|\mathcal S|\) denotes its
cardinality, and \(\binom{K}{N}\) denotes the number of \(N\)-element
combinations selected from a \(K\)-element set.
\begingroup\color{red}
The operator \(\gcd(\cdot)\) denotes the greatest common divisor.
\endgroup

%% file: sections/01_signal_model.tex
\section{Signal Model and Waveform Properties}
\label{sec:signal-model}

\subsection{Signal Model}

Consider a colocated MIMO--ISAC transmitter equipped with an \(N\)-element
uniform linear array (ULA) with half-wavelength spacing \(d=\lambda_c/2\),
where \(\lambda_c=c/f_c\) is the wavelength at the reference carrier frequency
\(f_c\). A coherent processing interval (CPI) contains \(P\) pulses, and each
pulse consists of \(V\) non-overlapping subpulses of duration \(\Delta t\),
yielding the pulse duration \(T=V\Delta t\). \begingroup\color{red}
For the \(p\)-th PRI, let \(t\in[0,T_{\mathrm r})\) denote the fast time
measured from the start of the PRI, and let
\(t_p\triangleq(p-1)T_{\mathrm r}\) denote its start time relative to the
beginning of the CPI. The corresponding absolute time is \(t_p+t\). The
transmit carriers and receiver downconversion references are assumed phase
coherent over the entire CPI.
\endgroup

For antenna \(n\) and subpulse \(v\), the selected frequency offset and CPM
symbol are denoted by \(f_{n,v}\in\mathcal F\) and
\(\alpha_{n,v}\in\mathcal A\), respectively. The frequency-offset pool is
\(\mathcal F=\{f_1,\ldots,f_K\}\), with
\(f_k=f_0+(k-1)\Delta f\), where \(f_0\) is the initial offset and
\(\Delta f\) is the frequency spacing. The RF carrier associated with the
\(k\)-th offset is \(\tilde f_k=f_c+f_k\). The CPM-symbol alphabet is
\(\mathcal A=\{\alpha_1,\ldots,\alpha_M\}\). The number of candidate frequency
offsets satisfies \(K\ge N\), and distinct antennas use distinct frequency
offsets within the same subpulse, i.e., \(f_{n,v}\ne f_{n',v}\) for
\(n\ne n'\).

\subsubsection{Element-wise CP-FH waveform}

Let
\(\boldsymbol{\alpha}_n=[\alpha_{n,1},\ldots,\alpha_{n,V}]^{\mathsf T}\) and
\(\boldsymbol f_n=[f_{n,1},\ldots,f_{n,V}]^{\mathsf T}\). The baseband waveform
transmitted by the \(n\)-th element is written as
\begin{equation}
	s_n(t;\boldsymbol{\alpha}_n,\boldsymbol f_n)
	=
	s_n^{\mathrm{cpm}}(t;\boldsymbol{\alpha}_n)
	s_n^{\mathrm{fh}}(t;\boldsymbol f_n),
	\label{eq:sm_sn_product}
\end{equation}
which consists of the CPM component
\(s_n^{\mathrm{cpm}}(t;\boldsymbol{\alpha}_n)\) and the FH--LFM component
\(s_n^{\mathrm{fh}}(t;\boldsymbol f_n)\).

The CPM component follows the standard continuous phase modulation model
\cite{Anderson1986DigitalPhaseModulation,Sundberg1986CPM} and is given by
\begin{equation}
	s_n^{\mathrm{cpm}}(t;\boldsymbol{\alpha}_n)
	=
	\exp\bigl\{
	j\phi_n^{\mathrm{cpm}}(t;\boldsymbol{\alpha}_n)
	\bigr\},
	\label{eq:sm_cpm_component}
\end{equation}
where
\begin{equation}
	\phi_n^{\mathrm{cpm}}(t;\boldsymbol{\alpha}_n)
	=
	2\pi\bar h
	\sum_{v=1}^{V}
	\alpha_{n,v}q\bigl(t-(v-1)\Delta t\bigr).
	\label{eq:sm_cpm_phase_single}
\end{equation}
Here, \(\bar h \in \mathbb Q\) is the CPM modulation index, and \(q(t)\) is the rectangular
full-response CPM phase pulse defined as
\begin{equation}
	q(t)
	=
	\begin{cases}
		0, & t<0,\\
		t/(2\Delta t), & 0\le t<\Delta t,\\
		1/2, & t\ge \Delta t.
	\end{cases}
	\label{eq:sm_q_def}
\end{equation}
Thus, each subpulse of duration \(\Delta t\) carries one CPM symbol.

The FH--LFM component of the \(n\)-th transmit waveform is written as
\begin{equation}
	s_n^{\mathrm{fh}}(t;\boldsymbol f_n)
	=
	\sum_{v=1}^{V}
	\exp\bigl\{
	j\phi_{n,v}^{\mathrm{fh}}(t;f_{n,v})
	\bigr\}w_v(t),
	\label{eq:sm_fh_sum}
\end{equation}
where \(w_v(t)\) denotes the \(v\)-th subpulse window:
\begin{equation}
	w_v(t)
	=
	\begin{cases}
		1, & t\in[(v-1)\Delta t,v\Delta t),\\
		0, & \mathrm{otherwise}.
	\end{cases}
	\label{eq:sm_window}
\end{equation}
Let \(t_v=t-(v-1)\Delta t\) denote the local time within the \(v\)-th subpulse.
The corresponding FH--LFM phase is
\begin{equation}
	\phi_{n,v}^{\mathrm{fh}}(t;f_{n,v})
	=
	2\pi
	\left(
	f_{n,v}t_v
	+
	\frac{1}{2}It_v^2
	+
	C_{n,v}
	\right).
	\label{eq:sm_fh_phase}
\end{equation}
Here, \(I\) is the chirp slope, and \(C_{n,v}\) denotes the accumulated FH--LFM
phase offset at the beginning of the \(v\)-th subpulse. It is chosen recursively
as
\begin{equation}
	C_{n,v}
	=
	\begin{cases}
		0, & v=1,\\[3pt]
		C_{n,v-1}+f_{n,v-1}\Delta t+\dfrac{1}{2}I(\Delta t)^2,
		& v\ge2 .
	\end{cases}
	\label{eq:sm_Cnv_recursion}
\end{equation}
With this recursion, the FH--LFM phase evolves continuously across subpulse
boundaries.

\begin{remark}[Roles of CP-FH components]
	The roles of the CP-FH components are summarized as follows.
	\begin{itemize}
		\item \emph{Information embedding:} The frequency offsets \(f_{n,v}\in\mathcal F\)
		and CPM symbols \(\alpha_{n,v}\in\mathcal A\) carry information through FH selection and CPM phase modulation, respectively.
		\item \emph{Radar resolution:} The chirp slope \(I\) gives the
		within-subpulse bandwidth $B_{\mathrm{chirp}}\!=\!|I|\Delta t$, which sets the
		coarse pulse-compression resolution. The frequency pool
		\(\mathcal F=\{f_1,\ldots,f_K\}\) provides the frequency aperture for
		fine-range discrimination, as characterized by \eqref{eq:R3dB_approx_simple_final}.
		
		\item \emph{Phase continuity and envelope:} The accumulated CPM phase and
		\(C_{n,v}\) preserve inter-subpulse phase continuity and reduce
		phase-jump-induced spectral leakage, while the unit-modulus CPM and FH--LFM
		factors yield a constant-envelope waveform.
		
	\end{itemize}
\end{remark}

\subsubsection{Directional Transmit Signal}

Based on the element-wise waveform in \eqref{eq:sm_sn_product}, the directional
transmit signal toward azimuth \(\theta\) is written as
\begin{equation}
	x(t;\theta)
	=
	\textcolor{red}{\boldsymbol a_{\mathrm{tx}}}^{\mathsf H}(\theta)
	\boldsymbol s(t),
	\label{eq:directional_tx_signal}
\end{equation}
where
\begin{equation}
	\boldsymbol s(t)
	\triangleq
	\left[
	s_1(t;\boldsymbol{\alpha}_1,\boldsymbol f_1),
	\ldots,
	s_N(t;\boldsymbol{\alpha}_N,\boldsymbol f_N)
	\right]^{\mathsf T},
	\label{eq:tx_waveform_vector}
\end{equation}
and
\begin{equation}
	\textcolor{red}{\boldsymbol a_{\mathrm{tx}}}(\theta)
	\triangleq
	\left[
	1,
	e^{-j\frac{2\pi d}{\lambda_c}\sin\theta},
	\ldots,
	e^{-j\frac{2\pi d}{\lambda_c}(N-1)\sin\theta}
	\right]^{\mathsf T}
	\label{eq:tx_steering_vector}
\end{equation}
is the transmit steering vector.
\textcolor{red}{The spatial phase is evaluated at the reference carrier
\(f_c\) under the narrowband array approximation, which requires
\(B\Delta\tau_{\mathrm{arr,max}}\ll1\)~\cite{McCormick2017WidebandMIMO}.
Here, \(\Delta\tau_{\mathrm{arr,max}}=(N-1)d/c\) is the maximum
propagation-time difference between array elements, and \(B\) is the
signal bandwidth accounting for the hopping span and intrabranch
modulation. Its magnitude is estimated using the reference band defined
in Section~\ref{subsubsec:spectral_leakage}.}

\emph{Frequency-indexed directional components:}
For later structure-matched processing, the directional transmit signal is
decomposed into frequency-indexed components as

\begin{equation}
	x_k(t;\theta)
	=
	\sum_{v=1}^{V}
	\sum_{n=1}^{N}
	s_n(t;\boldsymbol{\alpha}_n,\boldsymbol f_n)
	\mu_{n,v,k}(\theta)
	w_v(t),
	\label{eq:xk_directional_component}
\end{equation}
where
\begin{equation}
	\mu_{n,v,k}(\theta)
	=
	\begin{cases}
		\exp\left(
		j\frac{2\pi d}{\lambda_c}(n-1)\sin\theta
		\right),
		& f_{n,v}=f_k,\\
		0,
		& f_{n,v}\ne f_k .
	\end{cases}
	\label{eq:mu_no_p}
\end{equation}
Here, \(w_v(t)\) extracts the \(v\)-th subpulse segment, and
\(\mu_{n,v,k}(\theta)\) retains the antenna--subpulse component occupying
\(f_k\) while applying its transmit spatial phase.

\begingroup\color{red}
For CPI-level processing, the pulse-indexed waveforms \(s_{p,n}(t)\) and
\(x_{p,k}(t;\theta)\) are obtained by replacing \(f_{n,v}\) and
\(\mu_{n,v,k}(\theta)\) with \(f_{p,n,v}\) and
\(\mu_{p,n,v,k}(\theta)\), respectively.
\endgroup

\subsection{Waveform Properties}
\label{subsec:waveform_properties}

This subsection summarizes two basic properties of the proposed CP-FH signal:
the element-wise residual correlation and the averaged transmit beampattern.
The former determines the waveform orthogonality condition, while the latter
shows how the remaining residual correlations affect the spatial response.

For the \(v\)-th subpulse, the normalized correlation between the \(n\)-th and
\(m\)-th transmit elements is defined as
\begin{equation}
	\rho_{n,m}^{(v)}
	\triangleq
	\frac{1}{E_{\mathrm s}}
	\int_{(v-1)\Delta t}^{v\Delta t}
	s_n(t;\boldsymbol{\alpha}_n,\boldsymbol f_n)
	s_m^*(t;\boldsymbol{\alpha}_m,\boldsymbol f_m)
	\,\mathrm d t,
	\label{eq:rho_nm_v}
\end{equation}
where
\(E_{\mathrm s}\triangleq
\int_{(v-1)\Delta t}^{v\Delta t}
|s_n(t;\boldsymbol{\alpha}_n,\boldsymbol f_n)|^2\,\mathrm d t
=\Delta t\)
is the subpulse energy due to the unit-modulus property of the CP-FH waveform.

\subsubsection{Subpulse-Level Orthogonality Conditions}
\label{subsubsec:correlation_orthogonality_conditions}

Evaluating \eqref{eq:rho_nm_v} for the proposed CP-FH waveform yields the
following orthogonality conditions.

\begin{prop}
	\label{prop:qo_corr_bound}
	Letting
	\(
	\Delta f_{nm,v}\triangleq f_{n,v}-f_{m,v}
	\),
	\(
	\Delta\alpha_{nm,v}\triangleq \alpha_{n,v}-\alpha_{m,v}
	\) and \(\Delta\alpha_{\max}
	\triangleq
	\max_{\alpha,\alpha'\in\mathcal A}
	|\alpha-\alpha'|\),
	for \(n\ne m\) and \(v=1,\ldots,V\), the following results hold.
	\begin{itemize}
		\item \emph{Exact orthogonality} holds if
		\begin{equation}
			\Delta f_{nm,v}\Delta t
			+
			\frac{\bar h}{2}\Delta\alpha_{nm,v}
			\in
			\mathbb Z\setminus\{0\}.
			\label{eq:qo_exact_orthogonality_condition}
		\end{equation}
		
		\item \emph{Bounded residual correlation} with
		\(
		|\rho_{n,m}^{(v)}|\le\varepsilon
		\)
		is guaranteed by the sufficient spacing rule
		\begin{equation}
			\Delta f\Delta t
			\ge
			\frac{1}{\pi\varepsilon}
			+
			\frac{|\bar h|}{2}
			\Delta\alpha_{\max}.
			\label{eq:qo_min_spacing_rule_main}
		\end{equation}
	\end{itemize}
\end{prop}

\begin{IEEEproof}
	See Appendix~\ref{app:proof_prop_qo_corr_bound}.
\end{IEEEproof}

Proposition~\ref{prop:qo_corr_bound} shows that the inter-element correlation
is governed by the combined spacing term
\(\Delta f_{nm,v}\Delta t+\bar h\Delta\alpha_{nm,v}/2\). Increasing
\(\Delta f\Delta t\), or choosing \(\bar h\) and \(\mathcal A\) to avoid
cancellation of the FH spacing, reduces the residual correlation.

\subsubsection{Averaged Transmit-Beampattern Behavior}
\label{subsubsec:averaged_transmit_beampattern}

The orthogonality conditions above directly affect the spatial behavior of the
transmit beampattern after subpulse averaging.
 Let \(\overline P(\theta)\) denote
the normalized transmit beampattern averaged over \(V\) subpulses, and measure
its ripple by
\begin{equation}
	\Delta\overline P(\theta)
	\triangleq
	\overline P(\theta)-1 .
	\label{eq:pattern_avg_main}
\end{equation}

\begin{prop}
	\label{prop:tx_beampattern_bound}
	The averaged transmit beampattern has the following properties.
	\begin{itemize}
		\item \emph{Isotropic beampattern under exact orthogonality:}
		If \eqref{eq:qo_exact_orthogonality_condition} holds for all
		\(n\ne m\) and \(v=1,\ldots,V\), then the \(V\)-subpulse averaged
		transmit beampattern is isotropic, namely
		\begin{equation}
			\Delta\overline P(\theta)
			=
			0,
			\quad
			\forall\theta .
			\label{eq:exact_bp_isotropic}
		\end{equation}
		
		\item \emph{Ripple bound under random assignments:}
		Suppose that, in each subpulse, the antenna-wise frequency-index vector
		is uniformly drawn from all ordered \(N\)-tuples with distinct entries
		in \(\{1,\ldots,K\}\), and the CPM symbols assigned to different
		transmit elements are independently and uniformly drawn from
		\(\mathcal A\). Then
		\begin{equation}
			\sup_{\theta}
			\mathbb E
			\left\{
			\left|
			\Delta\overline P(\theta)
			\right|
			\right\}
			\le
			\min
			\left\{
			B_{\mathrm{tri}},
			B_{\mathrm{rms}}
			\right\},
			\label{eq:tx_bp_bound_main}
		\end{equation}
	where 
\begin{equation}
	\begin{aligned}
		&B_{\mathrm{tri}}
	\triangleq
		(N-1)\mu_1, \\
		&B_{\mathrm{rms}}
		\triangleq
		(N-1)
		\sqrt{
			\frac{\mu_2}{V}
			+
			\frac{V-1}{V}\beta_V
		}, \\
		&\mu_i
		\triangleq
		\mathbb E
		\left\{
		\left|
		\rho_{n,m}^{(v)}
		\right|^i
		\right\},
		\; i=1,2, \\
		&\beta_V
		\!\triangleq\!
		\frac{1}{V(V-1)}
		\sum_{v\ne u}\!
		\left|
		\mathbb E
		\left[
		\rho_{n,m}^{(v)}
		\left(
		\rho_{n,m}^{(u)}
		\right)^*
		\right]
		\right|,
		\; V>1 .
	\end{aligned}
	\label{eq:tx_bp_bound_defs}
\end{equation}
	with $\beta_V$ belonging to $[0, \mu_2]$
	\end{itemize}
\end{prop}

\begin{IEEEproof}
	See Appendix~\ref{app:proof_prop_tx_beampattern_bound}.
\end{IEEEproof}

It is worth noting that the factor \(N-1\) reflects the off-diagonal antenna-pair contribution after
beampattern normalization. The bound \(B_{\mathrm{tri}}\) is controlled by the
first absolute moment of the residual correlation, while \(B_{\mathrm{rms}}\) reflects subpulse averaging and cross-subpulse coherent residuals. For the frequency-dependent terms, Appendix~\ref{app:scaling_residual_moments}
shows that, when the FH spacing dominates the bounded CPM perturbation,
\begin{equation}
	\mu_1
	=
	\mathcal O
	\left(
	\frac{\ln K}{K\Delta f\Delta t}
	\right),
	\quad
	\mu_2
	=
	\mathcal O
	\left(
	\frac{1}{K(\Delta f\Delta t)^2}
	\right).
	\label{eq:mu_scaling_main}
\end{equation}
Thus, increasing either \(\Delta f\Delta t\) or \(K\) reduces the average residual
correlation and suppresses the resulting beampattern ripple.

\begin{remark}[Cross-subpulse residual correlation]
	The term \(\beta_V\) characterizes the residual second-order coupling
	between different subpulses. If the subpulse residual correlations are
	second-order uncorrelated, namely
	\begin{equation}
		\mathbb E
		\left[
		\rho_{n,m}^{(v)}
		\left(
		\rho_{n,m}^{(u)}
		\right)^*
		\right]
		=
		0,
		\; v\ne u,
	\end{equation}
	then \(\beta_V=0\), and \(B_{\mathrm{rms}}\) reduces to
	\((N-1)\sqrt{\mu_2/V}\). This case corresponds to incoherent averaging of
	the residual correlations across subpulses. In the general case, nonzero
	cross-subpulse residual coupling is retained through \(\beta_V\), which
	prevents the bound from reducing to the purely \(1/\sqrt{V}\)-averaged form.
\end{remark}

%% file: sections/02_communication.tex
\section{Communication Embedding and Projection-Energy Demodulation}
\label{sec:comm}

\subsection{Communication Information Embedding}

The proposed waveform embeds information through an ordered frequency--phase representation. Let
\(\Gamma=\{\tilde{\boldsymbol f}^{(1)},\ldots,\tilde{\boldsymbol f}^{(2^{N_F})}\}\)
denote the ordered frequency-combination codebook. Each entry
\(\tilde{\boldsymbol f}^{(\nu)}=[\tilde f_1^{(\nu)},\ldots,\tilde f_N^{(\nu)}]^{\mathsf T}\)
is selected from \(\mathcal F\) and satisfies
\(\tilde f_1^{(\nu)}<\cdots<\tilde f_N^{(\nu)}\). The strict ordering gives a canonical representation of the distinct-frequency assignment within one subpulse, and \(2^{N_F}\le\binom{K}{N}\). Let
\(\Phi=\{\boldsymbol{\Upsilon}_1,\ldots,\boldsymbol{\Upsilon}_{2^{N_\Phi}}\}\)
denote the CPM-symbol codebook, where
\(\boldsymbol{\Upsilon}_\mu\in\mathcal A^N\) and \(2^{N_\Phi}\le M^N\).

The codebooks are equipped with one-to-one bit-labeling maps
\(\mathcal L_F:\{0,1\}^{N_F}\rightarrow\Gamma\) and
\(\mathcal L_\alpha:\{0,1\}^{N_\Phi}\rightarrow\Phi\). For the \(v\)-th subpulse, the two bit blocks
\(\tilde{\boldsymbol c}_v\in\{0,1\}^{N_F}\) and
\(\boldsymbol c_v\in\{0,1\}^{N_\Phi}\) are mapped to
\begin{equation}
	\tilde{\boldsymbol f}_v
	=
	\mathcal L_F(\tilde{\boldsymbol c}_v),
	\tilde{\boldsymbol\alpha}_v
	=
	\mathcal L_\alpha(\boldsymbol c_v).
	\label{eq:comm_bit_labeling}
\end{equation}
The ordered codewords are assigned to the transmit antennas by a common permutation matrix \(\boldsymbol P_v\):
\begin{equation}
	\boldsymbol f_v^e
	=
	\boldsymbol P_v\tilde{\boldsymbol f}_v,
	\boldsymbol\alpha_v^e
	=
	\boldsymbol P_v\tilde{\boldsymbol\alpha}_v .
	\label{eq:comm_antenna_assignment}
\end{equation}
Thus, \(f_{n,v}=[\boldsymbol f_v^e]_n\) and \(\alpha_{n,v}=[\boldsymbol\alpha_v^e]_n\). The common permutation preserves the frequency--CPM-symbol pairing and is not counted as an information-bearing index. The matrices \(\{\boldsymbol P_v\}\) may be optimized for radar ambiguity and sidelobe control \cite{Eedara2020GA,Eedara2022CSK}. This optimization is not pursued here.

Each subpulse conveys \(N_F+N_\Phi\) bits, giving \(R_b=f_{\mathrm{PRF}}V(N_F+N_\Phi)\). For full codebooks, this becomes
\begin{equation}
	R_b
	=
	f_{\mathrm{PRF}}V
	\left(
	\left\lfloor
	\log_2\binom{K}{N}
	\right\rfloor
	+
	N\log_2M
	\right).
	\label{eq:comm_rate_full}
\end{equation}

\subsection{Projection-Energy Demodulation}
\begingroup\color{red}
After subpulse timing synchronization, the local template associated with
frequency \(f_k\) and CPM symbol \(\alpha_m\) is
\begin{equation}
\phi_{k,m}(t)=\exp\!\left\{j2\pi\left(f_kt+\frac12It^2+
\bar h\alpha_mq(t)\right)\right\},
\label{eq:comm_local_template}
\end{equation}
for \(0\le t<\Delta t\). The received segment in subpulse \(v\) is
\begin{equation}
\bar r_v(t)=\sum_{i=1}^{N}\chi_{i,v}
\phi_{k_{i,v},m_{i,v}}(t)+\bar n_v(t),
\label{eq:comm_local_rx_superposition}
\end{equation}
where \((k_{i,v},m_{i,v})\) denotes the frequency and symbol indices of
active component \(i\), and \(\bar n_v(t)\) is zero-mean complex Gaussian
noise. The unknown complex coefficient \(\chi_{i,v}\) absorbs the
propagation gain, array spatial phase, accumulated CPM phase, FH--LFM
compensation phase, and other terms that are constant within the subpulse.

For a candidate frequency combination \(\tilde{\boldsymbol f}^{(\nu)}\)
and CPM-symbol vector \(\boldsymbol\alpha\), let \(k_n^{(\nu)}\) and \(m_n\)
denote the frequency and symbol indices at position \(n\), respectively,
with \([\boldsymbol\alpha]_n=\alpha_{m_n}\).
The received signal is fitted using unknown complex coefficients:
\begin{equation}
\begin{aligned}
\Lambda(\nu,\boldsymbol\alpha)
=\min_{\boldsymbol\xi\in\mathbb C^{N\times 1}}
\int_0^{\Delta t}\!\left|\bar r_v(t)
-\sum_{n=1}^{N}\xi_n\phi_{k_n^{(\nu)},m_n}(t)\right|^2\!\mathrm dt,
\end{aligned}
\label{eq:comm_concentrated_ls}
\end{equation}
where \(\boldsymbol\xi=[\xi_1,\ldots,\xi_N]^{\mathsf T}\) contains the candidate
branch coefficients. Define the matched outputs as
\begin{equation}
z_{k,m,v}=\int_0^{\Delta t}\bar r_v(t)\phi_{k,m}^{*}(t)\,\mathrm dt.
\label{eq:comm_matched_output}
\end{equation}
The corresponding candidate output vector and template Gram matrix are
\begin{equation}
\begin{aligned}
[\boldsymbol z_{\nu,\boldsymbol\alpha,v}]_n
&=z_{k_n^{(\nu)},m_n,v},\\
[\boldsymbol G_{\nu,\boldsymbol\alpha}]_{n,r}
&=\int_0^{\Delta t}\phi_{k_r^{(\nu)},m_r}(t)
\phi_{k_n^{(\nu)},m_n}^{*}(t)\,\mathrm dt.
\end{aligned}
\label{eq:comm_candidate_gram}
\end{equation}
The diagonal entries of \(\boldsymbol G_{\nu,\boldsymbol\alpha}\) equal
the template energy \(\Delta t\), while its off-diagonal entries describe
correlations between different candidate branches.

Expanding the residual gives
\begin{equation}
\begin{aligned}
\Lambda(\nu,\boldsymbol\alpha)
\!=\!\min_{\boldsymbol\xi}\bigl\{\|\bar r_v\|^2
\!-\!2\Re(\boldsymbol\xi^{\mathsf H}\boldsymbol z_{\nu,\boldsymbol\alpha,v})
+\boldsymbol\xi^{\mathsf H}\boldsymbol G_{\nu,\boldsymbol\alpha}\boldsymbol\xi\bigr\},
\end{aligned}
\label{eq:comm_ls_expansion}
\end{equation}
where \(\|\bar r_v\|^2=\int_0^{\Delta t}|\bar r_v(t)|^2\,\mathrm dt\).
For a nonsingular Gram matrix, the least-squares solution and the
resulting minimum residual are
\begin{equation}
\begin{aligned}
\hat{\boldsymbol\xi}
&=\boldsymbol G_{\nu,\boldsymbol\alpha}^{-1}
\boldsymbol z_{\nu,\boldsymbol\alpha,v},\\
\Lambda(\nu,\boldsymbol\alpha)
&=\|\bar r_v\|^2
-\boldsymbol z_{\nu,\boldsymbol\alpha,v}^{\mathsf H}
\boldsymbol G_{\nu,\boldsymbol\alpha}^{-1}
\boldsymbol z_{\nu,\boldsymbol\alpha,v}.
\end{aligned}
\label{eq:comm_ls_solution}
\end{equation}
Since the received energy is independent of the candidate, minimizing
the residual is equivalent to maximizing the concentrated least-squares
(LS) metric
\begin{equation}
J_{\mathrm{LS}}(\nu,\boldsymbol\alpha)
=\boldsymbol z_{\nu,\boldsymbol\alpha,v}^{\mathsf H}
\boldsymbol G_{\nu,\boldsymbol\alpha}^{-1}
\boldsymbol z_{\nu,\boldsymbol\alpha,v}.
\label{eq:comm_ls_metric}
\end{equation}
The unknown branch coefficients are eliminated in this procedure, so
accurate branch-phase prior information is not required.

When the correlations between candidate branches are small relative to
the template energy, the off-diagonal Gram-matrix entries can be neglected.
This gives \(\boldsymbol G_{\nu,\boldsymbol\alpha}\approx\Delta t\,\boldsymbol I\)
and hence
\begin{equation}
J_{\mathrm{LS}}(\nu,\boldsymbol\alpha)
\approx\frac1{\Delta t}\sum_{n=1}^{N}
\left|z_{k_n^{(\nu)},m_n,v}\right|^2.
\label{eq:comm_joint_diag_metric}
\end{equation}
Dropping the common factor \(1/\Delta t\), which does not affect the
decision, gives the projection-energy demodulation (PED) metric
\begin{equation}
\begin{aligned}
Q_{k,m,v}&=|z_{k,m,v}|^2,\\
J_{\mathrm{PED}}(\nu,\boldsymbol\alpha)
&=\sum_{n=1}^{N}Q_{k_n^{(\nu)},m_n,v}.
\end{aligned}
\label{eq:comm_projection_energy}
\end{equation}

For a fixed frequency combination \(\nu\), the \(n\)-th term depends
only on its own symbol index \(m_n\). Each branch can therefore select
the symbol that maximizes its projection energy independently, giving
\begin{equation}
\begin{aligned}
\max_{\boldsymbol\alpha}J_{\mathrm{PED}}(\nu,\boldsymbol\alpha)
&=\max_{m_1,\ldots,m_N}\sum_{n=1}^{N}Q_{k_n^{(\nu)},m_n,v}\\
&=\sum_{n=1}^{N}\max_mQ_{k_n^{(\nu)},m,v}.
\end{aligned}
\label{eq:comm_search_separability}
\end{equation}
Thus, the diagonal joint search separates into frequency-combination
selection and branchwise symbol decisions with the same optimal metric value.

First, compute the maximum projection energy and the corresponding symbol
index at each frequency:
\begin{equation}
G_{k,v}=\max_mQ_{k,m,v},\quad
m_{k,v}^{\star}=\arg\max_mQ_{k,m,v}.
\label{eq:comm_frequency_component_metric}
\end{equation}
The frequency combination is recovered by summing the branch scores:
\begin{equation}
\begin{aligned}
\hat\nu_v=\arg\max_\nu\sum_{n=1}^{N}G_{k_n^{(\nu)},v},
\hat{\tilde{\boldsymbol f}}_v=\tilde{\boldsymbol f}^{(\hat\nu_v)}.
\end{aligned}
\label{eq:comm_frequency_decision}
\end{equation}
Next, retrieve the stored symbol indices at the selected frequencies:
\begin{equation}
\hat m_{n,v}=m_{k_n^{(\hat\nu_v)},v}^{\star},
[\hat{\tilde{\boldsymbol\alpha}}_v]_n=\alpha_{\hat m_{n,v}}.
\label{eq:comm_phase_decision}
\end{equation}
Finally, inverse labeling recovers the information bits:
\begin{equation}
\begin{aligned}
\hat{\tilde{\boldsymbol c}}_v
=\mathcal L_F^{-1}(\hat{\tilde{\boldsymbol f}}_v),
\hat{\boldsymbol c}_v
=\mathcal L_\alpha^{-1}(\hat{\tilde{\boldsymbol\alpha}}_v).
\end{aligned}
\label{eq:comm_inverse_labeling}
\end{equation}

\begin{remark}[Demodulation complexity]
Let \(L\) be the number of samples per subpulse. Computing the \(KM\)
matched outputs requires \(O(KML)\) operations; frequency-wise
maximization and frequency-codebook search require \(O(KM)\) and
\(O(|\Gamma|N)\) operations, respectively. The symbol decisions reuse
the stored maximizing indices, giving a total per-subpulse complexity of
\(O(KML+|\Gamma|N)\). In contrast, direct enumeration of the frequency
and symbol combinations under the diagonal metric requires
\(O(KML+|\Gamma|M^N N)\) operations.
\end{remark}
\endgroup

%% file: sections/03_radar.tex
\section{Structure-Matched Coherent Radar Processing and Performance Properties}
\label{sec:radar_processing}

\subsection{Receive Signal Model}
\label{subsec:rx_model}

The surveillance region contains \(Z\) point targets. The \(z\)-th target is
characterized by azimuth angle \(\theta_z\), \textcolor{red}{range \(R_z\) at the beginning
of the CPI}, radial velocity \(\nu_z\), and complex scattering coefficient
\(\rho_z\). \textcolor{red}{A positive \(\nu_z\) denotes radial motion toward
	the array.} Using the frequency-indexed directional transmit component in
\eqref{eq:xk_directional_component}, the receive-beamformed echo in the
\(p\)-th PRI is modeled as
\begin{equation}
	\label{eq:sR}
	\begin{aligned}
		s_{\mathrm R,p}(t)
		&=
		\boldsymbol{\omega}^{\mathsf H}
		\sum_{z=1}^{Z}
		\sum_{k=1}^{K}
		\rho_z
		\textcolor{red}{\boldsymbol a_{\mathrm{rx}}}(\theta_z)
		x_{p,k}\!\left(t-\tau_{p,z}(t);\theta_z\right) \\
		&\quad\times
		\exp\!\left(-j2\pi f_{\mathrm c}\tau_{p,z}(t)\right)
		+
		n_{\mathrm R,p}(t),
	\end{aligned}
\end{equation}
where \(\boldsymbol{\omega}\in\mathbb{C}^{N\times 1}\) is the receive
beamforming vector, \(n_{\mathrm R,p}(t)\) denotes the beamformed receiver
noise in the \(p\)-th PRI, and
\(\textcolor{red}{\boldsymbol a_{\mathrm{rx}}}(\theta)\in\mathbb{C}^{N\times 1}\) is the receive
steering vector, given by
\begin{equation}
	\textcolor{red}{\boldsymbol a_{\mathrm{rx}}}(\theta)
	\!=\!
	\textcolor{red}{\boldsymbol a_{\mathrm{tx}}}^{*}(\theta)
	\!=\!
	\left[
	1\!,\!
	e^{j\frac{2\pi d}{\lambda_c}\sin\theta},
	\ldots,
	e^{j(N-1)\frac{2\pi d}{\lambda_c}\sin\theta}
	\right]^{\mathrm T}\!\!\!.
	\label{eq:rx_steering_vector}
\end{equation}
\begingroup\color{red}
The instantaneous two-way propagation delay of target \(z\) in the \(p\)-th
PRI is
\begin{equation}
	\label{eq:tau_pz}
	\begin{aligned}
		\tau_{p,z}(t)
		&=
		\frac{2}{c}\left[R_z-\nu_z(t_p+t)\right] \\
		&=
		\tau_z-\frac{2\nu_z}{c}(t_p+t),
		\quad
		0\le t<T_{\mathrm r},
	\end{aligned}
\end{equation}
where \(\tau_z\triangleq 2R_z/c\) is the two-way propagation delay at the
beginning of the CPI.
\endgroup
\subsection{Proposed Structure-Matched Coherent Processor}
\label{subsec:proc}

\begin{figure*}[h]
	\centering
	\includegraphics[width=1\textwidth]{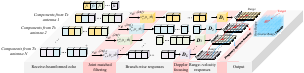}
	\caption{Processing flow of the proposed SMC processor.}
	\label{fig:smc_flow}
\end{figure*}

As shown in Fig.~\ref{fig:smc_flow}, the proposed SMC processor performs
space--time--frequency joint matched filtering, velocity-consistent Doppler
focusing, and frequency-aperture focusing under a given azimuth hypothesis
\(\tilde{\theta}\). Angular search is implemented by repeating this procedure
over the angle grid.

\subsubsection{Space--Time--Frequency Joint Matched Filtering}
\label{subsubsec:stf_jmf}

\begingroup\color{red}
For the \(k\)-th frequency branch in the \(p\)-th PRI, the receive-beamformed
echo is shifted to the corresponding baseband branch as
\begin{equation}
	\label{eq:downconvert}
	r_{p,k}(t)
	=
	s_{\mathrm R,p}(t)
	\exp\!\left[-j2\pi f_k(t_p+t)\right],
	0\le t<T_{\mathrm r}.
\end{equation}
The corresponding baseband reference waveform is
\begin{equation}
	\label{eq:xbb}
	x^{\textcolor{red}{\mathrm{b}}}_{p,k}(t;\tilde{\theta})
	\triangleq
	x_{p,k}(t;\tilde{\theta})
	\exp\!\left[-j2\pi f_k(t_p+t)\right].
\end{equation}
After sampling with period \(T_s\), the received branch is represented by
\(\tilde{\boldsymbol r}_{p,k}\in\mathbb C^{U\times 1}\), where
\(U=\lfloor T_{\mathrm r}/T_s\rfloor\). With
\(L_h=\lfloor T/T_s\rfloor\), define the sampled reference vector and its
energy as
\begin{equation}
	\begin{aligned}
		\boldsymbol x_{p,k}
		&\triangleq
		\bigl[
		x^{\textcolor{red}{\mathrm{b}}}_{p,k}(0;\tilde{\theta}),
		\ldots,
		x^{\textcolor{red}{\mathrm{b}}}_{p,k}((L_h-1)T_s;\tilde{\theta})
		\bigr]^{\mathsf T},\\
		\mathcal E_{p,k}
		&\triangleq
		\|\boldsymbol x_{p,k}\|_2^2,
	\end{aligned}
	\label{eq:branch_sample_vector_energy}
\end{equation}

\begin{remark}[Frequency-branch coverage under random data]
Let \(\Psi_{p,v}\subseteq\{1,\ldots,K\}\) denote the \(N\)-element
frequency-index set activated in the \(v\)-th subpulse of pulse \(p\).
Assume that all sets
\(\{\Psi_{p,v}\}_{p=1,v=1}^{P,V}\) are mutually independent and uniformly
selected from the \(\binom{K}{N}\) possible frequency combinations. For
\(L\) independent selections, the inclusion--exclusion principle gives the
probability that all \(K\) frequency branches are activated at least once as
\begin{equation}
	\mathcal C(L)
	\triangleq
	\sum_{i=0}^{K-N}
	(-1)^i
	\binom{K}{i}
	\left[
	\frac{\binom{K-i}{N}}{\binom{K}{N}}
	\right]^L.
	\label{eq:branch_coverage_probability}
\end{equation}
The probabilities that all frequency branches are activated at least once in
every pulse and within the entire CPI are, respectively,
\begin{equation}
	P_{\mathrm{PRI}}
	=
	\bigl[\mathcal C(V)\bigr]^P,
	\qquad
	P_{\mathrm{CPI}}
	=
	\mathcal C(PV).
\end{equation}
For given \(K\) and \(N\), the upper bound on the probability of
incomplete coverage decreases exponentially with the number of subpulses
\(V\). A large number of subpulses is typically used to achieve high
communication rates, which also favors complete frequency coverage.
For the typical configuration \(P=64\), \(V=30\), \(N=4\), and \(K=8\),
\(P_{\mathrm{PRI}}\approx0.9999995232\), while
\(P_{\mathrm{CPI}}\) is numerically indistinguishable from one.
\end{remark}

The branch reference for matched filtering is
\begin{equation}
\boldsymbol h_{p,k}=\boldsymbol x_{p,k}.
\label{eq:branch_matched_filter}
\end{equation}
If \(\mathcal E_{p,k}=0\), the reference vector and its matched output are zero.
The Toeplitz filtering matrix
\(\boldsymbol H_{p,k}\in\mathbb C^{U\times U}\) is constructed from shifted
copies of \(\boldsymbol h_{p,k}\) as
\begin{equation}
	\label{eq:Hpkt_def}
	[\boldsymbol H_{p,k}]_{u,u'}
	=
	\begin{cases}
		h_{p,k}[u-u'+1],
		& 0\le u-u'\le L_h-1,\\
		0,
		& \text{otherwise}.
	\end{cases}
\end{equation}
The branchwise matched response is finally obtained as
\begin{equation}
	\label{eq:yvec}
	\boldsymbol y_{p,k}
	\triangleq
	\boldsymbol H_{p,k}^{\mathsf H}
	\tilde{\boldsymbol r}_{p,k}
	\in\mathbb C^{U\times 1}.
\end{equation}
\endgroup

\begingroup\color{red}
Appendix~\ref{app:B1} shows that, under the conditions of
Proposition~\ref{prop:qo_corr_bound}, the normalized correlation between
distinct active frequency branches after branch-\(k\) downconversion
is bounded by \(\varepsilon_0\) at the matched delay under the
fixed-delay model.
\endgroup

As derived in Appendix~\ref{app:B2}, for a target with velocity
\(\tilde{\nu}\), \textcolor{red}{the matched-filter output of the
\((p,k)\)-th branch at the fixed delay \(\tilde{\tau}\), corresponding
to the target range at the beginning of the CPI,} can be approximated as
\begin{equation}
	\label{eq:peak}
	\begin{aligned}
		y_{p,k}(\textcolor{red}{\tilde{\tau}})
		&\approx
		A_{p,k}(\tilde{\nu})
		\exp\!\left(j\vartheta_p(\tilde{\nu})\right)\\
		&\quad\times
		\exp\!\left[
		-j2\pi
		\bigl(
		\tilde f_k-f_{d,k}(\tilde{\nu})
		\bigr)
		\textcolor{red}{\tilde{\tau}}
		\right]\\
		&\quad\times
		\textcolor{red}{\exp\!\left[j2\pi f_{d,k}(\tilde{\nu})t_p\right]}.
	\end{aligned}
\end{equation}
\begingroup\color{red}
where, under \(2\pi|f_{d,k}(\tilde{\nu})|T\ll1\),
\(A_{p,k}(\tilde{\nu})\approx|\tilde{\beta}\tilde{\rho}|\mathcal E_{p,k}\)
for the sampled implementation, as follows from
Appendix~\ref{app:B2}, and \(\vartheta_p(\tilde{\nu})\) is approximately
independent of \(k\).
\textcolor{red}{The power loss caused by residual phase differences in
subsequent coherent integration across pulses and frequency branches is
quantified in Appendix~\ref{app:phase_coherent_loss}.}
Here, \(f_{d,k}(\tilde{\nu})\triangleq2\tilde f_k\tilde{\nu}/c\).
After compensating \(\exp[j2\pi f_{d,k}(\tilde{\nu})t_p]\), the remaining
branch-dependent phase is used for frequency-aperture focusing.
\endgroup
For subsequent Doppler focusing, the branchwise matched responses are
collected as
\begin{equation}
	\label{eq:Yk}
	\boldsymbol Y_k
	\triangleq
	[
	\boldsymbol y_{1,k},
	\boldsymbol y_{2,k},
	\ldots,
	\boldsymbol y_{P,k}
	]
	\in\mathbb C^{U\times P}.
\end{equation}

\subsubsection{Velocity-Consistent Doppler Focusing}
\label{subsubsec:vcdf}

The velocity-consistent Doppler focusing stage performs slow-time coherent
integration on a common physical-velocity grid. Since different frequency
branches have different carrier frequencies, the same physical velocity
corresponds to different Doppler frequencies across branches.

For branch \(k\), with \(\tilde f_k=f_{\mathrm c}+f_k\),
the maximum unambiguous velocity is
\begin{equation}
	\label{eq:vmaxk}
	v_{\max,k}
	=
	\frac{c}{4T_{\mathrm r}\tilde f_k}.
\end{equation}
To ensure ambiguity-free processing for all branches, define
\(v_{\max}\triangleq\min_k v_{\max,k}\), and choose \(Q\) uniformly spaced
velocity bins as
\begin{equation}
	\label{eq:vq}
	\begin{aligned}
		v_q
		=
		\left(q \!-\!\frac{Q+1}{2}\right)\Delta v,
		\Delta v
		=
		\frac{2v_{\max}}{Q},
		q=1,2,\ldots,Q .
	\end{aligned}
\end{equation}
For branch \(k\), the Doppler frequency corresponding to \(v_q\) is
\begin{equation}
	\label{eq:fd}
	f_{d,k}(v_q)
	\triangleq
	\frac{2\tilde f_k}{c}v_q .
\end{equation}
Although \(f_{d,k}(v_q)\) depends on \(\tilde f_k\), the index \(q\)
corresponds to the same physical velocity \(v_q\) for all frequency branches.

Under this common velocity grid, define the Doppler focusing matrix
\(\boldsymbol D_k\in\mathbb C^{Q\times P}\) with entries
\begin{equation}
\label{eq:Dk}
[\boldsymbol D_k]_{q,p}
\triangleq
\textcolor{red}{
\exp\!\left[-j2\pi f_{d,k}(v_q)t_p\right]
}.
\end{equation}
The range--velocity response matrix of branch \(k\) is then
\begin{equation}
	\label{eq:Xk}
	\boldsymbol X_k
	=
	\boldsymbol Y_k\boldsymbol D_k^{\mathsf T}
	\in\mathbb C^{U\times Q},
\end{equation}
whose \((u,q)\)-th entry is
\begin{equation}
\label{eq:Xk_entry}
[\boldsymbol X_k]_{u,q}
=
\sum_{p=1}^{P}
[\boldsymbol Y_k]_{u,p}
\textcolor{red}{
\exp\!\left[-j2\pi f_{d,k}(v_q)t_p\right]
}.
\end{equation}

\begingroup\color{red}
For a target with $\tilde\nu\in[-v_{\max},v_{\max})$, the residual
slow-time phase $\phi_{p,k}$ at the nearest velocity bin $v_{\tilde q}$
satisfies
\begin{equation}
\label{eq:doppler_grid_phase_bound}
|\phi_{p,k}|
\le\frac{2\pi\tilde f_{\max}(P-1)T_{\mathrm r}\Delta v}{c}
\triangleq\varepsilon_{\mathrm d},\qquad\forall p,k,
\end{equation}
where $\tilde f_{\max}\triangleq\max_k\tilde f_k$.
Appendix~\ref{app:res_doppler} derives this bound and the corresponding
coherent-sum error bound, together with a sufficient condition on $Q$.
For $\varepsilon_{\mathrm d}\ll1$, the residual phase can be neglected,
and Doppler focusing of the peak model \eqref{eq:peak} gives
\endgroup
\begin{equation}
	\label{eq:X_approx}
	[\boldsymbol X_k]_{\tilde u,\tilde q}
	\approx
	\exp\!\left[
	-j2\pi
	\bigl(
	\tilde f_k-f_{d,k}(\tilde{\nu})
	\bigr)
	\tilde{\tau}
	\right]
	\sum_{p=1}^{P}
	c_{p,k}(\tilde{\nu}),
\end{equation}
where
\(
	\label{eq:cpk_def}
	c_{p,k}(\tilde{\nu})
	\triangleq
	A_{p,k}(\tilde{\nu})
	\exp\!\left(j\vartheta_p(\tilde{\nu})\right).
\)
Thus, velocity-consistent Doppler focusing coherently accumulates the target
over slow time while retaining the branch-dependent phase needed for
frequency-aperture focusing.

\subsubsection{Frequency-Aperture Focusing}
\label{subsubsec:faf}

After space--time--frequency joint matched filtering and velocity-consistent
Doppler focusing, each frequency branch provides a coarse range--velocity
response. The frequency-aperture focusing stage synthesizes these branch
responses to obtain fine-range focusing.

For a fixed velocity bin \(v_q\), define the effective frequency
\begin{equation}
	\label{eq:feff_def}
	f_{k,\mathrm{eff}}(v_q)
	\triangleq
	\tilde f_k-f_{d,k}(v_q).
\end{equation}
The fine-range focusing operation is implemented as
\begin{equation}
	\label{eq:Y_uq_tau}
	Y(u,q;\tau)
	=
	\sum_{k=1}^{K}
	[\boldsymbol X_k]_{u,q}
	\exp\bigl(j2\pi f_{k,\mathrm{eff}}(v_q)\tau\bigr).
\end{equation}
This operation compensates the branch-dependent phase under the delay
hypothesis \(\tau\).

If the branch frequencies are equally spaced, namely
\(f_k=f_1+(k-1)\Delta f\), then
\begin{equation}
	\label{eq:feff_affine}
	f_{k,\mathrm{eff}}(v_q)
	=
	f_{1,\mathrm{eff}}(v_q)
	+
	(k-1)\Delta f_{\mathrm{eff}}(v_q),
\end{equation}
where \(f_{1,\mathrm{eff}}(v_q)=\tilde f_1-f_{d,1}(v_q)\) and
\(\Delta f_{\mathrm{eff}}(v_q)=(1-2v_q/c)\Delta f\). Substituting
\eqref{eq:feff_affine} into \eqref{eq:Y_uq_tau} gives
\begin{equation}
	\label{eq:Y_factored}
	\begin{aligned}
		Y(u,q;\tau)
		&=
		\exp\!\bigl(j2\pi f_{1,\mathrm{eff}}(v_q)\tau\bigr)
		\sum_{k=1}^{K}
		[\boldsymbol X_k]_{u,q}\\
		&\times
		\exp\!\bigl(j2\pi(k-1)\Delta f_{\mathrm{eff}}(v_q)\tau\bigr).
	\end{aligned}
\end{equation}
Thus, apart from a common phase factor, fine-range focusing becomes a
DFT-like operation across the frequency-branch index.

For the usual radar velocity range with \(\Delta f_{\mathrm{eff}}(v_q)>0\),
choose an \(L\)-point delay grid with \(L\ge K\):
\begin{equation}
	\label{eq:tau_l}
	\begin{aligned}
		\tau_{q,\ell}
		&=
		\frac{\ell-1}{L\Delta f_{\mathrm{eff}}(v_q)},\\
		\ell
		&=
		1,2,\ldots,L .
	\end{aligned}
\end{equation}
Evaluating \eqref{eq:Y_factored} at \(\tau=\tau_{q,\ell}\) yields
\begin{equation}
	\label{eq:Y_l}
	\begin{aligned}
		Y(u,q;\tau_{q,\ell})
		&=
		\exp\!\bigl(j2\pi f_{1,\mathrm{eff}}(v_q)\tau_{q,\ell}\bigr)
		\sum_{k=1}^{K}
		[\boldsymbol X_k]_{u,q}\\
		&\times
		\exp\!\left(j2\pi\frac{(k-1)(\ell-1)}{L}\right).
	\end{aligned}
\end{equation}
Therefore, the samples \(Y(u,q;\tau_{q,\ell})\) form a three-dimensional
response cube over \((u,q,\ell)\) for the angle hypothesis
\(\tilde{\theta}\), and target detection is performed in this response domain.
If the carrier set is not equally spaced, \eqref{eq:Y_uq_tau} is evaluated as
a nonuniform Fourier sum.

\begingroup\color{red}
For a detected cell \((u,q,\ell)\), the coarse-range and fine-range
coordinates are
\[
R_{\mathrm c}=\frac{c\tau_u}{2},
\qquad
R_{\mathrm f}
=\operatorname{mod}\!\left(\frac{c\tau_{q,\ell}}{2},T_{\mu,q}\right),
\]
where \(\tau_u\) is the physical delay corresponding to the \(u\)-th
coarse-range cell, and \(\tau_{q,\ell}\) is defined in
\eqref{eq:tau_l}.
For a given velocity bin \(q\), the fundamental period of the
fine-range response is
\[
T_{\mu,q}
=\frac{c}{2g_\mu\Delta f_{\mathrm{eff}}(v_q)}
\approx\frac{c}{2g_\mu\Delta f},
\]
with the relevant definitions and derivation provided in
Section~\ref{subsubsec:range_resolution} and
Appendix~\ref{app:fine_range_periodicity}.

For improved ranging accuracy, the fine-range reconstruction is used as
the final physical-range estimate, with the coarse coordinate used to
select its cycle:
\begin{equation}
	\label{eq:coarse_fine_range_estimates}
	\widehat R^{(\mathrm f)}
	=R_{\mathrm f}+\widehat mT_{\mu,q},
\end{equation}
where
\begin{equation}
	\label{eq:range_cycle_selection}
	\widehat m
	=\underset{\substack{m\in\mathbb Z\\
			R_{\mathrm f}+mT_{\mu,q}\in\mathcal R_{\mathrm c}}}
	{\operatorname{arg\,min}}
	\left|
	R_{\mathrm c}-\left(R_{\mathrm f}+mT_{\mu,q}\right)
	\right|.
\end{equation}
Here, \(\mathcal R_{\mathrm c}\triangleq[0,cT_{\mathrm r}/2)\) is the
coarse-range unambiguous interval under the single-PRI assumption.
\endgroup
\begingroup\color{red}
\begin{remark}[Computational complexity and implementation]
	For one angle hypothesis, the direct implementation costs
\[
O(PKUL_h+KUQP+UQKL),
\]
where the three terms correspond to matched filtering, velocity-consistent
Doppler focusing, and frequency-aperture focusing, respectively.
FFT-based linear convolution and DFT-based focusing with equally spaced
effective frequencies and a DFT fine-delay grid reduce the cost to
\[
O(PKN_{\mathrm F}\log N_{\mathrm F}+KUQP+UQL\log L),
\]
where $N_{\mathrm F}$ is the convolution FFT length. Both costs scale by
$N_\theta$ for $N_\theta$ angle hypotheses.

Matched filtering and Doppler focusing are parallelizable across branches,
and frequency-aperture focusing across coarse-range--velocity cells.
Successive sliding sums over separable reference windows reduce background
power estimation to $O(UQL)$. Combined with pipelined background estimation
and threshold comparison, these implementations can reduce latency and
facilitate real-time processing.
\end{remark}
\endgroup

\subsection{Performance Properties}
\label{subsec:perf}
\FloatBarrier
This subsection summarizes three performance properties of the proposed SMC
processor: joint coarse--fine range discrimination, output-SNR behavior, and
effective receive beampattern. The detailed derivations are provided in Appendix~\ref{app:smc_perf_props}.

\subsubsection{Joint Coarse--Fine Range Discrimination}
\label{subsubsec:range_resolution}

\begingroup\color{red}
For equally spaced carriers, the joint power response is approximated by
\begin{equation}
\label{eq:Ajoint_final}
A_{\mathrm{joint}}^{(\mu)}(\Delta R)
\approx A_{\mathrm c}(\Delta R)A_{\mathrm f}^{(\mu)}(\Delta R),
\end{equation}
where \(\Delta R\) is the trial-range offset from the target range,
\(A_{\mathrm c}(\Delta R)\) is the normalized coarse-range power envelope,
and the fine-range response is
\begin{equation}
\label{eq:Af_weighted_final}
\begin{aligned}
A_{\mathrm f}^{(\mu)}(\Delta R)
=\left|\sum_{k=1}^{K}\mu_k
e^{j2\pi(k-1)\Delta R/T_f}\right|^2,
T_f=\frac{c}{2\Delta f}.
\end{aligned}
\end{equation}
The branch weights are defined by
\begin{equation}
\label{eq:mu_final}
\mu_k={|s_k|}/{\displaystyle\sum_{i=1}^{K}|s_i|},
 s_k=\sum_{p=1}^{P}c_{p,k}(\tilde\nu).
\end{equation}
Branch energy variations are reflected in \(\mu_k\), while a branch absent
throughout the CPI has \(\mu_k=0\).
For at least two active branches, the fundamental fine-range period is
\begin{equation}
\label{eq:Tmu_final}
\begin{aligned}
T_\mu=\frac{T_f}{g_\mu},
g_\mu=\gcd\{|k-\ell|:\mu_k\mu_\ell>0,\ k\ne\ell\},
\end{aligned}
\end{equation}
With only one active branch, \(A_{\mathrm f}^{(\mu)}(\Delta R)\equiv1\).

A larger weighted frequency variance gives a faster decrease of the
fine-range response near its peak. For small weight perturbations,
a decrease (increase) in the response at the descending half-power
point of a reference response corresponds to a narrower (wider)
3-dB mainlobe.

For \(K\ge2\) uniformly weighted branches, \(\mu_k=1/K\), the fine-range
response reduces to
\begin{equation}
\label{eq:Af_range_final}
A_{\mathrm f}(\Delta R)
=\left|\frac{\sin(\pi K\Delta R/T_f)}
{K\sin(\pi\Delta R/T_f)}\right|^2,
\end{equation}
with fundamental period \(T_f\).
\textcolor{red}{Frequency-aperture focusing therefore retains periodic
fine-range replicas. Multiplication by the coarse-range envelope can
attenuate these replicas but does not generally eliminate them.
The joint response below is evaluated along the physical-range mapping
defined in Section~\ref{subsubsec:faf}, rather than by adding the coarse
and fine coordinates.}
Using half the central peak power of the joint response as the threshold,
define the range-ambiguity region
\begin{equation}
\label{eq:joint_unresolved_region_final}
\mathcal U_{3\mathrm{dB}}
=\left\{\Delta R:A_{\mathrm{joint}}(\Delta R)\ge\frac12\right\}.
\end{equation}
Under the mainlobe approximation \(|\pi\Delta R/T_f|\ll1\) and an
approximately constant coarse-range envelope across the central fine-range
mainlobe, the central ambiguity interval has width
\begin{equation}
\label{eq:R3dB_approx_simple_final}
B_0\approx\frac{0.443c}{K\Delta f}.
\end{equation}

Still considering uniform weights, if the coarse-range envelope varies
slowly within each fine-range mainlobe, a nonzero-width ambiguity interval
forms around the periodic peak \(wT_f\) when
\(A_{\mathrm c}(wT_f)>1/2\).
If the coarse-range mainlobe is symmetric about zero and decreases
monotonically with \(|\Delta R|\), the number of ambiguity intervals
within this mainlobe, including the central interval, is
\begin{equation}
\label{eq:Nbranch_final}
\begin{aligned}
N_{\mathrm{amb}}=2W+1,
W=\max\{w\in\mathbb Z_{\ge0}:wT_f<\delta_{\mathrm c}\},
\end{aligned}
\end{equation}
where \(A_{\mathrm c}(\delta_{\mathrm c})=1/2\).
Using the same local mainlobe approximation around each periodic peak,
the width of the \(w\)-th ambiguity interval is
\begin{equation}
\label{eq:Bw_final}
B_w\approx\frac{2T_f x_w}{\pi K},
\end{equation}
where
\[
\left(\frac{\sin x_w}{x_w}\right)^2
=\frac{1}{2A_{\mathrm c}(wT_f)},
 0<x_w<\pi.
\]
For outer ambiguity intervals with \(1/2<A_{\mathrm c}(wT_f)<1\),
\(B_w<B_0\). These results are derived in
Appendix~\ref{app:joint_range_branch}.
\endgroup

\subsubsection{Output-SNR Comparison With PD Processing}
\label{subsubsec:output_snr}

This subsection compares the proposed SMC processor with the conventional
PD processor, i.e., direct pulse compression followed by
moving-target detection (MTD) \cite{Alabaster2012,Barton2004}.

\begin{prop}[SMC--PD signal and noise comparison]
	\label{prop:output_snr_behavior}
	At the correctly matched target cell
	\(\eta^\star=(\tilde{\tau},\tilde{\nu},\tilde{\theta})\), the SMC and PD
	output-SNR behavior can be decomposed into the following signal-side and
	noise-side comparisons.
	
	\begin{itemize}
		\item \emph{Signal-side comparison.}
		Let \(m_{p,k}^{(s)\star}\) denote the signal component of the
		\((p,k)\)-th branch matched output \textcolor{red}{at the fixed delay
		\(\tilde{\tau}\)}, and let \(w_{p,k}^{\star}\) denote
		the corresponding SMC branchwise phase-compensation coefficient. The SMC
		matched-cell signal statistic is
		\begin{equation}
			\label{eq:snr_main_smc_signal_stat}
			Z_{\mathrm{SMC}}^{(s)}(\eta^\star)
			=
			\sum_{p=1}^{P}
			\sum_{k=1}^{K}
			w_{p,k}^{\star}m_{p,k}^{(s)\star}.
		\end{equation}
		The conventional PD statistic admits the decomposition
		\begin{equation}
			\label{eq:snr_main_pd_signal_stat}
			Z_{\mathrm{PD}}^{(s)}(\eta^\star)
			=
			\sum_{p=1}^{P}
			\sum_{k=1}^{K}
			\gamma_{p,k}^{\star}
			w_{p,k}^{\star}m_{p,k}^{(s)\star},
		\end{equation}
	where \(\gamma_{p,k}^{\star}\) is a phase-only mismatch factor.
	\textcolor{red}{For conventional PD processing with a single Doppler
	reference \(f_{d,\mathrm{ref}}(\tilde{\nu})=2f_{\mathrm c}\tilde{\nu}/c\),
	the pulse-dependent term in the phase of \(\gamma_{p,k}^{\star}\) is}
	{\color{red}
	\[
	\begin{aligned}
	\Delta\phi_{p,k}^{\star,\mathrm{slow}}
	&=2\pi\bigl[f_{d,k}(\tilde{\nu})
	-f_{d,\mathrm{ref}}(\tilde{\nu})\bigr]t_p\\
	&=2\pi\frac{2\tilde{\nu}}{c}f_k t_p.
	\end{aligned}
	\]
	}
	\textcolor{red}{Thus, the residual slow-time phase is first order in
	\(\tilde{\nu}/c\) and depends on the carrier offset \(f_k\).}
		
	\item \emph{Noise-side comparison.}
	When the effective branch frequencies are equally spaced and the DFT
	fine-delay grid is used, the fine-delay-averaged SMC noise power and the PD
	output noise power satisfy
	\begin{equation}
		\label{eq:snr_main_noise_difference}
		\bar{\sigma}_{\mathrm{SMC}}^2(u,q)
		-
		\sigma_{\mathrm{PD}}^2(u,q)
		=
		-2N_0
		\Re
		\left\{
		\Lambda_{\mathrm{PD}}(u)
		\right\}.
	\end{equation}
	Here, \(\bar{\sigma}_{\mathrm{SMC}}^2(u,q)\) denotes the SMC noise power
	averaged over the fine-delay grid, and
	\[
		\label{eq:snr_main_pd_cross_term}
			\Lambda_{\mathrm{PD}}(u)
			\triangleq
			\sum_{p=1}^{P}
			\sum_{k<k'}
			\Gamma_{p;k,k'}(u)
			\exp\!\left[
			j2\pi(f_k-f_{k'})\textcolor{red}{\tau_u}
			\right]
		\]
	collects the residual inter-channel noise correlations in the PD output.
	The term \(\Gamma_{p;k,k'}(u)\) denotes the residual noise correlation between
	the \(k\)-th and \(k'\)-th pulse-compression templates at coarse-delay cell \(u\).

	\end{itemize}
\end{prop}

\begin{IEEEproof}
	See Appendix~\ref{app:snr_analysis}.
	\IEEEQEDoff
\end{IEEEproof}

Proposition~\ref{prop:output_snr_behavior} shows that the SMC--PD difference
has two sources. On the signal side, the single-reference PD chain leaves a
frequency-dependent Doppler mismatch across branches, whereas SMC explicitly
compensates the branch-dependent target phase. On the noise side, the
difference is determined by the residual inter-channel correlation term
\(\Lambda_{\mathrm{PD}}(u)\), and vanishes when the branch noise covariance is
diagonal.

\begingroup\color{red}
Equal branch energies are not required for this comparison. The actual
branch energies determine both the matched signal contributions and the
diagonal entries of the branch-noise covariance matrices. With independent
noise across pulses, orthogonal actual branch templates, rectangular
slow-time weights, and unit-modulus phase compensation, the SMC and PD
output noise powers are equal even for nonuniform energies and absent
branches. The energy-dependent expressions and conditions are given in
Appendix~\ref{app:snr_analysis}.
\endgroup

\subsubsection{Effective Receive Beampattern}
\label{subsubsec:beampattern}

The angular selectivity recovered by the SMC receiver is characterized next.

\begin{prop}[SMC-domain receive beampattern]
	\label{prop:smc_receive_beampattern}
	Under the colocated-array and matched-beamformer assumptions, when an echo
	arriving from azimuth \(\theta\) is processed using a template steered to
	\(\theta_0\), the normalized SMC-domain receive beampattern is
	\begin{equation}
		\label{eq:Gsmc_main}
		G_{\mathrm{SMC}}(\theta;\theta_0)
		=
		\left|
		\frac{
			\operatorname{sinc}\!\left(\frac{N}{2}\Delta\varphi\right)
		}{
			\operatorname{sinc}\!\left(\frac{1}{2}\Delta\varphi\right)
		}
		\right|^4,
	\end{equation}
	where
	\(
		\label{eq:Delta_varphi_def}
		\Delta\varphi
		=
		\frac{2\pi d}{\lambda_{\mathrm c}}
		\left(
		\sin\theta-\sin\theta_0
		\right).
	\)
\end{prop}

\begin{IEEEproof}
	See Appendix~\ref{app:beampattern_detail}.
	\IEEEQEDoff
\end{IEEEproof}

Proposition~\ref{prop:smc_receive_beampattern} indicates that SMC processing
restores the full-aperture coherent spatial factor at the receiver. Therefore,
the average transmit beampattern alone does not determine the final angular
selectivity of the complete receiver.

%% file: sections/04_simulation.tex
\section{Simulation Results}
\label{sec:simulation_results}

\subsection{Validation of CP-FH Waveform Properties}
\label{subsec:waveform_validation}

Unless otherwise specified, the simulations use
\(f_{\mathrm c}=10~\mathrm{GHz}\), \(N=4\), \(K=8\),
\(\Delta t=1~\mu\mathrm{s}\), \(V=30\), \(P=64\),
\(B_{\mathrm{chirp}}=1~\mathrm{MHz}\), and \(\epsilon=0.2\).
The baseline CPM parameters are \(M=4\), \(\bar h=0.5\), and
\(\mathcal A=\{-2,-1,1,2\}\), yielding \(\Delta\alpha_{\max}=4\).
In each parametric test, only the parameter shown on the horizontal axis is
varied, while the remaining parameters are kept at their baseline values unless
otherwise stated.

\subsubsection{Correlation of Frequency-Indexed Waveform Components}
\label{subsubsec:correlation_validation}

\begingroup\color{red}
Define the maximum subpulse and full-pulse correlations, respectively, as
\begin{gather}
\varepsilon_0^{\mathrm{emp}}
\triangleq\max_{p,v,\;k\ne k'}\left|\rho_{k,k'}^{(p,v)}\right|,
\label{eq:eps0_emp_def}\\
R_{\mathrm{full}}
\triangleq\max_{p,\;k\ne k'}\left|\rho_{k,k'}^{(p)}\right|.
\label{eq:Rfull_def}
\end{gather}
\endgroup

\begin{figure}[h]
	\centering
	\includegraphics[width=0.5\textwidth]{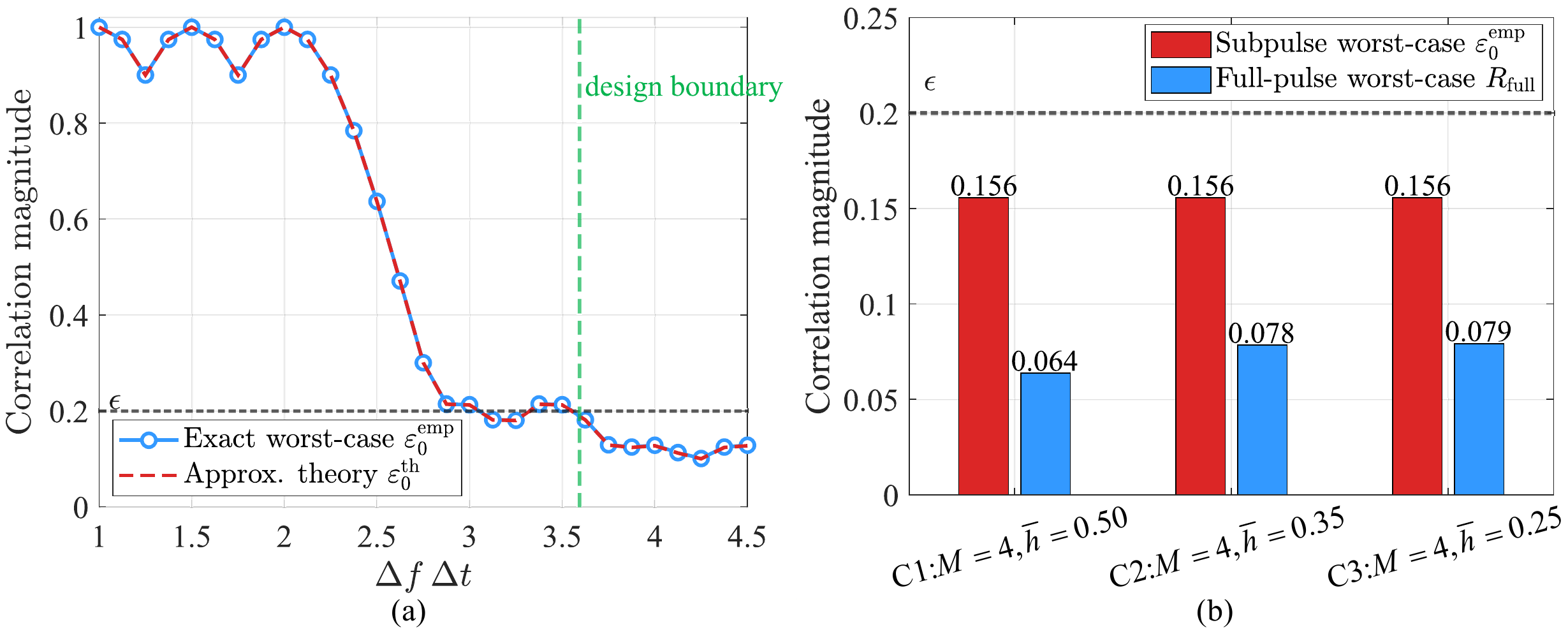}
	\caption{Correlation validation of the proposed CP-FH waveform:
		(a) subpulse-level residual correlation; (b) full-pulse correlation
		bound.}
	\label{fig:correlation_validation}
\end{figure}

\begingroup\color{red}
In Fig.~\ref{fig:correlation_validation}(a), the numerical subpulse
correlation agrees with Proposition~\ref{prop:qo_corr_bound}, confirming
its dependence on the effective normalized spacing
\(\Delta f_{nm,v}^{\mathrm{eff}}\Delta t\). As \(\Delta f\Delta t\)
increases, the residual correlation decreases and falls below the prescribed
tolerance beyond the design boundary.

Fig.~\ref{fig:correlation_validation}(b) presents Monte Carlo results for
random CPM-symbol sequences under different \((M,\bar h)\) settings.
The full-pulse correlation remains below the corresponding subpulse-level
bound and the prescribed tolerance, consistent with the full-pulse
inter-channel correlation analysis in Appendix~\ref{app:B1}.
Thus, subpulse-level correlation control maintains approximate
inter-channel separability over the full pulse.
\endgroup

\subsubsection{Transmit-Beampattern Characteristics}
\label{subsubsec:beampattern_characteristics}

\begin{figure}[h]
	\centering
	\includegraphics[width=0.5\textwidth]{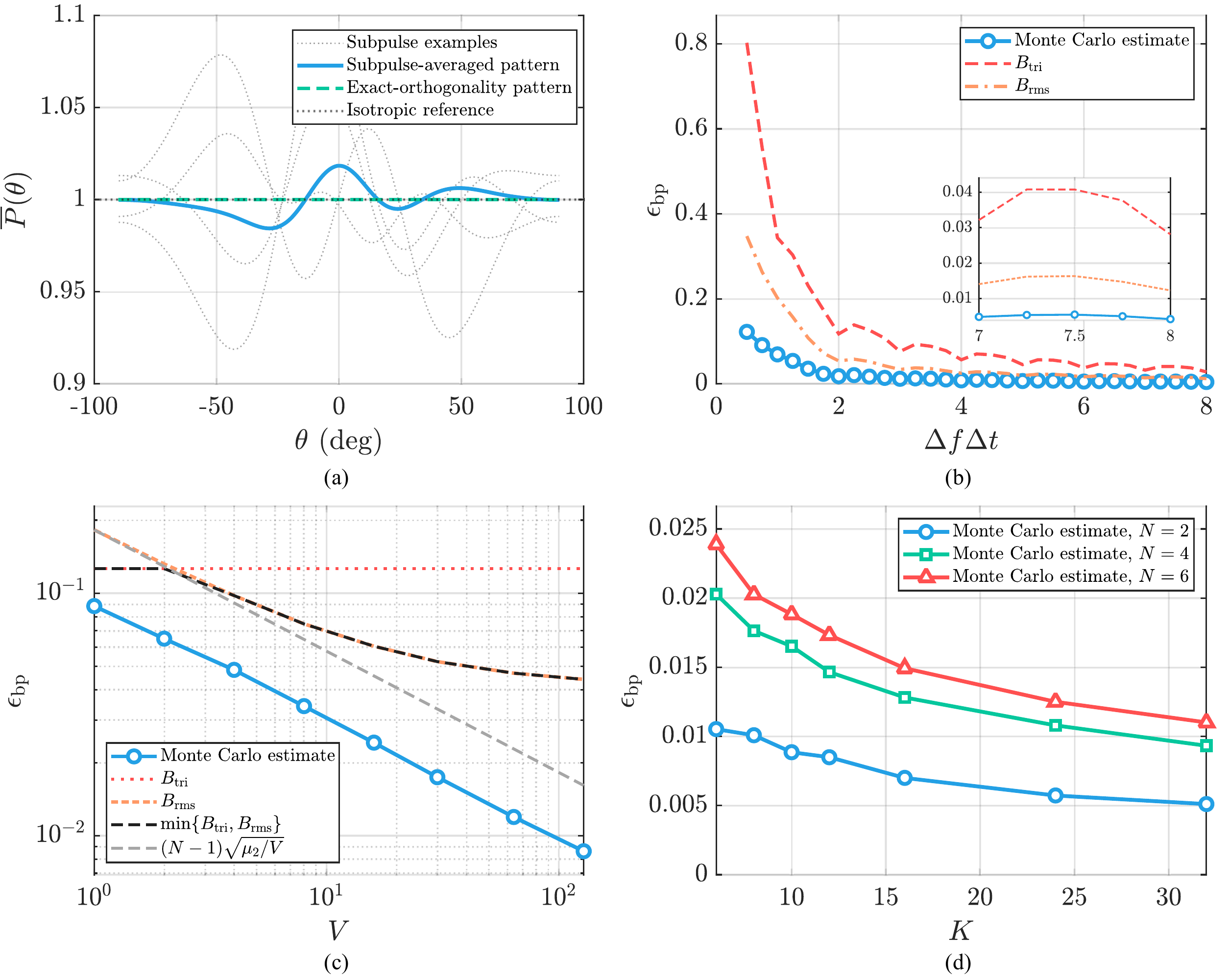}
	\caption{Validation of the transmit-beampattern behavior. (a) Beampattern
		examples under a generic random assignment and a special
		exact-orthogonality setting; (b) Beampattern deviation versus
		\(\Delta f\Delta t\), together with \(B_{\mathrm{tri}}\) and
		\(B_{\mathrm{rms}}\); (c) Beampattern deviation versus \(V\), together
		with \(B_{\mathrm{tri}}\), \(B_{\mathrm{rms}}\),
		\(\min\{B_{\mathrm{tri}},B_{\mathrm{rms}}\}\), and
		\((N-1)\sqrt{\mu_2/V}\); (d) Monte Carlo estimates versus \(K\) for
		different \(N\).}
	\label{fig:tx_beampattern_validation}
\end{figure}

\textcolor{red}{To examine the average transmit-beampattern behavior
characterized by Proposition~\ref{prop:tx_beampattern_bound}, numerical
results are compared with the theoretical bounds.} The deviation
from the isotropic transmit beampattern is measured by
\begin{equation}
	\epsilon_{\mathrm{bp}}
	\triangleq
	\max_{\theta}
	\left|
	\Delta \overline P(\theta)
	\right|.
	\label{eq:sim_bp_deviation_metric}
\end{equation}
The Monte Carlo estimates are obtained by generating sampled CP-FH baseband
waveforms and computing the transmit correlation matrices through numerical
integration. Each point in Figs.~\ref{fig:tx_beampattern_validation}(b)--(d)
is averaged over 300 independent randomized frequency--phase assignments.

Fig.~\ref{fig:tx_beampattern_validation}(a) compares a generic random
assignment with a special exact-orthogonality setting. In the random case,
subpulse-level residual correlations cause spatial ripples, while subpulse
averaging makes the response approach the isotropic reference. In the special
case with \(\mathcal A=\{-2,2\}\), \(\bar h=0.5\), and
\(\Delta f\Delta t=4\), the condition
\(\Delta f_{nm,v}^{\mathrm{eff}}\Delta t\in\mathbb Z\setminus\{0\}\) is
satisfied, so the residual correlations vanish and the beampattern becomes
isotropic.

Fig.~\ref{fig:tx_beampattern_validation}(b) shows that increasing
\(\Delta f\Delta t\) reduces the Monte Carlo estimate of
\(\epsilon_{\mathrm{bp}}\), which is consistent with the sinc-type residual
correlation in Proposition~\ref{prop:qo_corr_bound}. The two bounds
\(B_{\mathrm{tri}}\) and \(B_{\mathrm{rms}}\) remain conservative, and the
inset highlights their behavior in the large-spacing region.

Fig.~\ref{fig:tx_beampattern_validation}(c) shows the effect of the subpulse
number \(V\). The bound \(B_{\mathrm{tri}}\) is nearly independent of \(V\),
whereas \(B_{\mathrm{rms}}\) decreases with \(V\) when the cross-subpulse
residual is weak, so their pointwise minimum gives the effective upper bound.
The Monte Carlo estimate decreases at a rate close to the reference
\((N-1)\sqrt{\mu_2/V}\), indicating weak net cross-subpulse coherent residuals
under randomized assignments.

Fig.~\ref{fig:tx_beampattern_validation}(d) shows that a larger frequency pool
reduces the beampattern deviation, whereas a larger array introduces more
off-diagonal antenna-pair terms and increases the residual ripple. These
results agree with Proposition~\ref{prop:tx_beampattern_bound} and the moment
scaling in Appendix~\ref{app:scaling_residual_moments}.

\subsubsection{Spectral Leakage}
\label{subsubsec:spectral_leakage}

\begin{figure}[!htbp]
    \centering
    \includegraphics[width=0.98\linewidth]{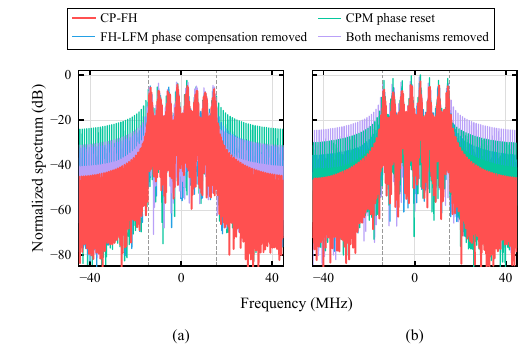}
    \caption{\textcolor{red}{Spectral comparison of different CP-FH
        phase-continuity variants at broadside: (a) \(\bar h=0.5\), \(M=4\);
        (b) \(\bar h=0.4\), \(M=2\).}}
    \label{fig:spectrum}
\end{figure}

\begingroup\color{red}
To examine the effect of phase continuity on the spectrum, consider the
transmitted signal in direction \(\theta\), \(x(t;\theta)\), with spectrum
\begin{equation}
X(f;\theta)\triangleq\int_{-\infty}^{\infty}x(t;\theta)e^{-j2\pi ft}\,\mathrm dt.
\label{eq:spec_spectrum_def}
\end{equation}
For each modulation configuration, define the reference band as
\begin{equation}
\mathcal B_{\mathrm{ref}}
=[f_{\min}-\delta_-,\;f_{\max}+B_{\mathrm{chirp}}+\delta_+],
\label{eq:spec_bin_def}
\end{equation}
where \(f_{\min}\) and \(f_{\max}\) bound the carrier pool, and
\(B_{\mathrm{chirp}}\) is the chirp bandwidth. For the adopted full-response
CPM, the instantaneous frequency offset is
\(\bar h\alpha q'(t)\)~\cite[Sec.~2.2.2]{Anderson1986DigitalPhaseModulation}.
Its maximum positive and negative excursions are defined as
\begin{equation}
\delta_\pm=\max_{\substack{\alpha\in\mathcal A\\0\le t<\Delta t}}
[\pm\bar h\alpha q'(t)]_+.
\label{eq:spec_cpm_margins}
\end{equation}
The same band is used for all compared waveforms and trials within each
configuration. The directional spectral leakage ratio is
\begin{equation}
\eta_{\mathrm{spec}}(\theta)=
\frac{\displaystyle\int_{f\notin\mathcal B_{\mathrm{ref}}}|X(f;\theta)|^2\,\mathrm df}
{\displaystyle\int_{-\infty}^{\infty}|X(f;\theta)|^2\,\mathrm df}.
\label{eq:spec_leakage_def}
\end{equation}
It measures the fraction of energy outside the reference band, including
sidelobes due to the finite pulse window.

To assess CPM phase accumulation and FH--LFM phase compensation,
four waveforms are compared:
\begin{enumerate}
    \item CP-FH: retains both CPM phase accumulation and FH--LFM phase
    compensation in \eqref{eq:sm_sn_product}--\eqref{eq:sm_Cnv_recursion}.
    \item CPM phase reset: resets the CPM phase in each subpulse while
    retaining FH--LFM phase compensation, with
    \(\phi_n^{\mathrm{cpm}}(t;\boldsymbol\alpha_n)
    =2\pi\bar h\alpha_{n,v}q(t-(v-1)\Delta t)\)
    for \(t\in[(v-1)\Delta t,v\Delta t)\).
    \item FH--LFM phase compensation removed: sets \(C_{n,v}\equiv0\)
    for all \(v\) while retaining CPM phase accumulation.
    \item Both mechanisms removed: removes both CPM phase accumulation
    and FH--LFM phase compensation.
\end{enumerate}

Rectangular CPM pulses and \(B_{\mathrm{chirp}}=1~\mathrm{MHz}\) are used
with the carrier pool \(\{-14,-10,\ldots,14\}~\mathrm{MHz}\); the remaining
common parameters follow the preceding setup. The modulation configurations
are (a) \(\bar h=0.5\), \(\mathcal A=\{-2,-1,1,2\}\), and
(b) \(\bar h=0.4\), \(\mathcal A=\{-1,1\}\). Their reference bands are
\([-14.5,15.5]~\mathrm{MHz}\) and \([-14.2,15.2]~\mathrm{MHz}\), respectively,
as marked by the dashed lines in Fig.~\ref{fig:spectrum}.
Within each configuration, the four waveforms share the carrier sequence,
CPM symbols, antenna assignment, and pulse window in every trial.
Spectra are evaluated at broadside \((\theta=0^\circ)\) with a sampling
rate of \(100~\mathrm{MHz}\), using unit-energy normalization and a common
dB reference.
\endgroup

\begingroup\color{red}
Fig.~\ref{fig:spectrum} shows one realization per configuration.
Over \(500\) Monte Carlo trials, the mean leakage ratios in the listed
order are \(1.49\%\),
\(6.63\%\), \(3.70\%\), and \(2.19\%\) in configuration (a), and
\(2.02\%\), \(3.47\%\), \(4.33\%\), and \(5.75\%\) in configuration (b).

In configuration (a), removing both mechanisms yields lower leakage than
either individual removal. When both mechanisms are removed, each antenna's
subpulse starts at zero phase. Since \(f_{n,v}\Delta t\) is an integer,
the phase increment within a subpulse satisfies
\[
\Delta\phi\equiv\pi I\Delta t^2+\pi\bar h\alpha\pmod{2\pi}.
\]
With \(I\Delta t^2=1\), \(\alpha=\pm2\) in configuration (a) makes the
two phase increments cancel modulo \(2\pi\), preserving continuity at
some boundaries and thereby helping to reduce spectral leakage.
In configuration (b), \(\Delta\phi\in\{0.6\pi,1.4\pi\}\), so both
symbols produce nonzero boundary phase jumps; correspondingly, the waveform
with both mechanisms removed has the highest mean spectral leakage in
the simulations.

\begin{samepage}
By combining CPM phase accumulation with FH--LFM phase compensation,
CP-FH ensures phase continuity at all subpulse boundaries and achieves
the lowest mean spectral leakage among the four waveforms in both
configurations.
\par\end{samepage}
\endgroup

\begin{table*}[!t]
	\centering\color{black}
	\caption{Structural comparison of representative FH-MIMO communication embedding schemes.}
	\label{tab:comm_summary}
	\footnotesize
	\setlength{\tabcolsep}{2.3pt}
	\renewcommand{\arraystretch}{1.12}
	\begin{tabular*}{\textwidth}{@{\extracolsep{\fill}}lcccccc@{}}
		\toprule
		Scheme
		& Rate / \((f_{\mathrm{PRF}}V)\)
		& \textcolor{red}{\shortstack{Bits per\\subpulse}}
		& Embedded resources
		& \shortstack{Channel-phase\\prior}
		& \shortstack{Phase\\continuity}
		& \textcolor{red}{\shortstack{Angular-averaged\\leakage ratio}} \\
		\midrule
		FH-PSK~\cite{Baxter2022JointRadarComFH}
		& \(N\log_2 M\)
		& \textcolor{red}{\(4\)}
		& Phase
		& Required
		& No
		& \textcolor{red}{\(7.51\pm1.47\%\)} \\
		
		FHCS~\cite{Baxter2022JointRadarComFH}
		& \(\left\lfloor \log_2 \binom{K}{N} \right\rfloor\)
		& \textcolor{red}{\(6\)}
		& Frequency combination
		& Not required
		& No
		& \textcolor{red}{\(1.51\pm0.30\%\)} \\
		
		FHSP~\cite{Baxter2022JointRadarComFH}
		& \(\left\lfloor \log_2\!\left(\binom{K}{N}N!\right) \right\rfloor\)
		& \textcolor{red}{\(10\)}
		& \textcolor{red}{\shortstack{Frequency combination\\+ permutation}}
		& Not required
		& No
		& \textcolor{red}{\(3.16\pm0.57\%\)} \\
		
		HIM~\cite{Xu2023HIM}
		& \(N+\left\lfloor \log_2\!\left(\binom{K}{N}N!\right)\right\rfloor\)
		& \textcolor{red}{\(14\)}
		& \textcolor{red}{\shortstack{Frequency + binary phase\\+ permutation}}
		& Not required
		& No
		& \textcolor{red}{\(4.63\pm0.70\%\)} \\
		
		POCS~\cite{Baxter2022JointRadarComFH}
		& \(\left\lfloor \log_2 \binom{K}{N} \right\rfloor + N\log_2 M\)
		& \textcolor{red}{\(10\)}
		& \textcolor{red}{\shortstack{Frequency combination\\+ phase}}
		& Required
		& No
		& \textcolor{red}{\(4.62\pm1.00\%\)} \\
		
		\textbf{CP-FH}
		& \(\left\lfloor \log_2 \binom{K}{N} \right\rfloor + N\log_2 M\)
		& \textcolor{red}{\(10\)}
		& \textcolor{red}{\shortstack{Frequency combination\\+ CPM symbols}}
		& \textbf{Not required}
		& \textbf{Yes}
		& \textcolor{red}{\(\mathbf{1.94\pm0.25\%}\)} \\
		\bottomrule
	\end{tabular*}
\end{table*}

\subsection{Communication Performance}
\begingroup\color{red}
This subsection compares CP-FH with frequency-hopping phase-shift keying
(FH-PSK), frequency-hopping code selection (FHCS), frequency-hopping spatial
permutation (FHSP)~\cite{Baxter2022JointRadarComFH}, hybrid index modulation
(HIM)~\cite{Xu2023HIM}, and PSK over code selection
(POCS)~\cite{Baxter2022JointRadarComFH}.
FH-PSK conveys information through phase symbols, FHCS through frequency
combinations, and FHSP jointly through frequency combinations and antenna
permutations. HIM additionally incorporates binary phase symbols.
POCS and CP-FH both convey part of the information through frequency
combinations and use PSK and CPM symbols, respectively, for the remaining
information. Table~\ref{tab:comm_summary} summarizes their communication
rates, embedding resources, channel-phase-prior requirements, phase
continuity, and spectral-leakage statistics.

The BER simulations use \(N=4\) transmit antennas and \(K=8\) candidate
frequencies. Each pulse contains \(V=100\) subpulses of duration
\(\Delta t=1~\mu\mathrm{s}\). The sampling rate is \(100~\mathrm{MHz}\),
the communication direction is \(15^\circ\), and the modulation order
is \(M=2\). The cross-scheme BER comparison uses a common frequency
spacing of \(4~\mathrm{MHz}\). CP-FH uses the symbol alphabet
\(\mathcal A=\{-1,1\}\), modulation index \(\bar h=0.7\), a rectangular
frequency pulse of duration \(\Delta t\), and chirp bandwidth
\(B_{\mathrm{chirp}}=1~\mathrm{MHz}\).

Under this configuration, the frequency-combination codebook retains
\(64\) of the \(\binom{8}{4}=70\) combinations, carrying \(6\) bits.
The codebook jointly encoding frequency combinations and antenna
permutations retains \(1024\) of the \(\binom{8}{4}4!=1680\) assignments,
carrying \(10\) bits. A phase- or CPM-symbol vector consists of four binary
symbols and has \(16\) possible values, carrying \(4\) bits.
Table~\ref{tab:comm_summary} lists the resulting bits per subpulse
alongside the general rate expressions. CP-FH, POCS, and FHSP each carry
\(10\) bits per subpulse, exceeding the \(4\) bits of FH-PSK and \(6\) bits
of FHCS, but below the \(14\) bits of HIM.

\begin{figure}[t]
	\centering
	\includegraphics[width=0.95\columnwidth]{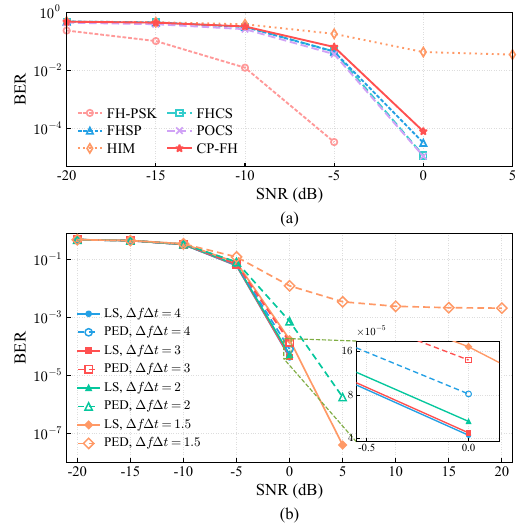}
	\caption{\textcolor{red}{Communication BER versus SNR:
		(a) comparison of FH-MIMO embedding schemes;
		(b) LS and PED at different normalized frequency spacings
		\(\Delta f\Delta t\).}}
	\label{fig:ber_snr}
\end{figure}

Fig.~\ref{fig:ber_snr}(a) compares the BER performance of the schemes.
Compared with FH-PSK and FHCS, CP-FH has a higher information rate but
also a higher BER in the evaluated configuration. At the same information rate, CP-FH and POCS achieve BERs of the same
order of magnitude at \(0~\mathrm{dB}\). While POCS requires accurate
channel-phase compensation or calibration, CP-FH enables noncoherent
reception without accurate channel-phase prior information. HIM supports a higher information rate, but
its BER is high at low SNR.

Fig.~\ref{fig:ber_snr}(b) compares PED with the least-squares (LS)
detector~\cite{Scharf1994MatchedSubspace}, which retains the full Gram
matrix, to assess the performance loss caused by the diagonal approximation.
The ranges in Table~\ref{tab:comm_gram_statistics} cover all \(1024\)
candidate templates formed by \(64\) frequency combinations and \(16\)
CPM-symbol vectors. Each SNR point uses \(10^6\)--\(5\times10^6\)
independent subpulses, with identical data and noise realizations for
the detectors. For normalized frequency spacings
\(\Delta f\Delta t=4,3,2\), the branch correlations are weak and the BERs
of both detectors decrease to low levels as SNR increases, although a
performance gap remains. At \(\Delta f\Delta t=1.5\), the maximum
correlation magnitude and Gram-matrix condition number increase to
\(0.234\) and \(2.29\), respectively, and the performance loss of
PED becomes markedly larger. At \(20~\mathrm{dB}\), PED has a BER of
\(2.10\times10^{-3}\), whereas LS has no observed errors in the simulations.
These analyses quantify the relationship among frequency spacing,
template correlation, and detection performance, providing a basis for
parameter selection when using the diagonal approximation.

\begin{table}[!ht]
    \centering
    \color{red}
    \caption{\textcolor{red}{Ranges of candidate-template correlation
        and Gram-matrix condition number.}}
    \label{tab:comm_gram_statistics}
    \footnotesize
    \renewcommand{\arraystretch}{1.15}
    \begin{tabular}{ccc}
        \toprule
        \(\Delta f\Delta t\)
        & \(|\rho_{nr}^{(\nu,\boldsymbol\alpha)}|\)
        & \(\kappa_2(\boldsymbol G_{\nu,\boldsymbol\alpha})\) \\
        \midrule
        \(4\) & \([0.000,\,0.078]\) & \([1.00,\,1.23]\) \\
        \(3\) & \([0.000,\,0.112]\) & \([1.00,\,1.33]\) \\
        \(2\) & \([0.000,\,0.198]\) & \([1.00,\,1.61]\) \\
        \(1.5\) & \([0.000,\,0.234]\) & \([1.00,\,2.29]\) \\
        \bottomrule
    \end{tabular}
    \par\smallskip
    \parbox{\columnwidth}{\raggedright
   Here \(\rho_{nr}=G_{nr}/\Delta t\) (\(n\ne r\)) and
   \(\kappa_2(G)=\lambda_{\max}(G)/\lambda_{\min}(G)\).
   Ranges are taken over all \(1024\) candidate matrices at each spacing.}
\end{table}

The bits-per-subpulse entries in Table~\ref{tab:comm_summary} correspond
to the BER configuration above. The spectral-leakage statistics use
the parameters of Fig.~\ref{fig:spectrum}(a), except that binary phase
or CPM symbols (\(M=2\), \(\mathcal A=\{-1,1\}\) for CP-FH) are used,
with a common reference band of \([-14.25,15.25]~\mathrm{MHz}\).
The angular-averaged leakage
\[
\bar\eta_{\mathrm{spec}}=\frac{1}{\pi}\int_{-\pi/2}^{\pi/2}
\eta_{\mathrm{spec}}(\theta)\,\mathrm d\theta
\]
is reported. The angular average is computed separately for each trial
by trapezoidal integration over \([-90^\circ,90^\circ]\) with a
\(0.1^\circ\) step. The reported values are the mean \(\pm\) sample
standard deviation across \(500\) trials.

CP-FH has a mean angular-averaged leakage ratio of \(1.94\%\),
lower than those of FH-PSK, FHSP, HIM, and POCS. FHCS yields lower
angular-averaged leakage. CP-FH combines CPM phase accumulation and FH--LFM
phase compensation to ensure inter-subpulse phase continuity.
Together with the controlled comparison in Fig.~\ref{fig:spectrum},
these leakage results are consistent with the suppression of spectral
spreading caused by boundary phase jumps through the continuous-phase design.

Overall, under the evaluated configurations, CP-FH maintains the same
relatively high information rate as POCS with comparable BER performance,
while reducing spectral leakage through its continuous-phase design and
enabling noncoherent reception without accurate channel-phase prior
information.
\endgroup

\subsection{Radar Processing Performance}
\label{subsec:radar_performance}

\subsubsection{Single-Target SMC-Domain Response}
\label{subsubsec:frv_response}

This experiment uses the waveform parameters in Section~\ref{subsec:waveform_validation}.
The target parameters are set as \(R_0=2000~\mathrm{m}\),
\(v_0=20~\mathrm{m/s}\), and \(\theta_0=15^\circ\), with the processing
template steered to \(\tilde{\theta}=15^\circ\). The frequency spacing is
\(\Delta f=\textcolor{red}{4}~\mathrm{MHz}\), and the pulse repetition
interval is \(T_{\mathrm r}=80~\mu\mathrm{s}\). The simulation uses noise-free echoes.

\begin{figure}[h]
	\centering
	\includegraphics[width=0.47\textwidth]{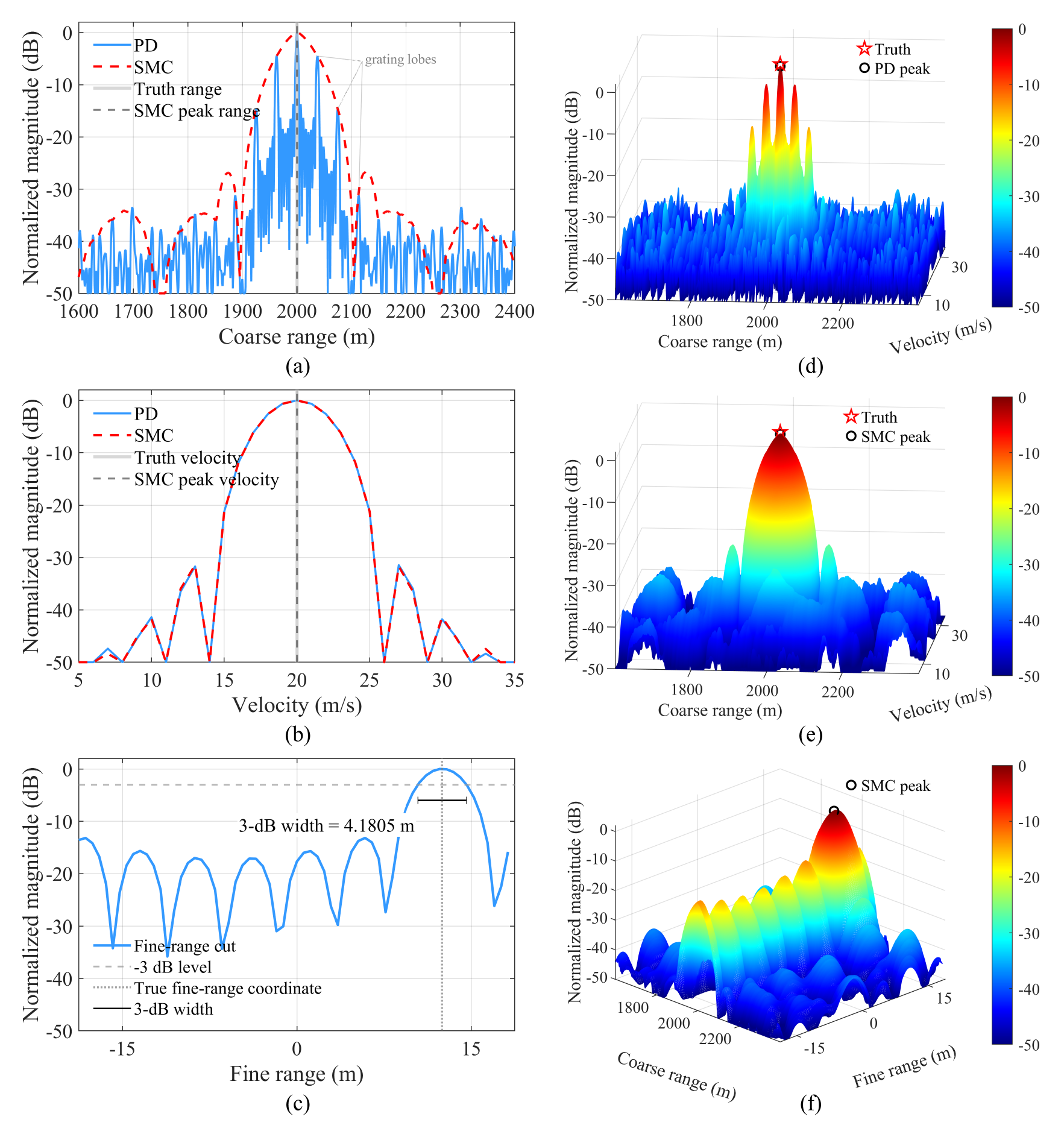}
	\caption{Single-target response comparison between conventional PD and
		the proposed SMC processor: (a) coarse-range cut; (b) velocity cut;
		(c) \textcolor{red}{within-cycle} fine-range cut with the measured 3-dB mainlobe width;
		(d) coarse range--velocity response of PD; (e) coarse range--velocity
		slice of the SMC output; (f) \((u,\ell)\)-slice of the SMC output at
		the target velocity bin.}
	\label{fig:comparison_STF_vs_WB}
\end{figure}

Figs.~\ref{fig:comparison_STF_vs_WB}(a), (b), (d), and (e) first examine the
coarse range--velocity localization of a single target. The SMC peak appears at
the same coarse range and velocity as the conventional PD peak, indicating that
the additional fine-delay focusing does not distort the original coarse
localization result. At the same time, the conventional PD coarse-range response
contains frequency-aperture-induced range grating lobes around the target,
whereas SMC preserves the coarse range--velocity envelope.

\begingroup\color{red}
The same SMC peak gives \(R_{\mathrm c}=2000.2500~\mathrm{m}\),
\(R_{\mathrm f}=12.3047~\mathrm{m}\), and \(v_q=20~\mathrm{m/s}\).
All eight frequency branches are active, so \(g_\mu=1\), and the
fine-range period is
\[
T_{\mu,q}
=\frac{c}{2(1-2v_q/c)\Delta f}
\approx37.500005~\mathrm{m}.
\]
The fine coordinate specifies a within-cycle position, with physical-range
candidates \(R_{\mathrm f}+mT_{\mu,q}\). Applying
\eqref{eq:range_cycle_selection} within the coarse-range unambiguous
interval \([0,12000)~\mathrm{m}\) selects \(\widehat m=53\), yielding
\[
\widehat R^{(\mathrm f)}
=R_{\mathrm f}+53T_{\mu,q}
\approx1999.8050~\mathrm{m}.
\]
Relative to \(R_0=2000~\mathrm{m}\), the absolute errors of the coarse
estimate \(\widehat R^{(\mathrm c)}=R_{\mathrm c}\) and the fine-range
reconstruction are \(0.2500~\mathrm{m}\) and \(0.1950~\mathrm{m}\),
respectively. In this scenario, the fine-range reconstruction yields a smaller range-estimation error.
\endgroup

The fine-range \textcolor{red}{focusing capability} is then validated in
Fig.~\ref{fig:comparison_STF_vs_WB}(c), which shows the fine-range cut through
the 3-D SMC peak. \textcolor{red}{Its horizontal axis shows the centered
within-cycle coordinate in \([-T_{\mu,q}/2,T_{\mu,q}/2)\), rather than
the reconstructed physical range.} The measured 3-dB mainlobe width is
\textcolor{red}{\(4.1805~\mathrm{m}\)}. For \(K=8\) and \(\Delta f=\textcolor{red}{4}~\mathrm{MHz}\), the
theoretical 3-dB fine-range resolution is
\[
\Delta R_{\mathrm{fine}}^{3\mathrm{dB}}=
\frac{c}{2}\cdot\frac{0.886}{K\Delta f}
\approx 4.1531~\mathrm{m}.
\]
The relative error between the measured and theoretical values is about
\textcolor{red}{\(0.66\%\), showing close agreement with the approximate
fine-range resolution analysis.}

\subsubsection{\textcolor{red}{Two-Target Responses and Discrimination}}
\label{subsubsec:multi_target_response}

\begin{figure}[h]
\centering
\includegraphics[width=\columnwidth]{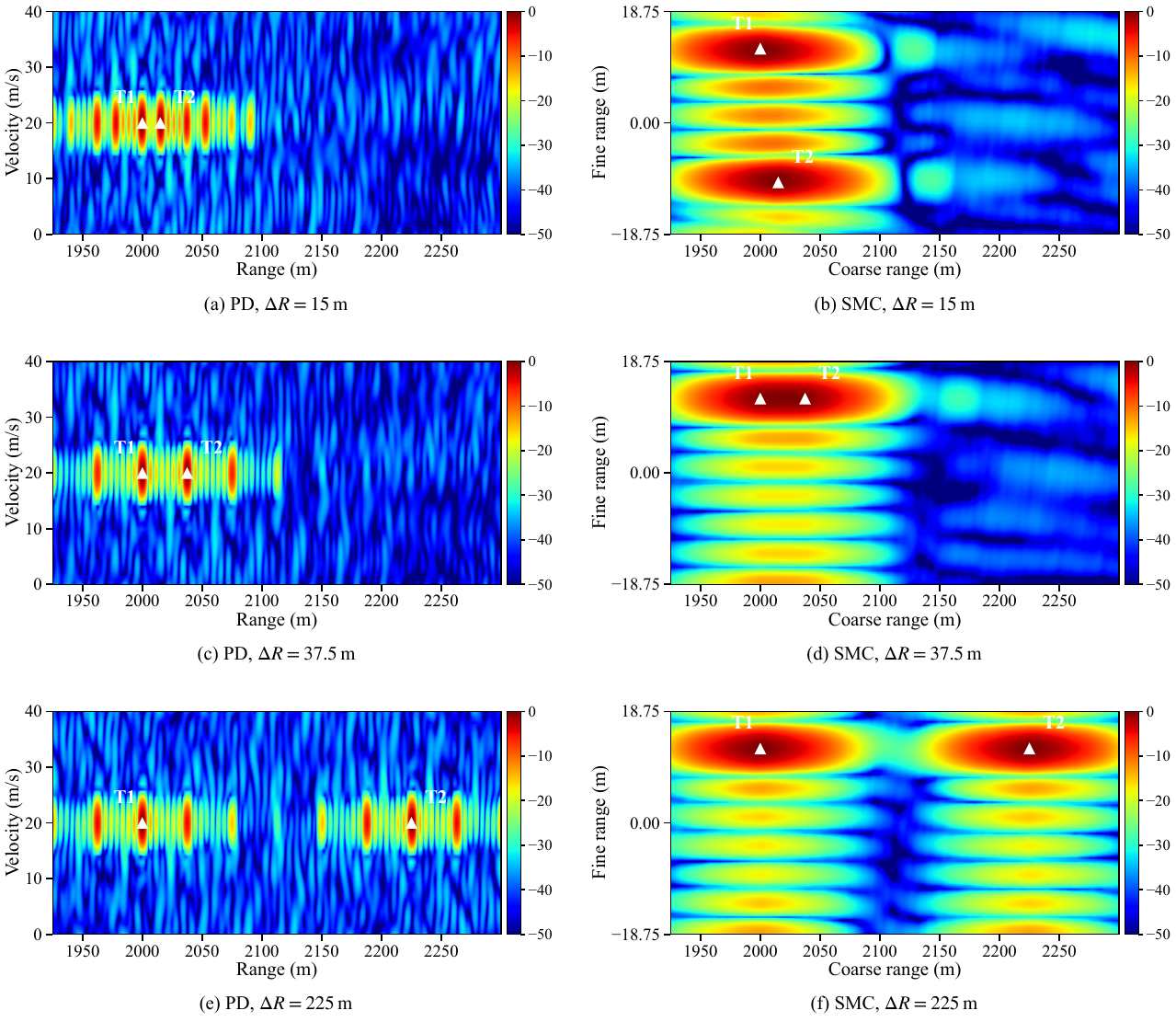}
\caption{\textcolor{red}{Two-target responses: (a), (b) Scenario 1;
(c), (d) Scenario 2; (e), (f) Scenario 3. The left column shows PD
range--velocity responses; the right column shows SMC coarse--fine
range responses at the common target velocity of \(20~\mathrm{m/s}\).}}
\label{fig:multi-target}
\end{figure}

\begingroup\color{red}
This experiment uses the same waveform and processing parameters as the
single-target experiment to examine the roles of coarse and fine range
in two-target discrimination. The two targets have equal strength,
velocity \(20~\mathrm{m/s}\), and angle \(15^\circ\). Their ranges in the
three scenarios are listed in Table~\ref{tab:multi_target_params}, and
the corresponding responses are shown in Fig.~\ref{fig:multi-target}.
Under \(\Delta f_{\mathrm{eff}}\approx\Delta f\), the fine-range period is
\(
T_f=\frac{c}{2\Delta f}=37.5~\mathrm m.
\)
\endgroup

\begin{table}[h]
\centering
\caption{\textcolor{red}{Target ranges for the three two-target scenarios, with \(\Delta R=R_2-R_1\).}}
\label{tab:multi_target_params}
\begingroup\color{red}
\setlength{\tabcolsep}{4pt}
\begin{tabular}{@{}ccccc@{}}
\toprule
Scenario & \(R_1\) (m) & \(R_2\) (m) & \(\Delta R\) (m) & \(\Delta R/T_f\) \\
\midrule
1 & 2000 & 2015 & 15 & 0.4 \\
2 & 2000 & 2037.5 & 37.5 & 1 \\
3 & 2000 & 2225 & 225 & 6 \\
\bottomrule
\end{tabular}
\endgroup
\end{table}

\begingroup\color{red}
Figs.~\ref{fig:multi-target}(a) and (b) show the responses for a target
separation of \(15~\mathrm m\). PD resolves the two targets but also
exhibits multiple periodic range grating lobes, whereas SMC produces
two separated peaks along fine range despite the overlapping
coarse-range responses.

When the separation increases to one fine-range period, PD in
Fig.~\ref{fig:multi-target}(c) still resolves the two targets, but the
main peak of one target coincides with a grating lobe of the other.
SMC in Fig.~\ref{fig:multi-target}(d) cannot resolve the targets because
their fine-range coordinates coincide and their coarse-range responses
remain unresolved.

Figs.~\ref{fig:multi-target}(e) and (f) further show the responses for a
separation of six fine-range periods. Both processors resolve the
targets, but PD still exhibits periodic range grating lobes. Although
the fine-range coordinates coincide, SMC distinguishes the targets
through their separated coarse-range responses.

These results show that SMC eliminates PD's periodic range grating lobes
that may be mistaken for additional targets, while retaining periodicity
in the fine-range coordinate.
However, target discrimination remains limited when the fine-range
coordinates coincide and the coarse-range responses cannot resolve
the targets.
\endgroup

\subsubsection{\textcolor{red}{Range-Domain Performance Analysis}}
\label{subsubsec:joint_range_resolution}

\begingroup\color{red}
The joint range profiles in this subsection are sampled along the physical
coordinate mapping \((R,\operatorname{mod}(R,T_f))\), using the
\(\Delta f_{\mathrm{eff}}\approx\Delta f\) approximation.
For a range offset \(\Delta R\), both coordinates correspond to the same
trial range \(R_0+\Delta R\). These target-referenced response profiles
evaluate local discrimination and periodic replicas; they do not establish
global cycle-identification performance.
\endgroup

\begingroup\color{red}
The two cases in Table~\ref{tab:joint_range_cases} have the same frequency
aperture \(K\Delta f=120~\mathrm{MHz}\) but different frequency spacings
\(\Delta f\), allowing joint range ambiguity to be compared at a similar
local resolution scale.
\endgroup

\begin{table}[h]
	\centering
	\caption{\textcolor{red}{Parameters for joint range-response validation.}}
	\label{tab:joint_range_cases}
	\setlength{\tabcolsep}{4pt}
	\begin{tabular}{@{}c c c c c c@{}}
		\toprule
		Case & \(K\) & \(\Delta f\) (MHz) & \(T_f\) (m) & \(\textcolor{red}{\delta_{\mathrm c}}\) (m) & \(T_f/\textcolor{red}{\delta_{\mathrm c}}\)  \\
		\midrule
		A & 12 & 10 & 15.00 & 11.84 & 1.27  \\
		B & 4  & 30 & 5.00  & 12.21 & 0.41  \\
		\bottomrule
	\end{tabular}
\end{table}

\begin{figure}[h]
	\centering
	\includegraphics[width=\columnwidth]{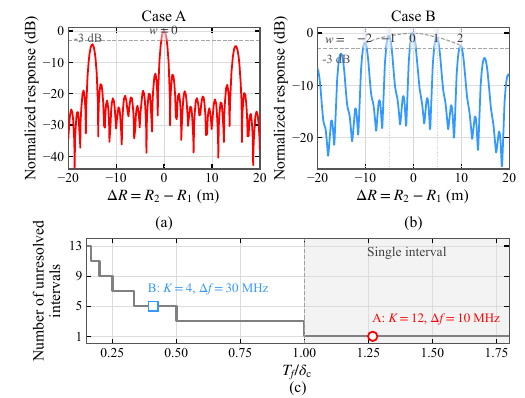}
	\caption{\textcolor{red}{Joint coarse--fine range responses and ambiguity:
		(a) Case A; (b) Case B; (c) number of 3-dB unresolved intervals
		versus \(T_f/\delta_{\mathrm c}\).}}
	\label{fig:joint_range_resolution}
\end{figure}

\begingroup\color{red}
Figs.~\ref{fig:joint_range_resolution}(a) and (b) show one and five 3-dB
unresolved intervals for Cases A and B, respectively. In Case B, the
intervals are centered near \(wT_f\), \(w=0,\pm1,\pm2\), and narrow away
from the central peak as the coarse-range envelope decreases.
Fig.~\ref{fig:joint_range_resolution}(c) summarizes the predicted stepwise
increase in interval count as \(T_f/\textcolor{red}{\delta_{\mathrm c}}\)
decreases. The two cases thus have comparable local widths but distinct
ambiguity structures.
\endgroup

\begin{figure}[!h]
	\centering
	\includegraphics[width=\columnwidth]{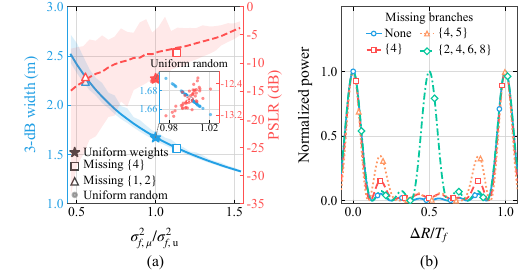}
	\caption{\textcolor{red}{Effects of branch-energy distribution and omissions on
		fine-range responses: (a) 3-dB width and PSLR versus normalized weighted
		frequency variance;
		(b) fine-range response under branch omissions.}}
	\label{fig:occupancy_range_validation}
\end{figure}

\begingroup\color{red}
Using the single-target SMC-domain response settings,
Fig.~\ref{fig:occupancy_range_validation}(a) evaluates random branch-energy
allocations at fixed total energy. Curves and shading give metric means
and sample ranges computed from \eqref{eq:app_Af_general_range}, grouped
by weighted frequency variance normalized to the uniform-weight value.
Positive weights are drawn uniformly subject to \(\sum_k\mu_k=1\) and
\(\mu_k\le1/N\), then mapped to CPI occupancy counts. Each normalized-variance
bin of width \(0.04\) over \([0.44,1.56]\) contains 384 allocations with
no missing branches.
PSLR excludes periodic main-peak replicas. Dots represent 50
noise-free waveform realizations with uniformly random frequency
combinations and CPM symbols. Mainlobe width generally decreases with
variance, consistent with the local response analysis, whereas PSLR
generally increases. The PSLR spread at comparable variances shows its
dependence on the specific energy allocation. The uniformly random
realizations remain close to the uniform-weight reference, with deviations
within \(1.23\%\) in 3-dB width and \(0.55~\mathrm{dB}\) in PSLR.

Fig.~\ref{fig:occupancy_range_validation}(b) tests CPI-wide branch omissions
at fixed total energy and approximately balanced active-branch occupancy.
Curves and open markers denote waveform simulations and theoretical
predictions, respectively.
Omitting either \(\{4\}\) or \(\{4,5\}\) preserves \(g_\mu=1\) and the
period \(T_f=15~\mathrm{m}\), whereas alternating omissions
\(\{2,4,6,8\}\) yield \(g_\mu=2\) and shorten the period to
\(T_f/2=7.5~\mathrm{m}\).
Waveform simulations agree with predictions based on the actual branch
energies, supporting the weighted-response and periodicity analysis in
Appendix~\ref{app:joint_range_branch}.
\endgroup

\begingroup\color{red}
\clubpenalty=10000
\subsubsection{Sensitivity of Coherent Integration}
\label{subsubsec:velocity_grid_sensitivity}

The waveform and scene settings follow Section~\ref{subsubsec:frv_response},
with $T_{\mathrm r}=80~\mu\mathrm{s}$. Noise-free responses at the true
target angle are evaluated under the fixed-delay model to examine grid
mismatch and residual phase differences arising from intrapulse Doppler.

In Fig.~\ref{fig:coherent_integration_sensitivity}(a), with $P=64$ and
a target velocity of $20~\mathrm{m/s}$, the maximum normalized coherent-sum
error across frequency branches under half-bin mismatch decreases with $Q$
and remains below the bound $\varepsilon_{\mathrm d}=\pi(P-1)/Q$ in
Appendix~\ref{app:res_doppler}.
Fig.~\ref{fig:coherent_integration_sensitivity}(b) compares nearest-bin
and true-velocity compensation for 2000 velocities drawn uniformly from
$[-v_{\max},v_{\max})$. The loss curves nearly coincide for different
$P$; at $Q/P=4$, the mean and sample maximum power losses are approximately
$0.074~\mathrm{dB}$ and $0.223~\mathrm{dB}$, respectively.

\begin{figure}[!ht]
    \centering
    \includegraphics[width=\columnwidth]{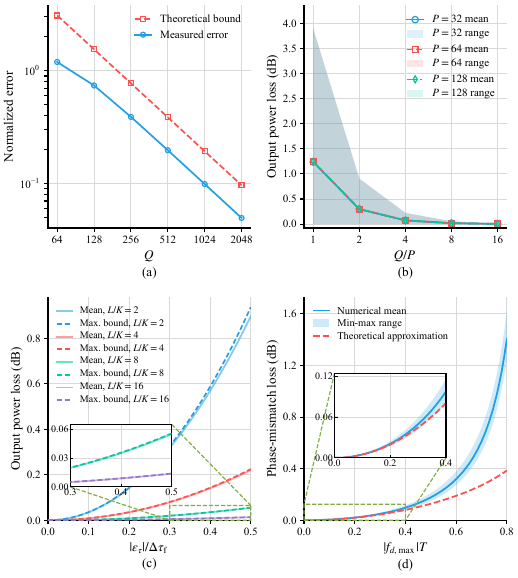}
    \caption{\textcolor{red}{Sensitivity of coherent integration:
        (a) velocity-grid error; (b) velocity-grid power loss;
        (c) fine-delay-grid power loss; (d) residual-phase coherent loss.}}
    \label{fig:coherent_integration_sensitivity}
    \label{fig:velocity_grid_sensitivity}
    \label{fig:phase_coherent_loss}
\end{figure}

Fig.~\ref{fig:coherent_integration_sensitivity}(c) evaluates fine-delay-grid
power loss for a stationary target with aligned coarse delay, using the
continuous fine-delay peak of the same response as the reference.
The grid spacing is $\Delta\tau_{\mathrm f}=1/(L\Delta f)$, with 32 random
waveform realizations and offsets of both signs up to half a grid interval.
The loss increases with offset magnitude and decreases with $L/K$, agreeing
with the theoretical bounds in Appendix~\ref{app:fine_delay_grid_mismatch}.
At half-bin offset, the mean losses for $L/K=4$ and $8$ are approximately
$0.221~\mathrm{dB}$ and $0.055~\mathrm{dB}$, respectively.

Fig.~\ref{fig:coherent_integration_sensitivity}(d) evaluates coherent loss using the desired-branch
matched-filter model with focusing at the true delay and velocity, relative
to an output with exact residual-phase compensation for each pulse and branch.
The same 200 random frequency assignments are used across normalized velocities
$x=2\tilde f_{\max}|\tilde\nu|T/c\in[0,0.8]$.
The approximation in Appendix~\ref{app:phase_coherent_loss} agrees with the
numerical results at low velocities, with increasing discrepancy at higher
velocities. For $x\leq0.2$, the maximum difference between the mean loss curves
is below $0.001~\mathrm{dB}$; this interval corresponds to radial speeds up to
approximately $99.3~\mathrm{m/s}$, and the maximum coherent-integration loss
over the tested assignments and velocity points is below $0.025~\mathrm{dB}$.

\endgroup

\subsubsection{Output-SNR Validation}

On the signal side, the evaluated
quantity is
\(10\log_{10}(E_{\mathrm{SMC}}^{(s)}/E_{\mathrm{PD}}^{(s)})\), where
\begin{equation}
	\label{eq:signal_energy_ratio_def}
	\begin{aligned}
		E_{\mathrm{SMC}}^{(s)}
		\triangleq
		\left|
		Z_{\mathrm{SMC}}^{(s)}(\eta^\star)
		\right|^2,
		E_{\mathrm{PD}}^{(s)}
		\triangleq
		\left|
		Z_{\mathrm{PD}}^{(s)}(\eta^\star)
		\right|^2 .
	\end{aligned}
\end{equation}
On the noise side, the evaluated quantity is the average output noise-power
ratio \(10\log_{10}(\bar E_{\mathrm{SMC}}^{(n)}/\bar E_{\mathrm{PD}}^{(n)})\),
where
\begin{equation}
	\bar E_{\mathrm{SMC}}^{(n)}
	\triangleq
	\frac{1}{UQL}
	\sum_{u=1}^{U}
	\sum_{q=1}^{Q}
	\sum_{\ell=1}^{L}
	\left|
	Y_{\mathrm{SMC}}^{(n)}(u,q,\ell)
	\right|^2,
	\label{eq:noise_energy_smc_sim}
\end{equation}
and
\begin{equation}
	\bar E_{\mathrm{PD}}^{(n)}
	\triangleq
	\frac{1}{UQ}
	\sum_{u=1}^{U}
	\sum_{q=1}^{Q}
	\left|
	Y_{\mathrm{PD}}^{(n)}(u,q)
	\right|^2.
	\label{eq:noise_energy_pd_sim}
\end{equation}

\begin{figure}[h]
	\centering
	\includegraphics[width=0.50\textwidth]{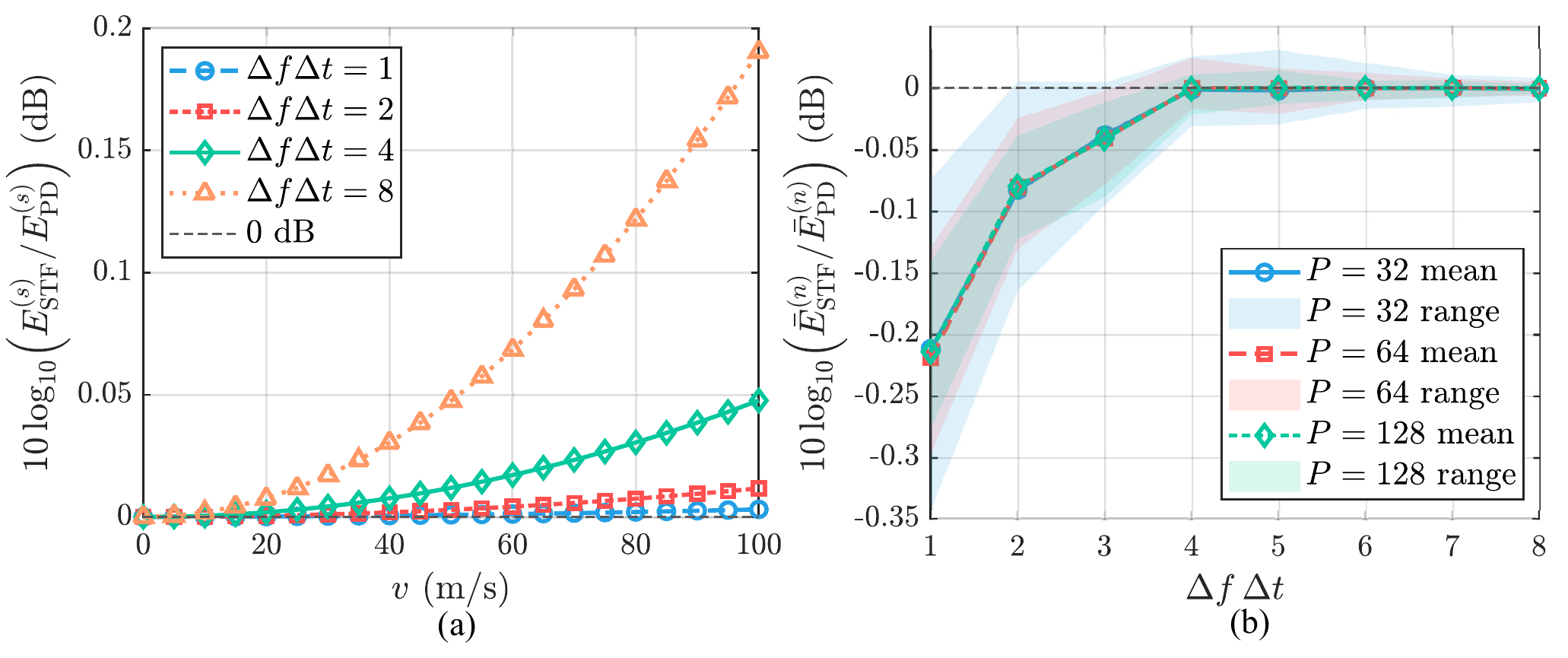}
	\caption{Output-SNR validation: (a) target-aligned signal-energy ratio
		\(10\log_{10}(E_{\mathrm{SMC}}^{(s)}/E_{\mathrm{PD}}^{(s)})\) versus target
		velocity \(v\) for \(P=64\); (b) average output noise-power ratio
		\(10\log_{10}(\bar E_{\mathrm{SMC}}^{(n)}/\bar E_{\mathrm{PD}}^{(n)})\) versus
		\(\Delta f\Delta t\) for different pulse numbers \(P\).}
	\label{fig:snr_behavior}
\end{figure}

Fig.~\ref{fig:snr_behavior}(a) shows that the signal-energy ratio is close to
\(0\) dB at \(v=0\), where the frequency-dependent Doppler mismatch of PD is
negligible. As \(v\) and \(\Delta f\Delta t\) increase, the single-reference PD
chain suffers a larger residual Doppler mismatch across frequency branches,
whereas SMC explicitly compensates the branch-dependent phase terms. Therefore,
the SMC signal-side advantage becomes more pronounced.

Fig.~\ref{fig:snr_behavior}(b) shows that the average output noise-power ratio
remains close to \(0\) dB. This agrees with the noise-side analysis, where the
SMC--PD noise-power difference is governed by the residual inter-channel
correlation term. Increasing \(P\) mainly reduces the fluctuation range of the
noise-power ratio, while its mean level remains nearly unchanged.

\raggedbottom
\subsubsection{\textcolor{red}{Detection Performance}}
\label{subsubsec:detection_performance}

\begingroup\color{red}
The preceding SNR analysis uses averaged output noise power, whereas
detection thresholds depend on background power estimated from local
reference cells. To further compare the detection performance of SMC and
PD, Monte Carlo simulations use the CP-FH waveform parameters of the
communication experiment, with \(P=64\), a pulse repetition interval of
\(T_{\mathrm r}=160~\mu\mathrm{s}\), and a target velocity of
\(20~\mathrm{m/s}\).
PD forms its matched-filter reference by combining
Hann-weighted frequency components, while SMC combines frequency branches
using the same weights; both receivers use a Hann window for coherent
integration across pulses. Each point comprises \(N_{\mathrm{MC}}=2000\)
trials.

The single-target range is uniformly distributed over
\([1995,2032.5)~\mathrm{m}\). The detection statistic is the ratio of the
output power at the cell nearest the true target coordinates to the mean
power of its local reference cells; detection is declared when this
statistic exceeds the threshold. Fig.~\ref{fig:detection_curves}(a) shows
the detection probabilities for per-cell false-alarm probabilities
\(p_{\mathrm{fa}}=10^{-2}\), \(10^{-3}\), and \(10^{-4}\), with the
thresholds for the two processors separately calibrated using noise-only
trials. At each specified \(p_{\mathrm{fa}}\), the SMC and PD detection
probability curves nearly coincide, indicating comparable target-cell
detection performance under this configuration.

\begin{figure}[!t]
    \centering
    \includegraphics[width=\columnwidth]{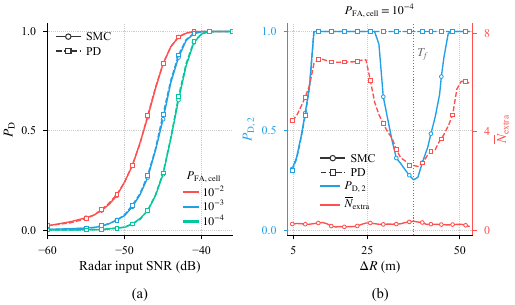}
    \caption{\textcolor{red}{Detection performance: (a) single-target
        detection versus radar input SNR; (b) two-target detection and mean extra detections
        versus \(\Delta R\).
        }}
    \label{fig:detection_curves}
\end{figure}

The two-target experiment considers two equal-strength targets with the
same velocity and a random relative phase. The radar input SNR\footnote{The radar input SNR is the mean per-target echo power over the active pulse duration divided by the input noise power.} is fixed at
\(-20~\mathrm{dB}\) per target to reduce noise-induced misses. The first
target's range remains uniformly distributed over
\([1995,2032.5)~\mathrm{m}\), and \(\Delta R\) varies from \(4.6875\) to \(52.5~\mathrm{m}\). Both methods use
\(p_{\mathrm{fa}}=10^{-4}\) and group neighbouring local peaks into
connected clusters, retaining the strongest peak in each cluster as a
detection point.

We count the detection points matched one-to-one to the true targets.
Matching requires the physical-range error for PD or the periodic
fine-range error for SMC to be below \(c/(2K\Delta f)=4.6875~\mathrm{m}\),
and the velocity error to be below
\(c/(2f_{\mathrm c}PT_{\mathrm r})\approx1.465~\mathrm{m/s}\).
SMC additionally requires a coarse-range error below \(31.5~\mathrm{m}\),
with this tolerance determined by the measured half-power half-width of
the windowed coarse-range response. Let \(N_i\) denote the number of
clustered detection points in trial \(i\), and \(M_i\) the maximum number
of one-to-one matches satisfying these conditions. The two-target
detection probability and the mean number of extra detections are then
defined as
\begin{equation}
\begin{aligned}
\widehat P_{\mathrm D,2}
&=\frac{1}{N_{\mathrm{MC}}}
\sum_{i=1}^{N_{\mathrm{MC}}}\mathbf{1}\{M_i=2\},\\
\overline N_{\mathrm{extra}}
&=\frac{1}{N_{\mathrm{MC}}}
\sum_{i=1}^{N_{\mathrm{MC}}}(N_i-M_i).
\end{aligned}
\label{eq:detection_two_target_metrics}
\end{equation}
\begingroup\interlinepenalty=10000\widowpenalty=10000\clubpenalty=10000
Here, \(\widehat P_{\mathrm D,2}\) is the fraction of trials in which
both true targets are matched to distinct detection points, and
\(\overline N_{\mathrm{extra}}\) is the mean number of detection points
not matched to true targets per trial.
\par\endgroup

In Fig.~\ref{fig:detection_curves}(b), the two-target detection
probabilities are similar and increase with separation in the
small-separation region. When the separation approaches the fine-range
period \(T_f=37.5~\mathrm{m}\), the targets have similar fine-range
coordinates and remain unresolved in coarse range, causing SMC's
detection probability to decrease. For PD, the true peak of one target
overlaps with a grating lobe of the other near the period; the combined
response can still satisfy the matching criteria near both true target
locations, maintaining a high two-target detection probability.
Approximately \(97.4\%\) of PD's extra detections lie within the matching
tolerances of nonzero periodic replicas of the true targets, indicating
that periodic grating lobes are the main source of its extra detections.
SMC produces very few such detections and substantially fewer total
extra detections than PD, showing that joint coarse--fine range
processing effectively suppresses grating-lobe-induced extra detections.

\endgroup

\subsubsection{Receive Beampattern}
\label{subsubsec:receive_beampattern}

The effective receive beampattern is evaluated in a noise-free single-target
setting with \(R=0\), \(v=0\), and steering angle \(\theta_0=0^\circ\). The
receive power is normalized by the peak response at \(\theta=\theta_0\), and
the angular scan range is \([-90^\circ,90^\circ]\).

\begin{figure}[h]
	\centering
	\includegraphics[width=0.30\textwidth]{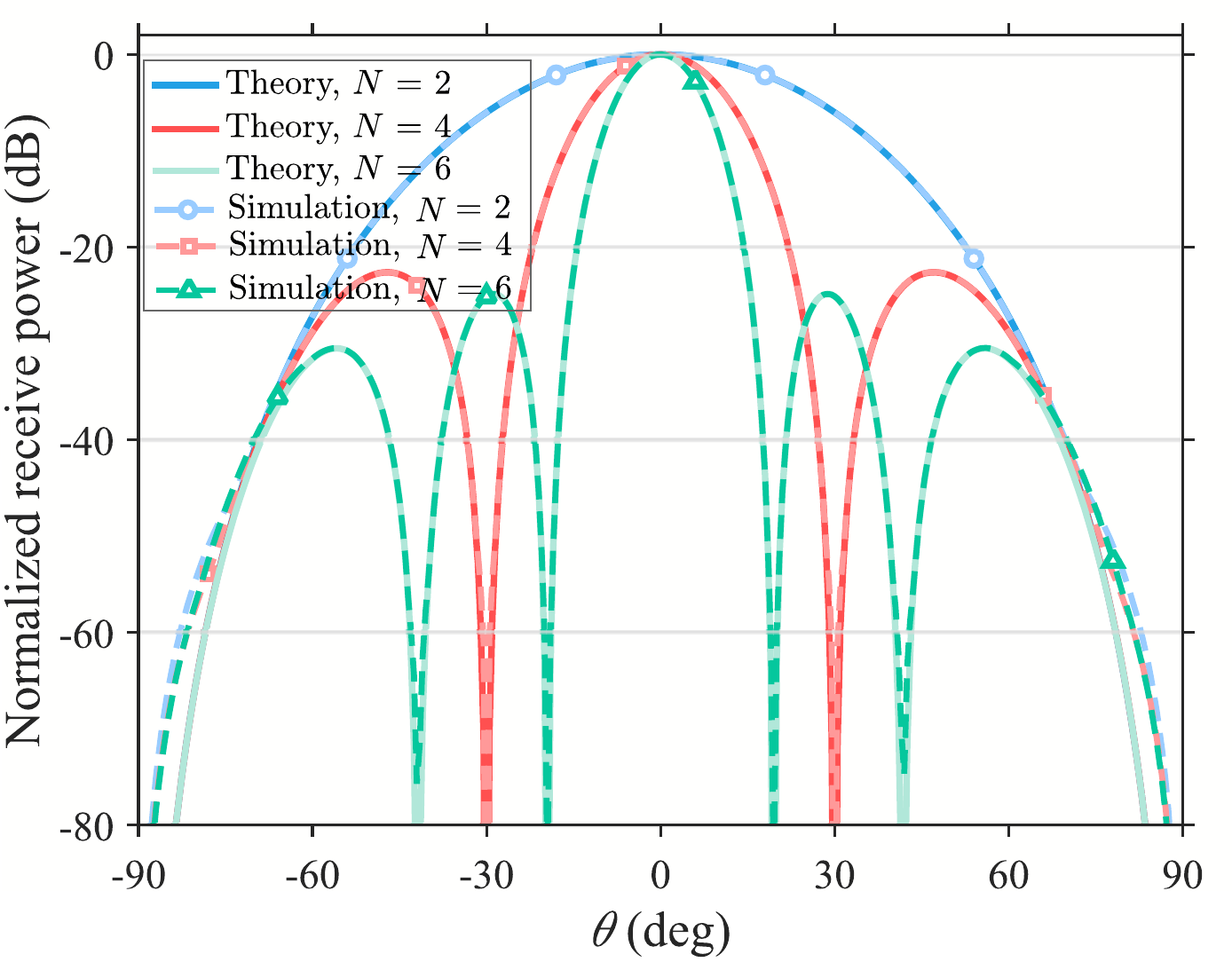}
	\caption{Theoretical and simulated receive beampatterns of the proposed
		SMC receiver for different array sizes \(N\).}
	\label{fig:receive_beampattern}
\end{figure}

\begingroup\color{red}
In Fig.~\ref{fig:receive_beampattern}, the full-chain simulations agree
with the theoretical receive beampatterns for different array sizes.
As \(N\) increases from \(2\) to \(6\), the mainlobe narrows, reflecting
the improved angular focusing provided by the larger array aperture.
\endgroup

%% file: sections/99_conclusion.tex
\section{Conclusion}
\label{sec:conclusion}
\begingroup\color{red}
This paper proposed a CP-FH MIMO waveform that jointly conveys frequency and CPM information while maintaining phase continuity and reducing spectral leakage. The analysis establishes orthogonality and residual cross-correlation conditions and bounds the averaged transmit-beampattern ripple. The proposed PED method approximates least-squares detection under weak inter-branch correlation without accurate channel-phase prior information; its separable search reduces complexity while preserving the joint PED decisions. The proposed SMC processor requires no target-range or velocity priors and combines frequency-aperture fine-range focusing with coarse-range constraints on periodic ambiguity. Under the analyzed colocated-array and matched-beamforming conditions, it recovers the full-aperture coherent spatial response. Simulations validate the theoretical results and demonstrate fewer extra target reports caused by periodic range responses in the evaluated scenarios.
\endgroup

%% file: sections/appendices.tex
\appendices
\section{Proofs for CP-FH Waveform Properties}
\label{app:waveform_properties}
\subsection{Proof of Proposition~\ref{prop:qo_corr_bound}}
\label{app:proof_prop_qo_corr_bound}

Let \(t_v=t-(v-1)\Delta t\in[0,\Delta t]\), and define
\(s_{n,v}(t_v) \triangleq s_n\bigl((v-1)\Delta t+t_v;\boldsymbol{\alpha}_n,\boldsymbol f_n\bigr)\).
From \eqref{eq:rho_nm_v} and \(E_{\mathrm s}=\Delta t\),
\begin{equation}
	\rho_{n,m}^{(v)}
	=
	\frac{1}{\Delta t}
	\int_{0}^{\Delta t}
	s_{n,v}(t_v)s_{m,v}^{*}(t_v)
	\,\mathrm d t_v .
	\label{eq:app_rho_start}
\end{equation}

For the adopted FH--LFM and rectangular full-response CPM phases, the
inter-element product admits the form
\begin{equation}
	s_{n,v}(t_v)s_{m,v}^{*}(t_v)
	=
	e^{j\Phi_{nm,v}}
	e^{j2\pi\Delta f_{nm,v}^{\mathrm{eq}}t_v},
	\label{eq:app_product_phase}
\end{equation}
where
\(\Phi_{nm,v} \triangleq 2\pi(C_{n,v}-C_{m,v}) + \pi\bar h \sum_{i=1}^{v-1} \Delta\alpha_{nm,i},\) and \(\Delta f_{nm,v}^{\mathrm{eq}} \triangleq \Delta f_{nm,v} + \frac{\bar h}{2\Delta t} \Delta\alpha_{nm,v}\).
In \eqref{eq:app_product_phase}, the common LFM quadratic term is canceled in
the inter-element product, while the current CPM-symbol difference contributes
the frequency-like term
\(\bar h\Delta\alpha_{nm,v}/(2\Delta t)\).

Substituting \eqref{eq:app_product_phase} into \eqref{eq:app_rho_start} gives
\begin{equation}
	\begin{aligned}
		\rho_{n,m}^{(v)}
		&=
		\frac{e^{j\Phi_{nm,v}}}{\Delta t}
		\int_{0}^{\Delta t}
		e^{j2\pi\Delta f_{nm,v}^{\mathrm{eq}}t_v}
		\,\mathrm d t_v \\
		&=
		e^{j\left(
			\Phi_{nm,v}
			+
			\pi\Delta f_{nm,v}^{\mathrm{eq}}\Delta t
			\right)}
		\operatorname{sinc}
		\left(
		\pi\Delta f_{nm,v}^{\mathrm{eq}}\Delta t
		\right).
	\end{aligned}
	\label{eq:app_rho_sinc_eq}
\end{equation}
Since the exponential factor has unit magnitude and
\(\Delta f_{nm,v}^{\mathrm{eq}}\Delta t = \Delta f_{nm,v}\Delta t + \frac{\bar h}{2}\Delta\alpha_{nm,v}\),
we obtain
\begin{equation}
	\left|
	\rho_{n,m}^{(v)}
	\right|
	=
	\left|
	\operatorname{sinc}
	\left[
	\pi
	\left(
	\Delta f_{nm,v}\Delta t
	+
	\frac{\bar h}{2}\Delta\alpha_{nm,v}
	\right)
	\right]
	\right| .
	\label{eq:app_rho_sinc_abs_direct}
\end{equation}

\emph{Exact orthogonality:}
If
\(\Delta f_{nm,v}\Delta t + \frac{\bar h}{2} \Delta\alpha_{nm,v} \in \mathbb Z\setminus\{0\}\),
then the sinc term in \eqref{eq:app_rho_sinc_abs_direct} is zero, and hence
\(\rho_{n,m}^{(v)}=0\). This proves
\eqref{eq:qo_exact_orthogonality_condition}.

\emph{Bounded residual correlation:}
For \(x\ne0\),
\(\left| \operatorname{sinc}(\pi x) \right| \le \frac{1}{\pi |x|}\).
Therefore, \(|\rho_{n,m}^{(v)}|\le\varepsilon\) is guaranteed if
\begin{equation}
	\left|
	\Delta f_{nm,v}\Delta t
	+
	\frac{\bar h}{2}
	\Delta\alpha_{nm,v}
	\right|
	\ge
	\frac{1}{\pi\varepsilon}.
	\label{eq:app_general_spacing_condition}
\end{equation}
For an equally spaced frequency pool with distinct assigned frequencies, let
\(\Delta f_{nm,v}=\ell_{nm,v}\Delta f\), where
\(\ell_{nm,v}\in\mathbb Z\setminus\{0\}\). Let
\(
\Delta\alpha_{\max}
\triangleq
\max_{\alpha,\alpha'\in\mathcal A}
|\alpha-\alpha'|.
\)
\begin{equation}
	\begin{aligned}
		\left|
		\Delta f_{nm,v}\Delta t
		\!+\!
		\frac{\bar h}{2}
		\Delta\alpha_{nm,v}
		\right|
		&\!\ge\!
		|\ell_{nm,v}|\Delta f\Delta t
		\!-\!
		\frac{|\bar h|}{2}\Delta\alpha_{\max} \\
		&\ge
		\Delta f\Delta t
		-
		\frac{|\bar h|}{2}\Delta\alpha_{\max}.
	\end{aligned}
	\label{eq:app_spacing_lower_bound}
\end{equation}
Thus, the sufficient rule
\(\Delta f\Delta t \ge \frac{1}{\pi\varepsilon} + \frac{|\bar h|}{2}\Delta\alpha_{\max}\)
implies \eqref{eq:app_general_spacing_condition} and guarantees
\(|\rho_{n,m}^{(v)}|\le\varepsilon\). This proves
\eqref{eq:qo_min_spacing_rule_main}.

\subsection{Proof of Proposition~\ref{prop:tx_beampattern_bound}}
\label{app:proof_prop_tx_beampattern_bound}

 Define the normalized correlation matrix of the $v$-th subpulse
as
\(\boldsymbol R_x^{(v)} \triangleq \frac{1}{E_{\mathrm s}} \int_{(v-1)\Delta t}^{v\Delta t} \boldsymbol s(t)\boldsymbol s^{\mathsf H}(t) \,\mathrm d t\).
Its diagonal entries are equal to one, and its off-diagonal entries are
\(\rho_{n,m}^{(v)}\). Averaging over \(V\) subpulses gives
\(\overline{\boldsymbol R}_x = \frac{1}{V} \sum_{v=1}^{V} \boldsymbol R_x^{(v)} = \boldsymbol I_N+\boldsymbol E\),
where \(\operatorname{diag}(\boldsymbol E)=\boldsymbol 0\) and
\([\boldsymbol E]_{n,m} = \overline{\rho}_{n,m} \triangleq \frac{1}{V} \sum_{v=1}^{V} \rho_{n,m}^{(v)},\) for \(n\ne m\).

For the normalized transmit steering vector \(\textcolor{red}{\boldsymbol a_{\mathrm{tx}}}(\theta)\),
\(\textcolor{red}{\boldsymbol a_{\mathrm{tx}}}^{\mathsf H}(\theta)
\textcolor{red}{\boldsymbol a_{\mathrm{tx}}}(\theta)=N\) and
\(|[\textcolor{red}{\boldsymbol a_{\mathrm{tx}}}(\theta)]_n|=1\). Therefore,
\begin{equation}
	\Delta\overline P(\theta)
	=
	\frac{
		\textcolor{red}{\boldsymbol a_{\mathrm{tx}}}^{\mathsf H}(\theta)
		\boldsymbol E
		\textcolor{red}{\boldsymbol a_{\mathrm{tx}}}(\theta)
	}{N}.
	\label{eq:app_deltaP_E}
\end{equation}

If the exact-orthogonality condition in
\eqref{eq:qo_exact_orthogonality_condition} holds for all \(n\ne m\) and all
\(v\), then \(\rho_{n,m}^{(v)}=0\). Hence \(\boldsymbol E=\boldsymbol 0\), and
\eqref{eq:app_deltaP_E} gives
\(\Delta\overline P(\theta)=0,\) for all \(\theta\).
This proves the isotropic-beampattern result.

Next consider the random-assignment case. From \eqref{eq:app_deltaP_E} and the
triangle inequality,
\(\left| \Delta\overline P(\theta) \right| \le \frac{1}{N} \sum_{n\ne m} \left| \overline{\rho}_{n,m} \right|\).
Taking expectation and using the identical distribution of the ordered
off-diagonal antenna-pair correlations yields
\begin{equation}
	\mathbb E
	\left\{
	\left|
	\Delta\overline P(\theta)
	\right|
	\right\}
	\le
	(N-1)
	\mathbb E
	\left\{
	\left|
	\overline{\rho}_{n,m}
	\right|
	\right\}.
	\label{eq:app_deltaP_pair_bound}
\end{equation}

It remains to bound
\(\mathbb E\{|\overline{\rho}_{n,m}|\}\). First, by the triangle inequality,
\(\mathbb E \left\{ \left| \overline{\rho}_{n,m} \right| \right\} \le \frac{1}{V} \sum_{v=1}^{V} \mathbb E \left\{ \left| \rho_{n,m}^{(v)} \right| \right\} = \mu_1\).
Combining \eqref{eq:app_deltaP_pair_bound} with this inequality gives
\begin{equation}
	\mathbb E
	\left\{
	\left|
	\Delta\overline P(\theta)
	\right|
	\right\}
	\le
	(N-1)\mu_1 .
	\label{eq:app_Btri_bound}
\end{equation}

Second, by Jensen's inequality,
\(\mathbb E \left\{ \left| \overline{\rho}_{n,m} \right| \right\} \le \sqrt{ \mathbb E \left\{ \left| \overline{\rho}_{n,m} \right|^2 \right\} }\).
Using
\(
\overline{\rho}_{n,m}
=
V^{-1}\sum_{v=1}^{V}\rho_{n,m}^{(v)}
\),
we have
\begin{equation}
	\begin{aligned}
		\mathbb E
		\left\{
		\left|
		\overline{\rho}_{n,m}
		\right|^2
		\right\}
		&=
		\frac{1}{V^2}
		\sum_{v=1}^{V}
		\mathbb E
		\left\{
		\left|
		\rho_{n,m}^{(v)}
		\right|^2
		\right\} \\
		&\quad
		+
		\frac{1}{V^2}
		\sum_{v\ne u}
		\mathbb E
		\left[
		\rho_{n,m}^{(v)}
		\left(
		\rho_{n,m}^{(u)}
		\right)^*
		\right].
	\end{aligned}
	\label{eq:app_rhobar_second_expand}
\end{equation}
By the definitions of \(\mu_2\) and \(\beta_V\),
\(\mathbb E \left\{ \left| \overline{\rho}_{n,m} \right|^2 \right\} \le \frac{\mu_2}{V} + \frac{V-1}{V}\beta_V\).
Combining \eqref{eq:app_deltaP_pair_bound}, Jensen's inequality, and this second-moment bound gives
\begin{equation}
	\mathbb E
	\left\{
	\left|
	\Delta\overline P(\theta)
	\right|
	\right\}
	\le
	(N-1)
	\sqrt{
		\frac{\mu_2}{V}
		+
		\frac{V-1}{V}\beta_V
	}.
	\label{eq:app_Brms_bound}
\end{equation}
Taking the tighter of \eqref{eq:app_Btri_bound} and
\eqref{eq:app_Brms_bound} proves the beampattern-ripple bound.

Finally, \(0\le\beta_V\) follows from its definition. For \(v\ne u\), the
Cauchy--Schwarz inequality gives
\[
	\left|
	\mathbb E
	\left[
	\rho_{n,m}^{(v)}
	\left(
	\rho_{n,m}^{(u)}
	\right)^*
	\right]
	\right|
	\le
	\sqrt{
		\mathbb E
		\left\{
		\left|
		\rho_{n,m}^{(v)}
		\right|^2
		\right\}
		\mathbb E
		\left\{
		\left|
		\rho_{n,m}^{(u)}
		\right|^2
		\right\}
	}
	=
	\mu_2 .
	\label{eq:app_beta_cauchy}
\]
Averaging over \(v\ne u\) yields
\(0\le\beta_V\le\mu_2\).
This completes the proof of Proposition~\ref{prop:tx_beampattern_bound}.

\subsection{Scaling of the Residual-Correlation Moments}
\label{app:scaling_residual_moments}

This subsection derives the scaling of the residual-correlation moments
\(\mu_1\) and \(\mu_2\). Under the random frequency--phase assignment,
\(\delta=k_{n,v}-k_{m,v}\) takes values in
\(\mathcal D_K=\{\pm1,\ldots,\pm(K-1)\}\), with
\(\Pr\{\delta\} = \frac{K-|\delta|}{K(K-1)}\).
The CPM symbols \(\alpha\) and \(\alpha'\) are independently and uniformly
drawn from \(\mathcal A\). From \eqref{eq:app_rho_sinc_abs_direct}, for
\(i=1,2\),
\[
	\mu_i
	\!= \!\!\!
	\sum_{\delta\in\mathcal D_K}\!\!\!
	\Pr\{\delta\}
	\mathbb E_{\alpha,\alpha'}
	\left\{
	\left|
	\operatorname{sinc}
	\left[
	\pi
	\left(
	\delta\Delta f\Delta t
	+
	\frac{\bar h}{2}(\alpha\!-\!\alpha')
	\right)
	\right]
	\right|^i
	\right\}.
	\label{eq:app_mu_finite_sum}
\]

When the FH spacing dominates the bounded CPM perturbation, namely
\(\Delta f\Delta t \ge |\bar h|\Delta\alpha_{\max}\),
then, for \(1\le|\delta|\le K-1\),
\begin{equation}
	\begin{aligned}
		\left|
		\delta\Delta f\Delta t
		+
		\frac{\bar h}{2}(\alpha-\alpha')
		\right|
		&\ge
		|\delta|\Delta f\Delta t
		-
		\frac{|\bar h|}{2}
		|\alpha-\alpha'| \\
		&\ge
		\frac{1}{2}
		|\delta|\Delta f\Delta t .
	\end{aligned}
	\label{eq:app_effective_spacing_lower}
\end{equation}
Using
\(
\left|
\operatorname{sinc}(\pi x)
\right|
\le
{1}/{\pi |x|},
 x\ne0,
\)
and \(|\mathcal A|^{-2}\sum_{\alpha,\alpha'\in\mathcal A}1=1\), the first
moment satisfies
\begin{equation}
	\begin{aligned}
		\mu_1
		&\le
		\frac{2}{\pi\Delta f\Delta t}
		\sum_{\substack{\delta=-(K-1)\\ \delta\ne0}}^{K-1}
		\frac{K-|\delta|}{K(K-1)}
		\frac{1}{|\delta|} \\
		&=
		\frac{4}{\pi\Delta f\Delta t}
		\frac{1}{K(K-1)}
		\sum_{r=1}^{K-1}
		\frac{K-r}{r}.
	\end{aligned}
	\label{eq:app_mu1_pre_scaling}
\end{equation}
The normalized summation is
\[
\begin{aligned}
		\frac{1}{K(K-1)}
		\sum_{r=1}^{K-1}
		\frac{K-r}{r}
		&=
		\frac{K H_{K-1}-(K-1)}{K(K-1)} \\
		&=
		\frac{H_{K-1}}{K-1}
		-
		\frac{1}{K},
	\end{aligned}
\]
where
\(
H_{K-1}
\triangleq
\sum_{r=1}^{K-1}
{1}/{r}
\)
is the \((K-1)\)-th harmonic number.
Using \(H_{K-1}=\mathcal O(\ln K)\) \cite{Graham1994ConcreteMathematics},
\(\mu_1 = \mathcal O \left( \frac{\ln K}{K\Delta f\Delta t} \right)\).

Similarly, the second moment satisfies
\begin{equation}
	\begin{aligned}
		\mu_2
		&\le
		\frac{4}{\pi^2(\Delta f\Delta t)^2}
		\sum_{\substack{\delta=-(K-1)\\ \delta\ne0}}^{K-1}
		\frac{K-|\delta|}{K(K-1)}
		\frac{1}{|\delta|^2} \\
		&=
		\frac{8}{\pi^2(\Delta f\Delta t)^2}
		\frac{1}{K(K-1)}
		\sum_{r=1}^{K-1}
		\frac{K-r}{r^2}.
	\end{aligned}
	\label{eq:app_mu2_pre_scaling}
\end{equation}
For the remaining summation,
\(\frac{1}{K(K-1)} \sum_{r=1}^{K-1} \frac{K-r}{r^2} = \frac{H_{K-1}^{(2)}}{K-1} - \frac{H_{K-1}}{K(K-1)}\),
where
\(
H_{K-1}^{(2)}
\triangleq
\sum_{r=1}^{K-1}
\frac{1}{r^2}.
\)
Since \(H_{K-1}^{(2)}=\mathcal O(1)\) and
\(H_{K-1}=\mathcal O(\ln K)\) \cite{Graham1994ConcreteMathematics},
\(\frac{1}{K(K-1)} \sum_{r=1}^{K-1} \frac{K-r}{r^2} = \mathcal O \left( \frac{1}{K} \right)\).
Therefore,
\(\mu_2 = \mathcal O \left( \frac{1}{K(\Delta f\Delta t)^2} \right)\).

\section{Auxiliary Derivations for the SMC Processor}
\label{app:smc_aux}
\subsection{Full-Pulse Inter-Channel Correlation Bound}
\label{app:B1}

This subsection derives the full-pulse inter-channel correlation bound used in
Section~\ref{subsubsec:stf_jmf}.
\begingroup\color{red}
Consider the branch-\(k\) receiver and a component on branch \(k'\), where
\(k'\ne k\). For a fixed propagation delay \(\tilde{\tau}\),
target-dependent scalar factors and time-invariant propagation phases do not
affect the normalized correlation magnitude. From \eqref{eq:sR} and
\eqref{eq:downconvert},
\[
\begin{aligned}
&
\int_{-\infty}^{\infty}
x^{\textcolor{red}{\mathrm{b}}}_{p,k'}(t-\tilde{\tau};\tilde{\theta})
x^{\textcolor{red}{\mathrm{b}}*}_{p,k}(t-\tilde{\tau};\tilde{\theta})\times
e^{j2\pi(f_{k'}-f_k)(t_p+t)}
\,\mathrm dt\\
&=
e^{j2\pi(f_{k'}-f_k)\tilde{\tau}}
\int_{-\infty}^{\infty}
x_{p,k'}(\xi;\tilde{\theta})
x_{p,k}^{*}(\xi;\tilde{\theta})
\,\mathrm d\xi .
\end{aligned}
\]
Since the phase factor has unit magnitude, it suffices to bound the correlation
integral on the right-hand side.
\endgroup

For pulse \(p\) and frequency branch \(k\), write the
\textcolor{red}{original frequency-indexed waveform component} as the sum of
its subpulse components:
\(\textcolor{red}{x_{p,k}(t;\tilde{\theta})} = \sum_{v=1}^{V} \textcolor{red}{x_{p,k,v}(t;\tilde{\theta})}\).
\begingroup\color{red}
where \(x_{p,k,v}(t;\tilde{\theta})\) is supported on the
\(v\)-th local subpulse interval
\(\mathcal I_v\triangleq[(v-1)\Delta t,v\Delta t)\). Since these intervals
are pairwise disjoint, for \(v\neq v'\),
\endgroup
\(\int \textcolor{red}{x_{p,k,v}(t;\tilde{\theta})} \textcolor{red}{x^{*}_{p,k',v'}(t;\tilde{\theta})} \mathrm dt = 0\).

Define
\(E^{(v)}_{p,k} \triangleq \int_{\textcolor{red}{\mathcal I_v}} \left| \textcolor{red}{x_{p,k,v}(t;\tilde{\theta})} \right|^2 \mathrm dt\),
and \(E_{p,k}=\sum_{v=1}^{V}E^{(v)}_{p,k}\). For subpulses with
\(E^{(v)}_{p,k}E^{(v)}_{p,k'}>0\), let
\begin{equation}
\rho^{(p,v)}_{k,k'}
\triangleq
\frac{
\int_{\textcolor{red}{\mathcal I_v}}
\textcolor{red}{x_{p,k,v}(t;\tilde{\theta})}
\textcolor{red}{x^{*}_{p,k',v}(t;\tilde{\theta})}
\mathrm dt
}{
\sqrt{
E^{(v)}_{p,k}
E^{(v)}_{p,k'}
}
}.
\end{equation}
\begingroup\color{red}
When branches \(k\) and \(k'\) are both active in subpulse \(v\), they
correspond to distinct transmit elements. Since the directional coefficients
have unit magnitude, applying the subpulse-level bound under the conditions
of Proposition~\ref{prop:qo_corr_bound} gives
\endgroup
\(\left| \rho^{(p,v)}_{k,k'} \right| \le \varepsilon_0\),
\begingroup\color{red}
The full-pulse cross-correlation therefore satisfies
\endgroup
\begin{equation}
\begin{aligned}
\left|
\int
\textcolor{red}{x_{p,k}(t;\tilde{\theta})}
\textcolor{red}{x^{*}_{p,k'}(t;\tilde{\theta})}
\mathrm dt
\right|
&\le
\varepsilon_0
\sum_{v=1}^{V}
\sqrt{
E^{(v)}_{p,k}
E^{(v)}_{p,k'}
}\\
&\le
\varepsilon_0
\sqrt{E_{p,k}E_{p,k'}},
\end{aligned}
\end{equation}
where the last inequality follows from the Cauchy--Schwarz inequality.
\begingroup\color{red}
For \(E_{p,k}E_{p,k'}>0\), the normalized full-pulse inter-channel correlation
\endgroup
\begin{equation}
\rho^{(p)}_{k,k'}
\triangleq
\frac{
\int
\textcolor{red}{x_{p,k}(t;\tilde{\theta})}
\textcolor{red}{x^{*}_{p,k'}(t;\tilde{\theta})}
\mathrm dt
}{
\sqrt{E_{p,k}E_{p,k'}}
}
\end{equation}
satisfies
\(\left| \rho^{(p)}_{k,k'} \right| \le \varepsilon_0\),
\textcolor{red}{If either branch has zero energy, the unnormalized
cross-correlation is zero, while the normalized coefficient is undefined.
For nonzero-energy branches, this establishes the stated inter-branch
correlation bound after branch-\(k\) downconversion at the matched delay
under the fixed-delay model. This result is not a bound on the full
cross-ambiguity function at arbitrary delay and Doppler offsets.}

\subsection{Pulse-Compression Peak Model}
\label{app:B2}

This subsection derives the matched-peak approximation used in
Section~\ref{subsubsec:stf_jmf}.
\begingroup\color{red}
Consider a single target characterized by
\((\tilde{\theta},\tilde{\tau},\tilde{\nu},\tilde{\rho})\), where
\(\tilde{\tau}\) is its two-way propagation delay at the beginning of the CPI.
Following \eqref{eq:tau_pz}, its instantaneous delay in the \(p\)-th PRI is
\(\tau_p(t) = \tilde{\tau}-\frac{2\tilde{\nu}}{c}(t_p+t),\) and \(0\le t<T_{\mathrm r}\).
We consider the desired-branch return fully contained in the receive window,
neglecting noise and inter-branch leakage.
\endgroup
For branch \(k\), the downconverted received echo is written as
\begin{equation}
	\label{eq:appD_rpk}
	r_{p,k}(t)
	=
	\tilde{\beta}\tilde{\rho}
	x^{\textcolor{red}{\mathrm{b}}}_{p,k}\bigl(t-\tau_p(t);\tilde{\theta}\bigr)
	\exp\!\left[-j2\pi\tilde f_k\tau_p(t)\right],
\end{equation}
where \(\tilde{\beta}\triangleq
\boldsymbol{\omega}^{\mathsf H}\textcolor{red}{\boldsymbol a_{\mathrm{rx}}}(\tilde{\theta})\) and
\(\tilde f_k\triangleq f_{\mathrm c}+f_k\).
\begingroup\color{red}
For an active branch with \(E_{p,k}>0\), the matched
filter is
\(h_{p,k}(t) = \bigl(x^{\textcolor{red}{\mathrm{b}}}_{p,k}(-t;\tilde{\theta})\bigr)^*\).
If \(E_{p,k}=0\), the branch is inactive in the \(p\)-th PRI and its output
is set to zero.
Here \(E_{p,k}\) is the continuous-time branch energy defined in
Appendix~\ref{app:B1}; under sampling, \(E_{p,k}\approx T_s\mathcal E_{p,k}\).
\endgroup
\textcolor{red}{The matched-filter output at the fixed delay
\(\tilde{\tau}\) is}
\[
	\label{eq:appD_yconv_peak}
	\begin{aligned}
		y_{p,k}(\textcolor{red}{\tilde{\tau}})
		&\!=\!
		\tilde{\beta}\tilde{\rho}
		\int_{-\infty}^{\infty}\!\!\!
		x^{\textcolor{red}{\mathrm{b}}}_{p,k}
		\bigl(\xi\!-\!\tau_p(\xi);\tilde{\theta}\bigr)
		\!\times\!
		\left(
		x^{\textcolor{red}{\mathrm{b}}}_{p,k}
		(\xi\!-\!\textcolor{red}{\tilde{\tau}};\tilde{\theta})
		\right)^* \\
		&\hspace{1em}\times
		\exp\!\left[-j2\pi\tilde f_k\tau_p(\xi)\right]
		\mathrm d\xi .
	\end{aligned}
\]
\textcolor{red}{When the delay variation over the CPI produces
negligible displacement and distortion of the complex baseband
branch waveform, we use}
\(x^{\textcolor{red}{\mathrm{b}}}_{p,k} \bigl(\xi-\tau_p(\xi);\tilde{\theta}\bigr) \approx x^{\textcolor{red}{\mathrm{b}}}_{p,k} \bigl(\xi-\textcolor{red}{\tilde{\tau}};\tilde{\theta}\bigr)\).
\begingroup\color{red}
Applying this approximation to the output integral while
\textcolor{red}{retaining the time-varying propagation phase} gives
\begin{equation}
	\label{eq:appD_peak_before_shift}
	\begin{aligned}
		y_{p,k}(\textcolor{red}{\tilde{\tau}})
		&\approx
		\tilde{\beta}\tilde{\rho}\,
		\textcolor{red}{\exp\!\left(-j2\pi\tilde f_k\tilde{\tau}\right)}
		\textcolor{red}{e^{j2\pi f_{d,k}(\tilde{\nu})t_p}}
		\\
		&\quad\times
		\int_{-\infty}^{\infty}
		\left|
		x^{\textcolor{red}{\mathrm{b}}}_{p,k}
		(\xi-\textcolor{red}{\tilde{\tau}};\tilde{\theta})
		\right|^2
		e^{j2\pi f_{d,k}(\tilde{\nu})\xi}
		\,\mathrm d\xi .
	\end{aligned}
\end{equation}
where \(f_{d,k}(\tilde{\nu})\triangleq2\tilde f_k\tilde{\nu}/c\).
For \(E_{p,k}>0\), define the normalized intrapulse Doppler response
\begin{equation}
	\label{eq:appD_Gpk}
	G_{p,k}(f)
	\triangleq
	\frac{1}{E_{p,k}}
	\int_{-\infty}^{\infty}
	\left|
	x^{\textcolor{red}{\mathrm{b}}}_{p,k}(\eta;\tilde{\theta})
	\right|^2
	\exp(j2\pi f\eta)
	\mathrm d\eta .
\end{equation}
Setting \(\eta=\xi-\tilde{\tau}\) in
\eqref{eq:appD_peak_before_shift} introduces the phase factor
\textcolor{red}{\(e^{j2\pi f_{d,k}(\tilde{\nu})\tilde{\tau}}\)}, yielding
\begin{equation}
	\label{eq:appD_peak_with_G}
	\begin{aligned}
	y_{p,k}(\textcolor{red}{\tilde{\tau}})
	&\approx
	\tilde{\beta}\tilde{\rho}
	E_{p,k}
	G_{p,k}\bigl(f_{d,k}(\tilde{\nu})\bigr)\\
	&\quad\times\exp\!\left[
	-j2\pi
	\bigl(
	\tilde f_k-f_{d,k}(\tilde{\nu})
	\bigr)
	\textcolor{red}{\tilde{\tau}}
	\right]\\
	&\quad\times
	\textcolor{red}{\exp\!\left[j2\pi f_{d,k}(\tilde{\nu})t_p\right]}.
	\end{aligned}
\end{equation}
\endgroup

\begingroup\color{red}
Define the energy centroid
\[
\bar{\eta}_{p,k}\triangleq
\frac{1}{E_{p,k}}\int_{-\infty}^{\infty}
\eta\left|x^{\textcolor{red}{\mathrm{b}}}_{p,k}(\eta;\tilde{\theta})\right|^2\,\mathrm d\eta.
\]
Since the waveform duration is on the order of \(T\), under
\(2\pi|f|T\ll1\),
\(e^{j2\pi f\eta} =1+j2\pi f\eta+O\!\left((fT)^2\right)\).
Substitution into \eqref{eq:appD_Gpk} yields
\[
\begin{aligned}
	G_{p,k}(f)
	&=1+j2\pi f\bar{\eta}_{p,k}
	+O\!\left((fT)^2\right)\\
	&=e^{j2\pi f\bar{\eta}_{p,k}}
	+O\!\left((fT)^2\right).
	\end{aligned}
\]
\endgroup
\begingroup\color{red}
Taking \(f=f_{d,k}(\tilde{\nu})\), let
\(\phi_0\triangleq\arg\{\tilde{\beta}\tilde{\rho}\}\). Then the coefficient in
\eqref{eq:appD_peak_with_G} can be written as
\begin{equation}
	\label{eq:appD_coeff_polar}
	\tilde{\beta}\tilde{\rho}
	E_{p,k}
	G_{p,k}\bigl(f_{d,k}(\tilde{\nu})\bigr)
	=
	A_{p,k}(\tilde{\nu})
	\exp\!\left(j\vartheta_{p,k}(\tilde{\nu})\right),
\end{equation}
\endgroup
where
\begin{equation}
	\label{eq:appD_A_theta_explicit}
	\begin{aligned}
	A_{p,k}(\tilde{\nu})
	&=
	\left|\tilde{\beta}\tilde{\rho}\right|E_{p,k}
	+
	O\!\left(
	\left|\tilde{\beta}\tilde{\rho}\right|E_{p,k}
		\bigl(f_{d,k}(\tilde{\nu})T\bigr)^2
		\right),\\
		\vartheta_{p,k}(\tilde{\nu})
		&=
		\phi_0
		+
		2\pi f_{d,k}(\tilde{\nu})\bar{\eta}_{p,k}
		+
		O\!\left(
		\bigl(f_{d,k}(\tilde{\nu})T\bigr)^2
		\right).
	\end{aligned}
\end{equation}

\begingroup\color{red}
Since \(\textcolor{red}{2\pi}|f_{d,k}(\tilde{\nu})|T\ll1\) and
\(\bar{\eta}_{p,k}=O(T)\), the branch-dependent perturbation
\(2\pi f_{d,k}(\tilde{\nu})\bar{\eta}_{p,k}\) is small. Thus,
\(\vartheta_{p,k}(\tilde{\nu})\) can be replaced by an approximately
branch-independent phase \(\vartheta_p(\tilde{\nu})\), yielding the
peak model in \eqref{eq:peak}.
\endgroup

\begingroup\color{red}
\subsection{Coherent Loss Due to Residual Phase Differences}
\label{app:phase_coherent_loss}

Under the desired-branch matched-filter model in
Appendix~\ref{app:B2}, consider focusing at the true delay and velocity.
Using the coefficients $A_{p,k}$ and $\vartheta_{p,k}$ in
\eqref{eq:appD_coeff_polar}, define the coherent loss relative to exact
residual-phase alignment across pulses and frequency branches as
\begin{equation}
\label{eq:app_phase_loss_def}
L_{\phi}=-10\log_{10}
\frac{\left|\sum_{p,k}A_{p,k}e^{j\vartheta_{p,k}}\right|^2}
{\left(\sum_{p,k}A_{p,k}\right)^2}.
\end{equation}

Define the normalized amplitude coefficients and weighted mean phase as
\(a_{p,k}\triangleq\frac{A_{p,k}}{\sum_{p,k}A_{p,k}},\) and \(\bar\vartheta\triangleq\sum_{p,k}a_{p,k}\vartheta_{p,k}\),
and let $\Delta\vartheta_{p,k}\triangleq\vartheta_{p,k}-\bar\vartheta$.
Since $|e^{j\bar\vartheta}|=1$, the loss can equivalently be written as
\begin{equation}
\label{eq:app_phase_centered_loss}
L_{\phi}=-10\log_{10}
\left|\sum_{p,k}a_{p,k}e^{j\Delta\vartheta_{p,k}}\right|^2.
\end{equation}

Define the amplitude-weighted phase variance as
\(\operatorname{Var}_{a}(\vartheta_{p,k}) \triangleq\sum_{p,k}a_{p,k}(\Delta\vartheta_{p,k})^2\).
For small residual phase differences, expanding the exponential to
second order and using $\sum_{p,k}a_{p,k}=1$ and
$\sum_{p,k}a_{p,k}\Delta\vartheta_{p,k}=0$ gives
\[
\label{eq:app_phase_expansion}
\begin{aligned}
\sum_{p,k}a_{p,k}e^{j\Delta\vartheta_{p,k}}
&\approx\sum_{p,k}a_{p,k}\left(1+j\Delta\vartheta_{p,k}
\!-\!\frac{(\Delta\vartheta_{p,k})^2}{2}\right)\\
&=1-\frac12\operatorname{Var}_{a}(\vartheta_{p,k}).
\end{aligned}
\]
Taking the squared magnitude and retaining terms through second order
in the phase differences yields
\begin{equation}
\label{eq:app_phase_power}
\left|\sum_{p,k}a_{p,k}e^{j\Delta\vartheta_{p,k}}\right|^2
\approx1-\operatorname{Var}_{a}(\vartheta_{p,k}).
\end{equation}
Using $-\ln(1-z)\approx z$, the coherent loss becomes
\begin{equation}
\label{eq:app_phase_loss_variance}
L_{\phi}\approx\frac{10}{\ln10}\operatorname{Var}_{a}(\vartheta_{p,k}).
\end{equation}

For $2\pi|f_{d,k}|T\ll1$, the intrapulse Doppler expansion in
\eqref{eq:appD_A_theta_explicit} gives
\(\vartheta_{p,k}\approx\phi_0+2\pi f_{d,k}\bar\eta_{p,k},\) and \(f_{d,k}=\frac{2\tilde f_k\tilde\nu}{c}\).
Let $x\triangleq2\tilde f_{\max}|\tilde\nu|T/c$ denote the normalized
velocity.
Substituting this phase expression into
\eqref{eq:app_phase_loss_variance}, the common phase $\phi_0$ cancels
in the variance, giving
\begin{equation}
\label{eq:app_phase_normalized_loss}
L_{\phi}\approx\frac{40\pi^2}{\ln10}\,x^2
\operatorname{Var}_{a}\!\left(
\frac{\tilde f_k}{\tilde f_{\max}}\frac{\bar\eta_{p,k}}{T}
\right)\quad\mathrm{dB}.
\end{equation}
This expression relates the coherent loss to the normalized velocity
and the amplitude-weighted variance of the carrier-weighted normalized
energy centroids. The coefficients $a_{p,k}$ are evaluated at the
corresponding target velocity.
\endgroup

\begingroup\color{red}
\setlength{\jot}{0pt}
\subsection{Grid-Mismatch Phase and Accumulation Error Bounds}
\label{app:res_doppler}

\subsubsection{Velocity-Grid Mismatch}
\label{app:velocity_grid_mismatch}

Consider a target with $\tilde\nu\in[-v_{\max},v_{\max})$ in the
coarse-delay cell $\tilde u$, and let $v_{\tilde q}$ be its nearest
velocity-grid point.
Substituting the peak model \eqref{eq:peak} into
\eqref{eq:Xk_entry}, while retaining the velocity-grid mismatch, gives
\begin{equation}
\label{eq:appD_Xk_phase}
\begin{aligned}
[\boldsymbol X_k]_{\tilde u,\tilde q}
&\approx e^{-j2\pi[\tilde f_k-f_{d,k}(\tilde\nu)]\tilde\tau}
S_k,\\
S_k&\triangleq\sum_{p=1}^{P}c_{p,k}(\tilde\nu)e^{j\phi_{p,k}},
\end{aligned}
\end{equation}
where $c_{p,k}(\tilde\nu)$ is the complex coefficient defined after
\eqref{eq:X_approx} and
\begin{equation}
\label{eq:appD_phi_pk}
\begin{aligned}
\phi_{p,k}
\triangleq2\pi[f_{d,k}(\tilde\nu)-f_{d,k}(v_{\tilde q})]t_p
=\frac{4\pi\tilde f_k}{c}(\tilde\nu-v_{\tilde q})t_p.
\end{aligned}
\end{equation}
Since $|\tilde\nu-v_{\tilde q}|\le\Delta v/2=v_{\max}/Q$ and
$0\le t_p\le(P-1)T_{\mathrm r}$, the phase is uniformly bounded by
\begin{equation}
\label{eq:appD_phi_uniform}
\begin{aligned}
|\phi_{p,k}|
&\le\frac{2\pi\tilde f_{\max}(P-1)T_{\mathrm r}\Delta v}{c}\\
&=\frac{4\pi\tilde f_{\max}v_{\max}(P-1)T_{\mathrm r}}{cQ}
=\varepsilon_{\mathrm d}.
\end{aligned}
\end{equation}
This proves \eqref{eq:doppler_grid_phase_bound}. To enforce a prescribed
phase tolerance $\bar\varepsilon_{\mathrm d}>0$, it suffices to choose
\begin{equation}
\label{eq:appD_Q_cond}
Q\ge\left\lceil
\frac{4\pi\tilde f_{\max}v_{\max}(P-1)T_{\mathrm r}}
{c\bar\varepsilon_{\mathrm d}}
\right\rceil,
\end{equation}
which ensures $\varepsilon_{\mathrm d}\le\bar\varepsilon_{\mathrm d}$.

Let $S_k^{(0)}\triangleq\sum_{p=1}^{P}c_{p,k}(\tilde\nu)$ denote
the coherent sum obtained by neglecting the grid-mismatch phase.
Using $|e^{j\phi}-1|\le|\phi|$ yields
\[
\label{eq:appD_phase_error}
\begin{aligned}
|S_k-S_k^{(0)}|
\le\sum_{p=1}^{P}|c_{p,k}(\tilde\nu)|\,|e^{j\phi_{p,k}}-1|
\le\varepsilon_{\mathrm d}\sum_{p=1}^{P}|c_{p,k}(\tilde\nu)|.
\end{aligned}
\]
\begin{samepage}
Thus, the approximation in \eqref{eq:X_approx} has a controlled
absolute error in the slow-time coherent sum under the peak model.\par
\end{samepage}

\subsubsection{Fine-Delay-Grid Mismatch}
\label{app:fine_delay_grid_mismatch}

Fix the correctly matched coarse-delay and velocity cell $(u_0,q_0)$,
and let $\tau_0$ be the true fine delay. Define
\(b_k\triangleq[\boldsymbol X_k]_{u_0,q_0} e^{j2\pi f_{k,\mathrm{eff}}(\tilde\nu)\tau_0}\).
For a fine-delay hypothesis $\tau$, let $\varepsilon_\tau=\tau_0-\tau$.
Using the equally spaced effective frequencies in \eqref{eq:feff_affine},
and removing a common phase factor that does not affect power, the
synthesis in \eqref{eq:Y_uq_tau} becomes
\begin{equation}
\label{eq:appD_fine_factored}
S_{\mathrm f}(\varepsilon_\tau)
=\sum_{k=1}^{K}b_k
e^{-j2\pi(k-1)\Delta f_{\mathrm{eff}}\varepsilon_\tau},
\end{equation}
where $\Delta f_{\mathrm{eff}}=\Delta f_{\mathrm{eff}}(\tilde\nu)$ is the
effective frequency spacing, and $b_k$ retains the actual complex branch
coefficient.

Let $\varepsilon_\tau^\star$ be the interior continuous main-peak
location of the power response, and define
\(S_\star=S_{\mathrm f}(\varepsilon_\tau^\star),\) and \(A_\star=|S_\star|>0\),
\(a_k=\frac{S_\star^*}{A_\star}b_k e^{-j2\pi(k-1)\Delta f_{\mathrm{eff}}\varepsilon_\tau^\star}\).
Then $|a_k|=|b_k|$ and $\sum_k a_k=A_\star$. Define
\(\bar k=\frac{\sum_k(k-1)|b_k|}{\sum_k|b_k|},\) and \(M_2=\sum_k|b_k|(k-1-\bar k)^2\).
For a grid-point offset $d_\tau$ from the continuous peak in the
$\varepsilon_\tau$ coordinate, construct
\begin{equation}
\label{eq:appD_fine_projection}
g(d_\tau)=\Re\!\left\{
\sum_k a_k e^{-j2\pi(k-1-\bar k)\Delta f_{\mathrm{eff}}d_\tau}
\right\}.
\end{equation}
The sum differs from $S_{\mathrm f}(\varepsilon_\tau^\star+d_\tau)$
only by a unit-modulus phase factor. Hence,
\(g(0)=A_\star,\) and \(|S_{\mathrm f}(\varepsilon_\tau^\star+d_\tau)|\ge g(d_\tau)\).

Differentiating term by term gives
\[
\begin{aligned}
g'(d_\tau)
&=\Re\!\Biggl\{-j2\pi\Delta f_{\mathrm{eff}}
\sum_k(k-1-\bar k)a_k\\
&\hspace{3em}\times
 e^{-j2\pi(k-1-\bar k)\Delta f_{\mathrm{eff}}d_\tau}\Biggr\},\\
g''(d_\tau)
&=-4\pi^2\Delta f_{\mathrm{eff}}^2
\Re\!\Biggl\{\sum_k(k-1-\bar k)^2a_k\\
&\hspace{3em}\times
 e^{-j2\pi(k-1-\bar k)\Delta f_{\mathrm{eff}}d_\tau}\Biggr\}.
\end{aligned}
\]
The power-stationarity condition at the continuous peak,
$\Re\{S_\star^*S_{\mathrm f}'(\varepsilon_\tau^\star)\}=0$,
yields
\[
\begin{aligned}
g'(0)
&=\frac{\Re\{S_\star^*S_{\mathrm f}'(\varepsilon_\tau^\star)\}}
{A_\star}
+\Re\{j2\pi\bar k\Delta f_{\mathrm{eff}}A_\star\}\\
&=0.
\end{aligned}
\]
Meanwhile, the triangle inequality and $|a_k|=|b_k|$ give
\[
\label{eq:appD_fine_curvature}
\begin{aligned}
|g''(d_\tau)|
&\le4\pi^2\Delta f_{\mathrm{eff}}^2
\sum_k(k-1-\bar k)^2|a_k|\\
&=4\pi^2\Delta f_{\mathrm{eff}}^2M_2.
\end{aligned}
\]

Taylor's formula with integral remainder therefore gives
\begin{equation}
\label{eq:appD_fine_amplitude_bound}
\begin{aligned}
g(d_\tau)
&=g(0)+d_\tau g'(0)
+d_\tau^2\int_0^1(1-t)g''(td_\tau)\,\mathrm dt\\
&\ge A_\star-4\pi^2\Delta f_{\mathrm{eff}}^2M_2d_\tau^2
\int_0^1(1-t)\,\mathrm dt\\
&=A_\star-2\pi^2\Delta f_{\mathrm{eff}}^2M_2d_\tau^2.
\end{aligned}
\end{equation}
This bound holds for both positive and negative offsets without
neglecting higher-order terms. When the right-hand side is positive,
\begin{equation}
\label{eq:appD_fine_power_ratio}
\frac{|S_{\mathrm f}(\varepsilon_\tau^\star+d_\tau)|^2}{A_\star^2}
\ge\left(1-
\frac{2\pi^2\Delta f_{\mathrm{eff}}^2M_2}{A_\star}d_\tau^2
\right)^2.
\end{equation}

The grid-peak power is no smaller than the power at the grid point
nearest the continuous peak. Taking $d_\tau$ as that nearest-point offset,
the grid-peak power loss relative to $A_\star^2$ satisfies
\begin{equation}
\label{eq:appD_fine_loss_bound}
L_\tau\le-20\log_{10}\!\left(
1-\frac{2\pi^2\Delta f_{\mathrm{eff}}^2M_2}{A_\star}d_\tau^2
\right),
\end{equation}
provided that the logarithm argument is positive.
For a uniform grid covering the peak neighborhood with spacing
$\Delta\tau_{\mathrm f}=1/(L\Delta f_{\mathrm{eff}})$,
$|d_\tau|\le\Delta\tau_{\mathrm f}/2$ further gives
\begin{equation}
\label{eq:appD_fine_loss_grid_bound}
L_\tau\le-20\log_{10}\!\left(
1-\frac{\pi^2M_2}{2A_\star L^2}\right),
\qquad\frac{\pi^2M_2}{2A_\star L^2}<1.
\end{equation}
\endgroup

\section{Derivations of SMC Performance Properties}
\label{app:smc_perf_props}

\subsection{Derivation of Joint Coarse--Fine Range Discrimination}
\label{app:joint_range_branch}

\begingroup\color{red}
\subsubsection{Joint Range Response}

Consider an isolated target focused at the correct coarse-delay and velocity
cell \((u_0,q_0)\), with fine delay \(\tau_0\). From
\eqref{eq:X_approx},
\begin{equation}
\label{eq:app_Xk_res_simple}
[\boldsymbol X_k]_{u_0,q_0}
\approx s_k(\tilde{\nu})
e^{-j2\pi f_{k,\mathrm{eff}}(\tilde{\nu})\tau_0},
\end{equation}
where
\(s_k(\tilde{\nu})\triangleq\sum_{p=1}^{P}c_{p,k}(\tilde{\nu})\).
Substitution into \eqref{eq:Y_uq_tau} gives
\begin{equation}
\label{eq:app_Y_res_simple}
Y(u_0,q_0;\tau)
\approx\sum_{k=1}^{K}s_k(\tilde{\nu})
e^{j2\pi f_{k,\mathrm{eff}}(\tilde{\nu})(\tau-\tau_0)}.
\end{equation}
Neglecting residual phase differences after compensation, define
\(\mu_k\triangleq \frac{|s_k(\tilde{\nu})|}{\sum_{i=1}^{K}|s_i(\tilde{\nu})|}, \mu_k\ge0,\) and \(\sum_{k=1}^{K}\mu_k=1\).
An absent pulse--branch contribution has \(c_{p,k}=0\); a branch absent
throughout the CPI has \(\mu_k=0\).

With \(\Delta R\triangleq c(\tau-\tau_0)/2\), adopting
\(f_{k,\mathrm{eff}}(\tilde{\nu})\approx\tilde f_k\) and equally spaced
carriers \(f_k=f_1+(k-1)\Delta f\), removal of the common carrier phase yields
the normalized fine-range power response
\begin{equation}
\label{eq:app_Af_general_range}
\begin{aligned}
A_{\mathrm f}^{(\mu)}(\Delta R)
=\left|\sum_{k=1}^{K}\mu_k
e^{j2\pi(k-1)\Delta R/T_f}\right|^2,
T_f\triangleq\frac{c}{2\Delta f}.
\end{aligned}
\end{equation}
When the normalized complex-amplitude range responses are approximately
identical across branches, the joint power response is
\begin{equation}
\label{eq:app_Ajoint_general}
A_{\mathrm{joint}}^{(\mu)}(\Delta R)
\approx A_{\mathrm c}(\Delta R)A_{\mathrm f}^{(\mu)}(\Delta R),
\end{equation}
where \(A_{\mathrm c}(\Delta R)\) is the normalized single-branch
coarse-range power response, with \(A_{\mathrm c}(0)=1\).
The corresponding 3-dB unresolved region is
\(\mathcal U^{(\mu)} =\left\{\Delta R:A_{\mathrm{joint}}^{(\mu)}(\Delta R)\ge\frac12\right\}\).

\subsubsection{Periodicity of the Fine-Range Response}
\label{app:fine_range_periodicity}

Let \(k_1,\ldots,k_{K_{\mathrm a}}\) denote the indices of the
\(K_{\mathrm a}\ge2\) active branches.
From \eqref{eq:app_Af_general_range}, a positive shift \(L\) is a period
if and only if the phase differences relative to branch \(k_1\) are
preserved, i.e.,
\((k_i-k_1)\frac{L}{T_f}\in\mathbb Z,\) for \(i=2,\ldots,K_{\mathrm a}\).
Define \(d_i\triangleq|k_i-k_1|\) and
\(g_\mu\triangleq\gcd(d_2,\ldots,d_{K_{\mathrm a}})\).
By B\'ezout's identity~\cite[Sec.~1.2]{Shoup2008NumberTheory},
\(g_\mu\) is an integer linear combination of the index differences:
there exist \(b_2,\ldots,b_{K_{\mathrm a}}\in\mathbb Z\) such that
\(g_\mu=\sum_{i=2}^{K_{\mathrm a}}b_i d_i\).
The period conditions therefore imply
\(\frac{g_\mu L}{T_f} =\sum_{i=2}^{K_{\mathrm a}}b_i\frac{d_iL}{T_f}\in\mathbb Z\).
Thus, every positive period is a positive integer multiple of \(T_f/g_\mu\).
Since each \(d_i\) is an integer multiple of \(g_\mu\), setting
\(L=T_f/g_\mu\) satisfies all the period conditions, giving the fundamental
period
\begin{equation}
\label{eq:app_Tmu}
T_\mu=\frac{T_f}{g_\mu}.
\end{equation}
With only one active branch, \(A_{\mathrm f}^{(\mu)}(\Delta R)\equiv1\),
so no fine-range discrimination is available.

\subsubsection{Branch-Energy Distribution and Mainlobe Width}
\label{app:branch_energy_mainlobe}

Define the weighted frequency mean and variance as
\(\bar f_\mu\triangleq\sum_{k=1}^{K}\mu_k f_k, \sigma_{f,\mu}^{2}\triangleq\sum_{k=1}^{K}\mu_k(f_k-\bar f_\mu)^2\).
Using
\(\sum_{k=1}^{K}\sum_{\ell=1}^{K} \mu_k\mu_\ell(f_k-f_\ell)^2=2\sigma_{f,\mu}^{2}\),
expansion of \eqref{eq:app_Af_general_range} about \(\Delta R=0\) gives
\begin{equation}
\label{eq:app_fine_response_local_expansion}
\begin{aligned}
A_{\mathrm f}^{(\mu)}(\Delta R)
&=\sum_{k=1}^{K}\sum_{\ell=1}^{K}\mu_k\mu_\ell
\cos\!\left[\frac{4\pi(f_k-f_\ell)\Delta R}{c}\right]\\
&=1-\frac{16\pi^2\sigma_{f,\mu}^{2}}{c^2}\Delta R^2
+O(\Delta R^4).
\end{aligned}
\end{equation}

Let \(A_{\mathrm{ref}}(\Delta R)\) denote the fine-range power response
for a fixed set of reference weights and define
\(\Delta A(\Delta R) \triangleq A_{\mathrm f}^{(\mu)}(\Delta R)-A_{\mathrm{ref}}(\Delta R)\).
Let \(r_{\mathrm h}\) be the smallest positive root of
\(A_{\mathrm{ref}}(r_{\mathrm h})=1/2\), with
\(A_{\mathrm{ref}}'(r_{\mathrm h})<0\) and reference 3-dB width
\(B_{\mathrm{ref}}=2r_{\mathrm h}\).
For small perturbations of the reference weights, the perturbed
3-dB width \(B_\mu\) satisfies
\begin{equation}
\label{eq:app_halfpower_perturbation}
\begin{aligned}
A_{\mathrm f}^{(\mu)}\!\left(\frac{B_\mu}{2}\right)
\approx\frac12+\frac{A_{\mathrm{ref}}'(r_{\mathrm h})}{2}
(B_\mu-B_{\mathrm{ref}})+\Delta A(r_{\mathrm h}).
\end{aligned}
\end{equation}
Hence,
\begin{equation}
\label{eq:app_fine_width_perturbation}
B_\mu-B_{\mathrm{ref}}
\approx-\frac{2\Delta A(r_{\mathrm h})}{A_{\mathrm{ref}}'(r_{\mathrm h})}.
\end{equation}

\subsubsection{Range Discrimination With Uniform Branch Weights}

Neglecting the residual intrapulse Doppler effect in
Appendix~\ref{app:B2}, the branch weights satisfy
\[
\mu_k\approx
{\displaystyle\sum_{p=1}^{P}E_{p,k}}/
{\displaystyle\sum_{i=1}^{K}\sum_{p=1}^{P}E_{p,i}}.
\]
When the accumulated energy over the CPI is identical across frequency
branches, the weights are uniform, \(\mu_k=1/K\).
For \(K\ge2\), the geometric sum in
\eqref{eq:app_Af_general_range} then yields
\begin{equation}
	\label{eq:app_Af_range}
	A_{\mathrm f}(\Delta R)
	=
	\left|
	\frac{\sin(\pi K\Delta R/T_f)}
	{K\sin(\pi\Delta R/T_f)}
	\right|^2.
\end{equation}
The fundamental period is \(T_f\), with peaks at
\(\Delta R=wT_f\), \(w\in\mathbb Z\).

The following analysis characterizes the 3-dB width of the central branch
and the number and widths of the ambiguity branches.
\paragraph{3-dB Width of the Central Branch}

For \(|\pi\Delta R/T_f|\ll1\), the approximation
\(\sin(\pi\Delta R/T_f)\approx\pi\Delta R/T_f\) gives
\begin{equation}
\label{eq:app_Afine_sinc_mainlobe}
A_{\mathrm f}(\Delta R)
\approx\left(\frac{\sin x}{x}\right)^2,
\qquad x\triangleq\frac{\pi K\Delta R}{T_f}.
\end{equation}
Let \(x_{3\mathrm{dB}}\approx1.3916\) be the smallest positive root of
\(\left(\frac{\sin x_{3\mathrm{dB}}}{x_{3\mathrm{dB}}}\right)^2=\frac12\).
The fine-range 3-dB full width is therefore
\begin{equation}
\label{eq:app_R_fine_3dB}
\Delta R_{\mathrm{fine},3\mathrm{dB}}
\approx\frac{2x_{3\mathrm{dB}}T_f}{\pi K}
\approx\frac{0.443c}{K\Delta f}.
\end{equation}

If the coarse-range envelope varies slowly across the central fine-range
mainlobe, then \(A_{\mathrm c}(\Delta R)\approx A_{\mathrm c}(0)=1\).
By \eqref{eq:app_Ajoint_general}, the central branch of the joint response
therefore has 3-dB full width
\(B_0\approx\Delta R_{\mathrm{fine},3\mathrm{dB}}\).

\paragraph{Number and Widths of Ambiguity Branches}

Near the \(w\)-th fine-range peak, write
\(\Delta R=wT_f+\varepsilon\). Periodicity gives
\(A_{\mathrm{joint}}(wT_f+\varepsilon) \approx A_{\mathrm c}(wT_f+\varepsilon)A_{\mathrm f}(\varepsilon)\).
If the coarse-range envelope is approximately constant within one
fine-range mainlobe, then
\(A_{\mathrm{joint}}(wT_f+\varepsilon) \approx A_{\mathrm c}(wT_f)A_{\mathrm f}(\varepsilon)\).

Assume the coarse-range mainlobe is symmetric about zero and decreases
monotonically with \(|\Delta R|\). Define its 3-dB half-width
\(\delta_{\mathrm c}\) by
\(A_{\mathrm c}(\delta_{\mathrm c})=\frac12\).
Under this local approximation, branches
whose peaks exceed the half-power threshold satisfy
\(A_{\mathrm c}(wT_f)>\frac12 \Longleftrightarrow |w|T_f<\delta_{\mathrm c}\).
Thus, the number of retained branches within the coarse-range mainlobe,
including the central branch, is
\(N_{\mathrm{branch}}=2W+1, W\triangleq \max\{w\in\mathbb Z_{\ge0}:wT_f<\delta_{\mathrm c}\}\).

Under this local approximation, the \(w\)-th retained branch is
approximately symmetric about \(wT_f\). With the threshold set to half
the central peak power of the joint response, its unresolved half-width
\(\delta_w\) satisfies
\(A_{\mathrm c}(wT_f)A_{\mathrm f}(\delta_w)\approx\frac12\).
Using \eqref{eq:app_Afine_sinc_mainlobe} and setting
\(x_w=\pi K\delta_w/T_f\) gives
\(\left(\frac{\sin x_w}{x_w}\right)^2 =\frac{1}{2A_{\mathrm c}(wT_f)},\) and \(0<x_w<\pi\).
The corresponding unresolved full width is
\(B_w\approx2\delta_w=\frac{2T_f x_w}{\pi K}\).
For an outer branch with \(1/2<A_{\mathrm c}(wT_f)<1\),
the right-hand side of the relation above exceeds \(1/2\).
Since \((\sin x/x)^2\) is strictly decreasing on \(0<x<\pi\),
\(x_w<x_{3\mathrm{dB}}\), giving, under the above approximations,
\(B_w<B_0\).
\endgroup

\subsection{Proof of Proposition~\ref{prop:output_snr_behavior}: Output-SNR Behavior}
\label{app:snr_analysis}
\subsubsection{Signal-Side Output Behavior}
\label{app:snr_analysis_signal}

\begingroup\color{red}
Consider the matched target cell
\(\eta^\star\triangleq(\tilde{\tau},\tilde{\nu},\tilde{\theta})\).
Under the fixed-delay approximation in Appendix~\ref{app:B2}, the
signal-only matched output of branch \((p,k)\) is
\[
\begin{aligned}
m_{p,k}^{(s)\star}
\triangleq{}
\int_{\mathcal I}
s_{\mathrm R,p}^{(s)}(t)
e^{-j2\pi f_k(t_p+t)}
\quad\times
x_{p,k}^{\textcolor{red}{\mathrm{b}}*}(t-\tilde{\tau};\tilde{\theta})
\,\mathrm dt,
\end{aligned}
\]
where \(\mathcal I\triangleq[0,T_{\mathrm r})\) and
\(s_{\mathrm R,p}^{(s)}(t)\) is the received signal component in PRI \(p\).
The SMC phase-compensation coefficient is
\begin{equation}
\label{eq:snrdec_target_weight}
w_{p,k}^{\star}
\triangleq
\exp\!\left\{
j2\pi\left[
\bigl(\tilde f_k-f_{d,k}(\tilde{\nu})\bigr)\tilde{\tau}
-f_{d,k}(\tilde{\nu})t_p
\right]\right\},
\end{equation}
\endgroup
\begingroup\color{red}
which combines Doppler and frequency-aperture compensation, giving
\begin{equation}
\label{eq:snrdec_smc_signal_stat}
Z_{\mathrm{SMC}}^{(s)}(\eta^\star)
=
\sum_{p=1}^{P}\sum_{k=1}^{K}
w_{p,k}^{\star}m_{p,k}^{(s)\star}.
\end{equation}

The conventional PD processor uses the full-band reference
\(x_p(t;\tilde{\theta})=\sum_{k=1}^{K}x_{p,k}(t;\tilde{\theta})\)
and a single Doppler reference
\textcolor{red}{\(f_{d,\mathrm{ref}}(\tilde{\nu})=2f_{\mathrm c}\tilde{\nu}/c\)},
with slow-time weights
\textcolor{red}{\(a_p^\star=e^{-j2\pi f_{d,\mathrm{ref}}(\tilde{\nu})t_p}\)}.
Using \eqref{eq:xbb}, its matched-cell statistic becomes
\begin{equation}
\label{eq:snrdec_pd_signal_stat}
\begin{aligned}
Z_{\mathrm{PD}}^{(s)}(\eta^\star)
&=
\sum_{p=1}^{P}a_p^\star
\int_{\mathcal I}
s_{\mathrm R,p}^{(s)}(t)
x_p^*(t-\textcolor{red}{\tilde{\tau}};\tilde{\theta})
\,\mathrm dt\\
&=
\sum_{p=1}^{P}\sum_{k=1}^{K}
\beta_k^\star a_p^\star m_{p,k}^{(s)\star},
\end{aligned}
\end{equation}
where \textcolor{red}{\(\beta_k^\star=e^{j2\pi f_k\tilde{\tau}}\)}.
Relative to the SMC coefficient, the PD mismatch factor
\(\gamma_{p,k}^\star\triangleq\beta_k^\star a_p^\star/w_{p,k}^\star\)
is
\begin{equation}
\label{eq:snrdec_pd_mismatch_exact}
\begin{aligned}
\gamma_{p,k}^{\star}
&=\exp\!\left[
j2\pi\bigl(f_{d,k}(\tilde{\nu})-f_{\mathrm c}\bigr)\tilde{\tau}
\right]\\
&\quad\times\exp\!\left[
j2\pi
\bigl(f_{d,k}(\tilde{\nu})-f_{d,\mathrm{ref}}(\tilde{\nu})\bigr)
\textcolor{red}{t_p}
\right],
\end{aligned}
\end{equation}
which gives \eqref{eq:snr_main_pd_signal_stat}.
The first factor is pulse-independent, whereas the residual slow-time phase is
\begin{equation}
\label{eq:snrdec_pd_phase_mismatch}
\begin{aligned}
\Delta\phi_{p,k}^{\star,\mathrm{slow}}
&=2\pi\bigl[f_{d,k}(\tilde{\nu})
-f_{d,\mathrm{ref}}(\tilde{\nu})\bigr]\textcolor{red}{t_p}\\
&=\frac{4\pi f_k\tilde{\nu}t_p}{c}.
\end{aligned}
\end{equation}
Thus, SMC compensates the branch-dependent Doppler phases, whereas
single-reference PD retains a mismatch proportional to the frequency
offset, target velocity, and slow time.
\endgroup

\begingroup\color{red}
To make the role of frequency occupancy explicit, let
\(E_{p,k}=\int_{\mathbb R} |x_{p,k}^{\textcolor{red}{\mathrm{b}}}(t;\tilde\theta)|^2\,\mathrm dt\)
denote the actual continuous-time branch energy. This is the same energy
used in Appendix~\ref{app:B2}. An absent branch has \(E_{p,k}=0\).
Under the fixed-delay approximation, negligible inter-branch leakage,
and \(2\pi|f_{d,k}(\tilde\nu)|T\ll1\), neglecting the residual
intrapulse phase gives
\(w_{p,k}^{\star}m_{p,k}^{(s)\star} \approx\tilde\beta\tilde\rho E_{p,k}\).
Consequently,
\[
\begin{aligned}
Z_{\mathrm{SMC}}^{(s)}(\eta^\star)
&\approx\tilde\beta\tilde\rho\sum_{p,k}E_{p,k},\\
Z_{\mathrm{PD}}^{(s)}(\eta^\star)
&\approx\tilde\beta\tilde\rho\sum_{p,k}E_{p,k}\gamma_{p,k}^{\star}.
\end{aligned}
\]
These expressions retain unequal and zero branch energies without
requiring complete coverage or equal branch amplitudes.
\endgroup

\subsubsection{Noise-Side Output Behavior}
\label{app:snr_analysis_noise}

\begingroup\color{red}
At coarse-delay cell \(u\), with delay \(\tau_u\), the branch-noise output is
\[
\begin{aligned}
\zeta_{p,k}^{(n)}(u)
\triangleq{}&
\int_{\mathcal I}
n_p(t)e^{-j2\pi f_k(t_p+t)}\\
&\quad\times
x_{p,k}^{\textcolor{red}{\mathrm{b}}*}(t-\tau_u;\tilde{\theta})
\,\mathrm dt,
\end{aligned}
\]
where the receiver noise satisfies
\begin{equation}
\label{eq:snrdec_noise_model}
\mathbb{E}_n[n_p(t)n_{p'}^*(t')]
=N_0\delta_{p,p'}\delta(t-t').
\end{equation}
\endgroup
\begingroup\color{red}
For \(\boldsymbol{\zeta}_p^{(n)}(u)
=[\zeta_{p,1}^{(n)}(u),\ldots,\zeta_{p,K}^{(n)}(u)]^{\mathsf T}\),
\eqref{eq:snrdec_noise_model} gives
\begin{equation}
\label{eq:snrdec_noise_cov}
\begin{aligned}
\mathbb{E}_n\!\left[
\boldsymbol{\zeta}_p^{(n)}(u)
\boldsymbol{\zeta}_{p'}^{(n)\mathsf H}(u)\right]
&=\delta_{p,p'}\boldsymbol{\Sigma}_p(u),\\
[\boldsymbol{\Sigma}_p(u)]_{k,k'}
&=N_0\Gamma_{p;k,k'}(u),
\end{aligned}
\end{equation}
where
\begin{equation}
\label{eq:snrdec_gamma_def}
\begin{aligned}
\Gamma_{p;k,k'}(u)
\triangleq{}&
\int_{\mathcal I}
\textcolor{red}{e^{-j2\pi(f_k-f_{k'})(t_p+t)}}\\
&\quad\times x_{p,k}^{\textcolor{red}{\mathrm{b}}*}
(t-\textcolor{red}{\tau_u};\tilde{\theta})
x_{p,k'}^{\textcolor{red}{\mathrm{b}}}
(t-\textcolor{red}{\tau_u};\tilde{\theta})
\,\mathrm dt.
\end{aligned}
\end{equation}

The direct PD pulse-compression noise output is
\begin{equation}
\label{eq:snrdec_wb_pc_noise}
\begin{aligned}
y_{\mathrm{PD},p}^{(n)}(u)
&=\int_{\mathcal I}
n_p(t)x_p^*(t-\textcolor{red}{\tau_u};\tilde{\theta})
\,\mathrm dt\\
&=\sum_{k=1}^{K}
e^{j2\pi f_k\textcolor{red}{\tau_u}}\zeta_{p,k}^{(n)}(u)
=\boldsymbol v_p^{\mathsf H}(u)\boldsymbol{\zeta}_p^{(n)}(u),
\end{aligned}
\end{equation}
where \([\boldsymbol v_p(u)]_k=e^{-j2\pi f_k\textcolor{red}{\tau_u}}\).
Unit-modulus MTD weights and the cross-PRI covariance in
\eqref{eq:snrdec_noise_cov} give
\begin{equation}
\label{eq:snrdec_wb_noise_power}
\sigma_{\mathrm{PD}}^2(u,q)
=\sum_{p=1}^{P}
\boldsymbol v_p^{\mathsf H}(u)
\boldsymbol{\Sigma}_p(u)\boldsymbol v_p(u).
\end{equation}

For SMC, the output noise sample is
\begin{equation}
\label{eq:snrdec_smc_noise_output}
Y_{\mathrm{SMC}}^{(n)}(u,q,\ell)
=\sum_{p=1}^{P}
\boldsymbol s_p^{\mathsf H}(q,\ell)
\boldsymbol{\zeta}_p^{(n)}(u),
\end{equation}
with
\([\boldsymbol s_p(q,\ell)]_k \triangleq d_{p,k}^*(q)b_{\ell,k}(q),\) and \(d_{p,k}(q) \triangleq\textcolor{red}{e^{-j2\pi f_{d,k}(v_q)t_p}},\) and \(b_{\ell,k}(q) \triangleq e^{-j2\pi f_{k,\mathrm{eff}}(v_q)\tau_{q,\ell}}\).
Its noise power is therefore
\begin{equation}
\label{eq:snrdec_smc_noise_power}
\sigma_{\mathrm{SMC}}^2(u,q,\ell)
=\sum_{p=1}^{P}
\boldsymbol s_p^{\mathsf H}(q,\ell)
\boldsymbol{\Sigma}_p(u)\boldsymbol s_p(q,\ell).
\end{equation}

For the equally spaced effective frequencies in \eqref{eq:feff_affine}
and the DFT fine-delay grid in \eqref{eq:tau_l}, \(L\ge K\) gives
\(\frac{1}{L}\sum_{\ell=1}^{L} b_{\ell,k}(q)b_{\ell,k'}^*(q)=\delta_{k,k'}\).
Together with \(|d_{p,k}(q)|=1\), this gives
\(\frac{1}{L}\sum_{\ell=1}^{L} \boldsymbol s_p(q,\ell)\boldsymbol s_p^{\mathsf H}(q,\ell) =\boldsymbol I_K\).
Hence, the fine-delay-averaged SMC noise power is
\begin{equation}
\label{eq:snrdec_smc_lavg_final}
\begin{aligned}
\bar{\sigma}_{\mathrm{SMC}}^2(u,q)
&\triangleq\frac{1}{L}\sum_{\ell=1}^{L}
\sigma_{\mathrm{SMC}}^2(u,q,\ell)\\
&=\sum_{p=1}^{P}\mathrm{tr}[\boldsymbol{\Sigma}_p(u)]\\
&=N_0\sum_{p=1}^{P}\sum_{k=1}^{K}\Gamma_{p;k,k}(u).
\end{aligned}
\end{equation}

For PD, define the inter-branch correlation term
\begin{equation}
\label{eq:snrdec_pd_cross_term}
\begin{aligned}
\Lambda_{\mathrm{PD}}(u)
\triangleq{}&
\sum_{p=1}^{P}\sum_{k<k'}
\Gamma_{p;k,k'}(u)\times
e^{j2\pi(f_k-f_{k'})\textcolor{red}{\tau_u}}.
\end{aligned}
\end{equation}
Pairing the conjugate off-diagonal terms in
\eqref{eq:snrdec_wb_noise_power} gives
\begin{equation}
\label{eq:snrdec_pd_fixedu_final}
\begin{aligned}
\sigma_{\mathrm{PD}}^2(u,q)
\!=\! N_0\sum_{p=1}^{P}\sum_{k=1}^{K}\Gamma_{p;k,k}(u)
\!+\! 2N_0\Re\{\Lambda_{\mathrm{PD}}(u)\}.
\end{aligned}
\end{equation}
Subtracting this from \eqref{eq:snrdec_smc_lavg_final} yields
\begin{equation}
\label{eq:snrdec_fixedu_difference}
\bar{\sigma}_{\mathrm{SMC}}^2(u,q)-\sigma_{\mathrm{PD}}^2(u,q)
=-2N_0\Re\{\Lambda_{\mathrm{PD}}(u)\}.
\end{equation}
If every \(\boldsymbol{\Sigma}_p(u)\) is diagonal,
\(\Lambda_{\mathrm{PD}}(u)=0\), and
\(\bar{\sigma}_{\mathrm{SMC}}^2(u,q)=\sigma_{\mathrm{PD}}^2(u,q)\).
\endgroup

\begingroup\color{red}
For delayed templates fully contained in the receive window,
\(\Gamma_{p;k,k}(u)=E_{p,k},\) and \([\boldsymbol\Sigma_p(u)]_{k,k}=N_0E_{p,k}\).
Thus, unequal branch energies produce unequal matched-filter noise
variances. An absent branch has a zero template and contributes a zero
row and column to the covariance matrix. White receiver noise alone
does not imply uncorrelated matched-filter branch outputs; the
off-diagonal entries depend on the actual template inner products.

If these templates are orthogonal within each pulse, the covariance
matrices are diagonal. Together with the independent noise across pulses,
rectangular slow-time weights, and unit-modulus combining coefficients
used above, this gives, at each fine-delay cell,
\[
\sigma_{\mathrm{SMC}}^2(u,q,\ell)
=\sigma_{\mathrm{PD}}^2(u,q)
=N_0\sum_{p,k}E_{p,k}.
\]
Under these conditions and the signal approximations stated above,
the matched-target output SNRs are
\[
\begin{aligned}
\mathrm{SNR}_{\mathrm{SMC}}
&\approx\frac{|\tilde\beta\tilde\rho|^2}{N_0}\sum_{p,k}E_{p,k},\\
\mathrm{SNR}_{\mathrm{PD}}
&\approx
\frac{|\tilde\beta\tilde\rho|^2
|\sum_{p,k}E_{p,k}\gamma_{p,k}^{\star}|^2}
{N_0\sum_{p,k}E_{p,k}}.
\end{aligned}
\]
Equal branch energies are therefore not necessary for equal output
noise powers. The equality also permits zero-energy branches, provided
the stated covariance and combining conditions hold. For approximately
orthogonal templates, the noise powers are approximately equal; their
difference is determined by the retained inter-branch correlation terms.
In the general correlated case, the fine-delay-averaged SMC noise power
must not be substituted for the noise power at a particular target cell.
\endgroup

\subsection{Proof of Proposition~\ref{prop:smc_receive_beampattern}: SMC-Domain Receive Beampattern}
\label{app:beampattern_detail}
This subsection derives the normalized SMC-domain receive beampattern in
\eqref{eq:Gsmc_main}. The key point is that the structure-matched channel
synthesis restores the full transmit-aperture spatial factor, while the receive
beamformer contributes another full-aperture factor.

Let
\(b_{\mathrm R}(\theta) \triangleq \boldsymbol{\omega}^{\mathsf H} \textcolor{red}{\boldsymbol a_{\mathrm{rx}}}(\theta)\)
denote the receive beamforming factor. Consider an echo arriving from azimuth
\(\theta\) and processed using a template steered to \(\theta_0\). For a fixed
subpulse \((p,v)\), let \(\Sigma_{p,v,k}\subseteq\{1,2,\ldots,N\}\) denote the
set of transmit antennas occupying frequency branch \(k\), and let
\(M_{p,v,k}=|\Sigma_{p,v,k}|\). The active branch set is
\(\mathcal K_{p,v} \triangleq \{k:M_{p,v,k}>0\}\).
Under the CP-FH frequency-assignment rule, the nonempty antenna sets form a
disjoint partition of the transmit array:
\(\bigcup_{k\in\mathcal K_{p,v}}\!\!\! \Sigma_{p,v,k} = \{1,\ldots,N\}, \Sigma_{p,v,k}\cap\Sigma_{p,v,k'}=\varnothing,\;k\neq k'\).

Define the spatial phase mismatch between the arrival direction and the
template direction as
\(\Delta\varphi \triangleq \frac{2\pi d}{\lambda_{\mathrm c}} \left( \sin\theta-\sin\theta_0 \right)\).
For an active branch \(k\in\mathcal K_{p,v}\), the normalized transmit-spatial
correlation over the antennas assigned to this branch is
\(\rho_{p,v,k}(\theta,\theta_0) \triangleq \frac{1}{M_{p,v,k}} \sum_{n\in\Sigma_{p,v,k}} \exp[j(n-1)\Delta\varphi]\).
After coherent channel synthesis, the subpulse-level waveform spatial factor
is the active-element-weighted combination
\(\rho_{p,v}(\theta,\theta_0) \triangleq \sum_{k\in\mathcal K_{p,v}} \frac{M_{p,v,k}}{N} \rho_{p,v,k}(\theta,\theta_0)\).
Substituting the branch correlation into the weighted combination and using the
partition property above gives
\begin{equation}
	\label{eq:appE_full_sum}
	\begin{aligned}
		\rho_{p,v}(\theta,\theta_0)
		&=
		\frac{1}{N}
		\sum_{k\in\mathcal K_{p,v}}
		\sum_{n\in\Sigma_{p,v,k}}
		\exp[j(n-1)\Delta\varphi]\\
		&=
		\frac{1}{N}
		\sum_{n=1}^{N}
		\exp[j(n-1)\Delta\varphi].
	\end{aligned}
\end{equation}
Therefore, the magnitude of the transmit-side SMC spatial factor is
\begin{equation}
	\label{eq:appE_A}
	A(\theta;\theta_0)
	\triangleq
	\left|
	\rho_{p,v}(\theta,\theta_0)
	\right|
	=
	\left|
	\frac{
		\operatorname{sinc}\!\left(\frac{N}{2}\Delta\varphi\right)
	}{
		\operatorname{sinc}\!\left(\frac{1}{2}\Delta\varphi\right)
	}
	\right|.
\end{equation}
This factor is independent of the particular frequency assignment because the
active branch sets form a partition of the full transmit aperture.

For colocated transmit and receive arrays with a receive beamformer matched to
\(\theta_0\), the normalized receive beamforming factor has the same magnitude:
\(\left| \frac{ b_{\mathrm R}(\theta) }{ b_{\mathrm R}(\theta_0) } \right| = A(\theta;\theta_0)\).
Hence, the normalized SMC-domain field response contains one transmit-side
factor and one receive-side factor. The corresponding power beampattern is
\begin{equation}
	\label{eq:appE_Gsmc}
	G_{\mathrm{SMC}}(\theta;\theta_0)
	=
	A^4(\theta;\theta_0)
	=
	\left|
	\frac{
		\operatorname{sinc}\!\left(\frac{N}{2}\Delta\varphi\right)
	}{
		\operatorname{sinc}\!\left(\frac{1}{2}\Delta\varphi\right)
	}
	\right|^4.
\end{equation}
This proves \eqref{eq:Gsmc_main}.

%% file: main.bbl
\begin{thebibliography}{10}
\providecommand{\url}[1]{#1}
\csname url@samestyle\endcsname
\providecommand{\newblock}{\relax}
\providecommand{\bibinfo}[2]{#2}
\providecommand{\BIBentrySTDinterwordspacing}{\spaceskip=0pt\relax}
\providecommand{\BIBentryALTinterwordstretchfactor}{4}
\providecommand{\BIBentryALTinterwordspacing}{\spaceskip=\fontdimen2\font plus
\BIBentryALTinterwordstretchfactor\fontdimen3\font minus
  \fontdimen4\font\relax}
\providecommand{\BIBforeignlanguage}[2]{{%
\expandafter\ifx\csname l@#1\endcsname\relax
\typeout{** WARNING: IEEEtran.bst: No hyphenation pattern has been}%
\typeout{** loaded for the language `#1'. Using the pattern for}%
\typeout{** the default language instead.}%
\else
\language=\csname l@#1\endcsname
\fi
#2}}
\providecommand{\BIBdecl}{\relax}
\BIBdecl

\bibitem{Liu2018TowardDFRC}
F.~Liu, L.~Zhou, C.~Masouros, A.~Li, W.~Luo, and A.~Petropulu, ``Toward
  dual-functional radar-communication systems: Optimal waveform design,''
  \emph{IEEE Trans. Signal Process.}, vol.~66, no.~16, pp. 4264--4279, Aug.
  2018.

\bibitem{Liu2022ISAC6G}
F.~Liu \emph{et~al.}, ``Integrated sensing and communications: Toward
  dual-functional wireless networks for {6G} and beyond,'' \emph{IEEE J. Sel.
  Areas Commun.}, vol.~40, no.~6, pp. 1728--1767, Jun. 2022.

\bibitem{Mao2022WaveformMillimeterTHz}
T.~Mao, J.~Chen, Q.~Wang, C.~Han, Z.~Wang, and G.~K. Karagiannidis, ``Waveform
  design for joint sensing and communications in millimeter-wave and low
  terahertz bands,'' \emph{IEEE Trans. Commun.}, vol.~70, no.~10, pp.
  7023--7039, Oct. 2022.

\bibitem{Zhang2021OverviewJCR}
J.~A. Zhang \emph{et~al.}, ``An overview of signal processing techniques for
  joint communication and radar sensing,'' \emph{IEEE J. Sel. Topics Signal
  Process.}, vol.~15, no.~6, pp. 1295--1315, Nov. 2021.

\bibitem{Liu2022FundamentalLimitsISAC}
A.~Liu \emph{et~al.}, ``A survey on fundamental limits of integrated sensing
  and communication,'' \emph{IEEE Commun. Surveys Tuts.}, vol.~24, no.~2, pp.
  994--1034, 2022.

\bibitem{Zhou2022ISACWaveformSurvey}
W.~Zhou, R.~Zhang, G.~Chen, and W.~Wu, ``Integrated sensing and communication
  waveform design: A survey,'' \emph{IEEE Open J. Commun. Soc.}, vol.~3, pp.
  1930--1949, 2022.

\bibitem{Blunt2010IntrapulseREC}
S.~D. Blunt, P.~Yatham, and J.~Stiles, ``Intrapulse radar-embedded
  communications,'' \emph{IEEE Trans. Aerosp. Electron. Syst.}, vol.~46, no.~3,
  pp. 1185--1200, Jul. 2010.

\bibitem{Ciuonzo2015IntrapulseMOO}
D.~Ciuonzo, A.~{De Maio}, G.~Foglia, and M.~Piezzo, ``Intrapulse radar-embedded
  communications via multiobjective optimization,'' \emph{IEEE Trans. Aerosp.
  Electron. Syst.}, vol.~51, no.~4, pp. 2960--2974, Oct. 2015.

\bibitem{Hassanien2016SidelobeWaveformDiversity}
A.~Hassanien, M.~G. Amin, Y.~D. Zhang, and F.~Ahmad, ``Dual-function
  radar-communications: Information embedding using sidelobe control and
  waveform diversity,'' \emph{IEEE Trans. Signal Process.}, vol.~64, no.~8, pp.
  2168--2181, Apr. 2016.

\bibitem{Yang2020DualUseAF}
J.~Yang, G.~Cui, X.~Yu, and L.~Kong, ``Dual-use signal design for radar and
  communication via ambiguity function sidelobe control,'' \emph{IEEE Trans.
  Veh. Technol.}, vol.~69, no.~9, pp. 9781--9794, Sep. 2020.

\bibitem{Sturm2011WaveformFusion}
C.~Sturm and W.~Wiesbeck, ``Waveform design and signal processing aspects for
  fusion of wireless communications and radar sensing,'' \emph{Proc. IEEE},
  vol.~99, no.~7, pp. 1236--1259, Jul. 2011.

\bibitem{Braun2014OFDMRadarAlgorithms}
M.~Braun, ``{OFDM} radar algorithms in mobile communication networks,'' Ph.D.
  dissertation, Karlsruhe Institute of Technology, Karlsruhe, Germany, 2014.

\bibitem{Cui2022NRWaveformISAC}
Y.~Cui, X.~Jing, and J.~Mu, ``Integrated sensing and communications via {5G}
  {NR} waveform: Performance analysis,'' in \emph{Proc. IEEE Int. Conf.
  Acoustics, Speech and Signal Processing ({ICASSP})}, May 2022, pp.
  8747--8751.

\bibitem{Xiao2022FullDuplexISAC}
Z.~Xiao and Y.~Zeng, ``Waveform design and performance analysis for full-duplex
  integrated sensing and communication,'' \emph{IEEE J. Sel. Areas Commun.},
  vol.~40, no.~6, pp. 1823--1837, Jun. 2022.

\bibitem{Li2007MIMORadarColocated}
J.~Li and P.~Stoica, ``{MIMO} radar with colocated antennas,'' \emph{IEEE
  Signal Process. Mag.}, vol.~24, no.~5, pp. 106--114, Sep. 2007.

\bibitem{Yu2022SpatioSpectral}
X.~Yu, X.~Yao, J.~Yang, L.~Zhang, L.~Kong, and G.~Cui, ``Integrated waveform
  design for {MIMO} radar and communication via spatio-spectral modulation,''
  \emph{IEEE Trans. Signal Process.}, vol.~70, pp. 2293--2305, 2022.

\bibitem{Yao2023IntrapulseIM}
X.~Yao, J.~Yang, K.~Xiong, X.~Yu, G.~Cui, and X.~Nie, ``Integrated signal
  design for {MIMO} {DFRC} with intrapulse index modulation,'' \emph{IEEE
  Trans. Aerosp. Electron. Syst.}, vol.~60, no.~2, pp. 1490--1504, Apr. 2024.

\bibitem{Yao2024NarrowbandOrthogonal}
X.~Yao, Y.~Wang, B.~Pan, X.~Yu, G.~Cui, and L.~Su, ``Integrated narrowband
  orthogonal signal set design for {MIMO} {DFRC} system,'' \emph{IEEE Trans.
  Veh. Technol.}, vol.~73, no.~10, pp. 14\,633--14\,646, Oct. 2024.

\bibitem{Chen2008FHMIMOAmbiguity}
C.-Y. Chen and P.~P. Vaidyanathan, ``{MIMO} radar ambiguity properties and
  optimization using frequency-hopping waveforms,'' \emph{IEEE Trans. Signal
  Process.}, vol.~56, no.~12, pp. 5926--5936, Dec. 2008.

\bibitem{Baxter2018FHCS}
W.~Baxter, E.~Aboutanios, and A.~Hassanien, ``Dual-function {MIMO}
  radar-communications via frequency-hopping code selection,'' in \emph{Proc.
  52nd Asilomar Conf. Signals, Systems, and Computers}, Pacific Grove, CA, USA,
  Oct. 2018, pp. 1126--1130.

\bibitem{Wu2021ChannelEstimation}
K.~Wu, J.~A. Zhang, X.~Huang, Y.~J. Guo, and R.~W. Heath Jr., ``Waveform design
  and accurate channel estimation for frequency-hopping {MIMO} radar-based
  communications,'' \emph{IEEE Trans. Commun.}, vol.~69, no.~2, pp. 1244--1258,
  Feb. 2021.

\bibitem{Baxter2022JointRadarComFH}
W.~Baxter, E.~Aboutanios, and A.~Hassanien, ``Joint radar and communications
  for frequency-hopped {MIMO} systems,'' \emph{IEEE Trans. Signal Process.},
  vol.~70, pp. 729--742, 2022.

\bibitem{Wu2022Overview}
K.~Wu, J.~A. Zhang, X.~Huang, and Y.~J. Guo, ``Frequency-hopping {MIMO}
  radar-based communications: An overview,'' \emph{IEEE Aerosp. Electron. Syst.
  Mag.}, vol.~37, no.~4, pp. 42--54, Apr. 2022.

\bibitem{Baxter2020DictionarySelection}
W.~Baxter, H.~Nosrati, and E.~Aboutanios, ``A study on the performance of
  symbol dictionary selection for the frequency hopped {DFRC} scheme,'' in
  \emph{Proc. IEEE Radar Conf. ({RadarConf})}, Florence, Italy, Sep. 2020, pp.
  1--6.

\bibitem{Aboutanios2023ImprovedFHCS}
E.~Aboutanios, W.~Baxter, and Y.~D. Zhang, ``Improved implementation of the
  frequency hopped code selection {DFRC} scheme,'' in \emph{Proc. IEEE Radar
  Conf. ({RadarConf})}, San Antonio, TX, USA, May 2023, pp. 1--6.

\bibitem{Wu2021Reliable}
K.~Wu, J.~A. Zhang, X.~Huang, Y.~J. Guo, and J.~Yuan, ``Reliable
  frequency-hopping {MIMO} radar-based communications with multi-antenna
  receiver,'' \emph{IEEE Trans. Commun.}, vol.~69, no.~8, pp. 5502--5513, Aug.
  2021.

\bibitem{Wu2022Secure}
K.~Wu, J.~A. Zhang, X.~Huang, and Y.~J. Guo, ``Integrating secure
  communications into frequency hopping {MIMO} radar with improved data rate,''
  \emph{IEEE Trans. Wireless Commun.}, vol.~21, no.~7, pp. 5392--5405, Jul.
  2022.

\bibitem{Eedara2019DPSK}
I.~P. Eedara and M.~G. Amin, ``Dual function {FH} {MIMO} radar system with
  {DPSK} signal embedding,'' in \emph{Proc. 27th European Signal Processing
  Conf. (EUSIPCO)}, A {Coru\~na}, Spain, Sep. 2019, pp. 1--5.

\bibitem{Eedara2022CSK}
I.~P. Eedara, M.~G. Amin, A.~Hoorfar, and B.~K. Chalise, ``Dual-function
  frequency-hopping {MIMO} radar system with {CSK} signaling,'' \emph{IEEE
  Trans. Aerosp. Electron. Syst.}, vol.~58, no.~3, pp. 1501--1513, Jun. 2022.

\bibitem{Xu2023HIM}
J.~Xu, X.~Wang, E.~Aboutanios, and G.~Cui, ``Hybrid index modulation for
  dual-functional radar communications systems,'' \emph{IEEE Trans. Veh.
  Technol.}, vol.~72, no.~3, pp. 3186--3200, Mar. 2023.

\bibitem{Liu2025CRM}
J.~Liu, W.~Jiang, W.~Yang, T.~Su, and J.~Chen, ``Enhancing the communication
  bandwidth of {FH}-{MIMO} {DFRC} systems through constellation rotation
  modulation,'' \emph{Remote Sens.}, vol.~17, no.~6, Mar. 2025, {Art. no.}
  1058.

\bibitem{Wang2020PhaseModulatedFHMIMO}
X.~Wang and A.~Hassanien, ``Phase modulated communications embedded in
  correlated {FH}-{MIMO} radar waveforms,'' in \emph{Proc. IEEE Radar Conf.
  ({RadarConf})}, Florence, Italy, Sep. 2020, pp. 1--6.

\bibitem{Eedara2021TargetDetection}
I.~P. Eedara, M.~G. Amin, and G.~A. Fabrizio, ``Target detection in frequency
  hopping {MIMO} dual-function radar-communication systems,'' in \emph{Proc.
  IEEE Int. Conf. Acoustics, Speech and Signal Processing ({ICASSP})}, Toronto,
  ON, Canada, Jun. 2021, pp. 8458--8462.

\bibitem{Chen2025Processing}
W.~Chen, Z.~Xie, J.~Hu, Y.~Zhang, and Z.~Chen, ``Waveform and processing
  algorithm design for frequency-hopping {MIMO} radar-communication integration
  utilizing multi-parameter modulation,'' \emph{IEEE Trans. Commun.}, vol.~73,
  no.~11, pp. 12\,114--12\,130, Nov. 2025.

\bibitem{Eedara2020GA}
I.~P. Eedara, M.~G. Amin, and A.~Hoorfar, ``Optimum code design using genetic
  algorithm in frequency hopping dual function {MIMO} radar communication
  systems,'' in \emph{Proc. IEEE Radar Conf. ({RadarConf})}, Florence, Italy,
  Sep. 2020, pp. 1--6.

\bibitem{Eedara2020Clutter}
I.~P. Eedara, M.~G. Amin, and A.~Hassanien, ``Controlling clutter modulation in
  frequency hopping {MIMO} dual-function radar communication systems,'' in
  \emph{Proc. IEEE Int. Radar Conf. ({RADAR})}, Washington, DC, USA, Apr. 2020,
  pp. 466--471.

\bibitem{Alabaster2012}
C.~Alabaster, \emph{Pulse Doppler Radar: Principles, Technology,
  Applications}.\hskip 1em plus 0.5em minus 0.4em\relax Stevenage, U.K.: IET,
  2012.

\bibitem{Barton2004}
D.~K. Barton, \emph{Radar System Analysis and Modeling}.\hskip 1em plus 0.5em
  minus 0.4em\relax Norwood, MA, USA: Artech House, 2004.

\bibitem{Wang2025HybridFHPC}
Y.~Wang, X.~Yu, J.~Yang, G.~Cui, and Z.~Cui, ``Design of {ISAC} system based on
  hybrid frequency-hopping and phase-continuous signal,'' in \emph{Proc. IEEE
  Radar Conf. ({RadarConf})}, 2025, pp. 882--887.

\bibitem{Li2024InterPulseAgile}
Y.~Li \emph{et~al.}, ``Inter-pulse amplitude-frequency-phase agile design for
  cognitive radar,'' \emph{Signal Process.}, vol. 223, Oct. 2024, {Art. no.}
  109557.

\bibitem{Lellouch2015SteppedOFDM}
G.~Lellouch, A.~K. Mishra, and M.~Inggs, ``Stepped {OFDM} radar technique to
  resolve range and {Doppler} simultaneously,'' \emph{IEEE Trans. Aerosp.
  Electron. Syst.}, vol.~51, no.~2, pp. 937--950, Apr. 2015.

\bibitem{Anderson1986DigitalPhaseModulation}
J.~B. Anderson, T.~Aulin, and C.-E. Sundberg, \emph{Digital Phase
  Modulation}.\hskip 1em plus 0.5em minus 0.4em\relax New York, NY, USA: Plenum
  Press, 1986.

\bibitem{Sundberg1986CPM}
C.-E. Sundberg, ``Continuous phase modulation,'' \emph{IEEE Commun. Mag.},
  vol.~24, no.~4, pp. 25--38, Apr. 1986.

\bibitem{McCormick2017WidebandMIMO}
P.~M. McCormick, S.~D. Blunt, and J.~G. Metcalf, ``Wideband {MIMO}
  frequency-modulated emission design with space-frequency nulling,''
  \emph{IEEE J. Sel. Topics Signal Process.}, vol.~11, no.~2, pp. 363--378,
  Mar. 2017.

\bibitem{Scharf1994MatchedSubspace}
L.~L. Scharf and B.~Friedlander, ``Matched subspace detectors,'' \emph{IEEE
  Trans. Signal Process.}, vol.~42, no.~8, pp. 2146--2157, Aug. 1994.

\bibitem{Graham1994ConcreteMathematics}
R.~L. Graham, D.~E. Knuth, and O.~Patashnik, \emph{Concrete Mathematics: A
  Foundation for Computer Science}, 2nd~ed.\hskip 1em plus 0.5em minus
  0.4em\relax Reading, MA, USA: Addison-Wesley, 1994.

\bibitem{Shoup2008NumberTheory}
V.~Shoup, \emph{A Computational Introduction to Number Theory and Algebra},
  2nd~ed.\hskip 1em plus 0.5em minus 0.4em\relax Cambridge, U.K.: Cambridge
  University Press, 2008.

\end{thebibliography}
